\documentclass[aps,amssymb,amsmath,prc,twocolumn,
showpacs,preprintnumbers,superscriptaddress,nofootinbib,floatfix]{revtex4-1}

\usepackage{graphicx}
\usepackage{bm}
\usepackage{mathrsfs}
\usepackage{latexsym}
\usepackage{color}
\usepackage[normalem]{ulem} 
\usepackage{dcolumn}
\usepackage{xspace}
\usepackage{isotope}
\usepackage[colorlinks=true,citecolor=blue,urlcolor=blue]{hyperref}
\usepackage[usenames,dvipsnames]{xcolor}
\usepackage[caption=false]{subfig}
\usepackage{hyperref}
\graphicspath{{./figures/}}

\newcommand{\beq}{\begin{equation}}
\newcommand{\eeq}{\end{equation}}
\newcommand{\ba}{\begin{array}}
\newcommand{\ea}{\end{array}}
\newcommand{\bea}{\begin{eqnarray}}
\newcommand{\eea}{\end{eqnarray}}
\newcommand{\bc}{\begin{center}}
\newcommand{\ec}{\end{center}}

\newcommand{\De}{\Delta}
\newcommand{\ep}{\varepsilon}

\newcommand{\La}{\Lambda}

\newcommand{\nb}{n_{\rm B}}
\newcommand{\nsat}{n_{\rm sat}}
\newcommand{\esat}{\ep_{\rm sat}}
\newcommand{\pmat}{P_{\rm m}}
\newcommand{\nmat}{n_{\rm m}}
\newcommand{\emat}{\varepsilon_{\rm m}}
\newcommand{\Msolar}{{\rm M}_{\odot}}

\newcommand{\ncent}{n_{\rm cent}}

\newcommand{\csq}{c^2_{s}}
\newcommand{\csqmat}{c^2_{s, {\rm match}}}

\newcommand{\Mmax}{M_{\rm max}}

\newcommand{\Rmax}{R_{\Mmax}}

\newcommand{\nmax}{n_{\rm max}}

\newcommand{\Rtyp}{R_{1.4}}
\newcommand{\Rtwo}{R_{2.0}}

\newcommand{\Mchirp}{{\mathcal M}}

\newcommand\Eqn[1]{Eq.~(\ref{#1})}  

\newcommand{\etal}{\textit{et~al.}\xspace}

\newcommand{\km}{\, \text{km}}

\newcommand{\fmiq}{\, \text{fm}^{-3}}
\newcommand{\keV}{\, \text{keV}}
\newcommand{\MeV}{\, \text{MeV}}

\newcommand{\NNLO}{\ensuremath{{\rm N}{}^2{\rm LO}}\xspace}
\newcommand{\NNNLO}{\ensuremath{{\rm N}{}^3{\rm LO}}\xspace}

\newcommand{\chiEFT}{$\chi$EFT\xspace}
\newcommand{\normal}{\mathcal{N}}

\newcommand{\kf}{\ensuremath{k_{\scriptscriptstyle\textrm{F}}}}
\DeclareMathOperator{\cov}{cov}

\begin{document}
\preprint{INT-PUB-20-035, N3AS-20-004}
\title{Limiting masses and radii of neutron stars and 
their implications}

\author{Christian~Drischler}
\email{cdrischler@berkeley.edu}
\affiliation{Department of Physics, University of California, Berkeley, CA~94720, USA}
\affiliation{Nuclear Science Division, Lawrence Berkeley National Laboratory,
Berkeley, CA~94720, USA}

\author{Sophia~Han}
\email{sjhan@berkeley.edu}
\affiliation{Department of Physics, University of California, Berkeley, CA~94720, USA}
\affiliation{Institute for Nuclear Theory, University of Washington, Seattle, WA~98195, USA}
\affiliation{Department of Physics and Astronomy, Ohio University,
Athens, OH~45701, USA}

\author{James~M.~Lattimer}
\email{james.lattimer@stonybrook.edu}
\affiliation{Department of Physics and Astronomy, Stony Brook University,
Stony Brook, NY~11794, USA}

\author{Madappa~Prakash}
\email{prakash@ohio.edu}
\affiliation{Department of Physics and Astronomy, Ohio University,
Athens, OH~45701, USA}

\author{Sanjay~Reddy}
\email{sareddy@uw.edu}
\affiliation{Institute for Nuclear Theory, University of Washington, Seattle, WA~98195, USA}

\author{Tianqi~Zhao}
\email{tianqi.zhao@stonybrook.edu}
\affiliation{Department of Physics and Astronomy, Stony Brook University, 
Stony Brook, NY~11794, USA}

\date{April 20, 2021}

\begin{abstract}

We combine the equation of state of dense matter up to twice nuclear saturation density $n_{\rm sat}$ obtained using chiral effective field theory ($\chi$EFT), and recent observations of neutron stars to gain insights about the high-density matter encountered in their cores. A key element in our study is the recent Bayesian analysis of correlated EFT truncation errors based on order-by-order calculations up to next-to-next-to-next-to-leading order in the $\chi$EFT expansion. We refine the bounds on the maximum mass imposed by causality at high densities, and provide stringent limits on the maximum and minimum radii of $\sim1.4\,{\rm M}_{\odot}$ and $\sim2.0\,{\rm M}_{\odot}$ stars. Including $\chi$EFT predictions from $n_{\rm sat}$ to $2\,n_{\rm sat}$ reduces the permitted ranges of the radius of a $1.4\,{\rm M}_{\odot}$ star,  $R_{1.4}$, by $\sim3.5\, \text{km}$. If observations indicate $R_{1.4}<11.2\, \text{km}$, our study implies that  either the squared speed of sound  $c^2_{s}>1/2$  for densities above $2\,n_{\rm sat}$,  or that $\chi$EFT breaks down below $2\,n_{\rm sat}$. 
We also comment on the nature of the secondary compact object in GW190814 with mass $\simeq 2.6\,{\rm M}_{\odot}$,  and discuss the implications of massive neutron stars $>2.1 \,{\rm M}_{\odot}\,(2.6\,{\rm M}_{\odot})$ in future radio and gravitational-wave searches. Some form of strongly interacting  matter with $c^2_{s}>0.35\, (0.55)$ must be realized in the cores of such massive neutron stars. In the absence of phase transitions below $2\,n_{\rm sat}$, the small tidal deformability inferred from GW170817 lends support for the relatively small pressure predicted by $\chi$EFT for the baryon density $n_{\rm B}$ in the range $1-2\,n_{\rm sat}$. Together they imply that the rapid stiffening required to support a high maximum mass should  occur only when $n_{\rm B} \gtrsim 1.5-1.8\,n_{\rm sat}$.

\end{abstract}


\maketitle

\section{Introduction} 

The maximum mass, $\Mmax$, and radii of neutron stars (NSs) are related to each other by the equation of state (EOS) of dense matter and both can be accessed by observations. 
Primary constraints on $\Mmax$ come from observations and have a number of astronomical and physical implications. $\Mmax$ is predominately determined by the EOS at densities higher than three times nuclear saturation density, $\nsat\simeq0.16\fmiq$~\cite{Gandolfi:2011xu}, and is therefore a probe of the nature of high-density matter. 
Pinning down $\Mmax$
enables the exploration of the phases of cold and dense 
matter in the strongly coupled region of 
quantum chromodynamics (QCD) as well as the determination of the pressure vs energy density relation 
(or the EOS) of such phases. 
The radii of canonical NSs with masses $\simeq 1.4~\Msolar$, on the other hand, are largely determined by the EOS at densities less than $3\,\nsat$~\cite{Lattimer:2000nx}.

$\Mmax$ also fixes the minimum mass of a stellar mass $\mathcal{O} (\Msolar)$ black hole (BH). 
It is therefore a crucial factor in determining the final fate of 
core-collapse supernovae and 
binary neutron star (BNS) mergers. 
In core-collapse supernovae, the formation of a BH will depend on the amount of fall-back matter and will be sensitive to the nature of the progenitor and neutrino emission after the initial formation of a proto-neutron star. In BNS mergers, the formation of a BH depends on the total inspiralling mass, mass ejection, and the extent of rotational and magnetohydrodynamic support~\cite{Margalit:2017dij,Shibata:2019ctb}.  
Now that at least a few mergers involving NS have been detected through  gravitational-wave (GW) radiation, and many more are anticipated in the near future, improved constraints on $\Mmax$ will become available. 
As the high-frequency capabilities of GW detectors are improved, the detection of post-merger radiation will profoundly influence our knowledge of $\Mmax$. Already,  
knowledge of $\Mmax$ would determine the nature of
the components of the recently 
observed 
mergers GW190425 and GW190814, 
both of which show indications of having a component with a mass larger than $2\,\Msolar$ that either could be a heavy NS or a light BH.
If concomitant electromagnetic (EM)
signals are also detected from future GW events, as they were in the BNS merger GW170817~\cite{LIGO:2017qsa,Abbott:2018wiz,Abbott:2018exr}, 
additional information about $\Mmax$ becomes available~\cite{Margalit:2017dij,Shibata:2019ctb}. 

On the theoretical front, 
$\Mmax$ plays a crucial role in determining both the minimum and maximum radius 
as a function of 
the NS mass $M$. 
Therefore, 
besides the important contributions from radio and X-ray binary pulsar observations that have accurately measured several NS masses and provided a lower bound $\Mmax\gtrsim2\,\Msolar$~\cite{Demorest:2010bx,Antoniadis:2013pzd,Fonseca:2016tux,Arzoumanian:2017puf,Cromartie:2019kug,Romani:2021xmb}, 
GW and X-ray data that can simultaneously determine NS masses and radii offer important constraints. 
So far, the radii inferred from X-ray observations (see Ref.~\cite{Ozel:2016oaf} for a review) of quiescent low-mass X-ray binaries (QLMXBs)~\cite{Rybicki:2005id}, photospheric radius expansion bursts (PREs)~\cite{Ozel:2008kb}, and pulse-profiles from rotation-powered millisecond pulsars~\cite{Bogdanov:2006zd}, 
together with 
the first GW detection of the BNS merger GW170817~\cite{LIGO:2017qsa,Abbott:2018wiz}, have 
mostly 
been of NSs with 
canonical 
masses 
around $1.4\,\Msolar$. 
Consequently, the \emph{Neutron Star Interior Composition ExploreR} ({NICER}) proposal~\cite{Miller:2016kae} to measure the radii of relatively massive NSs such as PSR~J1614-2230 ($M\simeq 1.91\,\Msolar$~\cite{Demorest:2010bx,Fonseca:2016tux,Arzoumanian:2017puf}) and PSR~J0740+6620 ($M\simeq2.14 \,\Msolar$~\cite{Cromartie:2019kug})
is 
of considerable interest. 
The same is true of future radio observations using the 
\emph{Square Kilometre Array} (SKA) telescope~\cite{Watts:2014tja}, etc. 
from binary pulsars 
that could reveal 
even more 
massive NSs.

The purpose of this paper is to explore the interplay between $\Mmax$ and NS radii and to confront theoretical expectations with currently available observational constraints. 
An earlier study~\cite{Lattimer:2000nx} 
showed that the radii of $\simeq 1.4\,\Msolar$ NSs 
are 
strongly correlated with the 
pressure 
of matter in the density range $1-3\,\nsat$. 
In the important regime $\lesssim 2\,\nsat$, chiral effective field theory (\chiEFT) with pion and nucleon
degrees of freedom~\cite{Epelbaum:2008ga, Machleidt:2011zz, Hammer:2019poc, Tews:2020hgp} has become the dominant microscopic approach to describing nuclear interactions.
\chiEFT has enabled significant progress in predicting the EOS of infinite nuclear matter and the structure of neutron stars with quantifiable theoretical uncertainties (see Refs.~\cite{Hebeler:2015hla, Drischler:2019xuo, Sammarruca:2019ncy,Drischler:2021kxf} for recent reviews). 
An important step toward the full uncertainty quantification of the EOS has been achieved recently. 
The \emph{Bayesian Uncertainty Quantification: Errors in Your EFT} (BUQEYE) collaboration~\cite{BUQEYEgithub} has introduced a Bayesian framework~\cite{Drischler:2020hwi,Drischler:2020yad} for quantifying and propagating correlated EFT truncation errors in infinite-matter calculations using Gaussian Processes (GPs). They also conducted a statistical analysis of the zero-temperature EOS based on \chiEFT nucleon-nucleon (NN) and three-nucleon (3N) interactions and inferred posterior distributions for nuclear saturation properties as well as key quantities for neutron stars, including the nuclear symmetry energy and its density dependence.
This study was motivated by recent advances in many-body perturbation theory (MBPT)~\cite{Drischler:2017wtt} that have enabled improved \chiEFT predictions of the pure neutron matter (PNM) EOS and first order-by-order calculations in symmetric nuclear matter (SNM) up to next-to-next-to-next-to-leading order (\NNNLO) in the chiral expansion~\cite{Drischler:2017wtt, Leonhardt:2019fua,Drischler:2020hwi}.

In this paper, we use BUQEYE's analysis of the EOS in the limits of PNM and SNM 
at baryon densities $\nb \leq 2\,\nsat$ 
to construct the EOS of charge neutral and beta-stable neutron-star matter (NSM). This is coupled to a standard NS crust for $\nb \lesssim0.5\,\nsat$ and extrapolations for $\nb \gtrsim2.0\,\nsat$ to assess the overall impact on NS structure. 
One goal of this study is to address quantitatively the extent to which EOS knowledge
at 
$\sim 2.0\,\nsat$ can inform us about the NS maximum mass, and how it can be combined with observations of massive NSs to constrain the properties of matter encountered at the highest densities in their cores. Another goal is to derive model-independent bounds on the radii of NSs with masses in the range $1-2\,\Msolar$. 

As the squared speed of sound $\csq$ reflects the stiffness of the EOS, we probe both maximum and minimum radius bounds by 
matching the \NNNLO results, including possible extrapolations up to $3\,\nsat$, with a constant sound speed beyond a matching density $\nmat$.  
The existence of nuclei, observations of accreting NSs that implicate the presence of neutron-rich nuclei in the NS crust, and heavy ion collisions (HICs) 
at intermediate energies 
together provide compelling circumstantial evidence to indicate that $\nmat > \nsat$, and in this work we consider 
$\nmat = 1-3 \,\nsat$. 
The use of the maximally stiff EOS with $\csq=1$ (the causal limit) for $\nb>\nmat$ establishes firm upper bounds both on $\Mmax$ and the radius as a function of mass. In addition, we also consider energy density discontinuities at $\nmat$ to refine minimum bounds on radii as functions of mass for specified values of $\Mmax$. We also explore models with 
smaller $\csq$ at high density to ascertain maximum possible sound speeds from values of $\Mmax$ and mass-radius ($M$--$R$) observations.

The discovery of a massive secondary compact object with mass $\sim 2.6\,\Msolar$ through GW observations of the binary merger GW190814 generated a flurry of articles addressing if this object can be a NS, and, 
if so, 
its possible implications~\cite{Tan:2020ics,Lim:2020zvx,Tews:2020ylw,Essick:2020ghc,Tsokaros:2020hli,Fattoyev:2020cws,Godzieba:2020tjn,Kanakis-Pegios:2020jnf}. 
Our results complement earlier studies, but go beyond in several aspects. Most significantly,
\begin{itemize}
\item[(i)]  we consistently include statistically meaningful EFT truncation errors in the EOS of
NSM up to \NNNLO, 
and determine 
its range of applicability, to provide a framework for constraining 
$\Mmax$ and NS radii, 

\item [(ii)] we identify correlations of NS radii and tidal deformabilities with $\Mmax$, together with 
their 
possible 
implications for the EOS at 
$\nb\gtrsim2\,\nsat$, and

\item[(iii)] we 
show how 
these correlations 
and 
future observations can tighten current bounds on NS masses and radii.
\end{itemize}

This paper is organized as follows. 
Section~\ref{sec:EOSs} contains details of the various EOSs used along with the rationale for their choice. Our results and their discussion in light of the current observational constraints and possible future findings are presented in Sec.~\ref{sec:Results}.
An overall discussion and  comparison with pertinent recent works are contained in Sec.~\ref{sec:Discs}. Our concluding remarks are given in Sec.~\ref{sec:Concs}. 
Appendix \ref{sec:Bounds} examines the most conservative bounds and the scaling relations for
the masses and radii of NSs imposed by causality.  
The current shortcomings and prognosis for future improvements to \chiEFT 
are discussed in Appendix \ref{app:chiral_interactions}.
Appendix \ref{sec:Sensitivity} quantifies the density ranges for which $R_{1.4}$ and $R_{2.0}$, the radii of $1.4\,\Msolar$ and $2.0\,\Msolar$ stars, respectively, and the neutron star maximum mass $M_{\rm max}$ are most sensitive.
We use natural units in which $\hbar=c=1$ unless explicitly specified.

\section{Construction of the EOS}
\label{sec:EOSs}
\subsection{General considerations}

Since the pressure-energy density relation, 
which we call the EOS, completely determines the neutron star $M$--$R$ relation through the general relativistic TOV equations, bounds of the allowed $M$--$R$ space are determined by assumptions concerning the EOS.  From the perspective of this paper, the three most important regions
for the EOS of a NS are the crust, outer core, and inner core. 
The EOS up to the outer core-crust boundary at $n_{cc}\approx 0.5\,\nsat$ is generally considered to be well-understood~\cite{Negele:1971vb,Baym:1971pw}. Because nucleons contribute $\lesssim 10\%$ to the crust pressure, uncertainties in the NN potential only 
weakly propagate into the crust EOS\@. The proton fraction $x$ in the uniform nucleonic matter at densities higher than $n_{cc}$ in the outer core is relatively small, so that the EOS in the vicinity of 
$\nsat$ is dominated by that of PNM\@. The admixture of protons and leptons produces small corrections, which are effectively minimized because of the requirement that NSM be in beta equilibrium; that is, the total energy is minimized with respect to $x$. 

A causal maximum radius bound $R_{\rm max,c}(M)$, 
as detailed in Appendix~\ref{sec:Bounds}, 
can be obtained by assuming a causal EOS at densities greater than that of a fiducial density $n_0$, generally greater than that of the core-crust boundary, while  below $n_0$, the pressure and energy density are taken to be zero. $R_{\rm max,c}$ will depend on the values of the associated fiducial energy density, $\ep_0$. 
This calculation explicitly ignores the existence of a crust. Appendix~\ref{sec:Bounds} also highlights the important role $\Mmax$ plays in determining bounds on the radii of neutron stars. 
In the \sout{most} extreme case, in which only causality is assumed with the EOS $\ep=\ep_0+P$, absolute upper bounds on $\Mmax\simeq4.09\,\Msolar$ and $\Rmax\simeq17.1 \km$ exist as long as $\ep_0>\esat$ (see \Eqn{eq:mmax0} and \Eqn{eq:rmax0}). 
Firm lower bounds on $R_{\rm min}(M)$ and $\Rmax$ that scale with $\Mmax$ can also be established.
For the case that $\Mmax=2.0\,\Msolar$, $R_{\rm min}(1.4\,\Msolar)=8.2$ km and $\Rmax=8.4$ km. 

However, there are no observations that indicate the absence of a crust.  More realistic bounds to the allowed $M$--$R$ space (as well as for other relations such as $M$--$\Lambda$ for the tidal deformability or $M$--$\bar I$ for the moment of inertia) are obtained by including the presence of the neutron star crust and also imposing theoretical limits to the properties of neutron-rich matter in the outer core up to a matching density $\nmat$, with associated energy density $\emat$ and pressure $\pmat$ (which replace $\ep_0$ and $P_0=0$, respectively).  Above the matching density, in now what is effectively the inner core, a constant sound speed EOS is assumed, and the maximum radius bounds occur when this sound speed is the speed of light.

Initially, we will explore radius bounds assuming the validity of theoretical studies up to the transition density $\nmat=2.0\,\nsat$ and imposing the causal EOS at higher densities.   If the causal EOS is imposed exactly at $\nmat$, one obtains maximum radius contours $R_{\rm max}(M)$ and the greatest value for $M_{\rm max}$.  The artificial introduction of a first-order phase transition between the two densities $\nmat$ and $n_u>\nmat$ with the imposition of the causal EOS for $\nb>n_u$, on the other hand, results in a smaller value of $M_{\rm max}$ but a minimum radius contour $R_{\rm min}(M)$ unique to that $M_{\rm max}$ (or $n_u$).  We will also explore how $R_{\rm min}$ and $R_{\rm max}$ change if the value of $\nmat$ is changed, or if the EOS above $\nmat$ or $n_u$ is assumed to have a subluminal sound speed.

\subsection{The EOS 
of the outer 
core 
}

\label{sec:eos_nsm}

\begin{figure*}[htbp]
\begin{center}
\includegraphics[width=\textwidth]{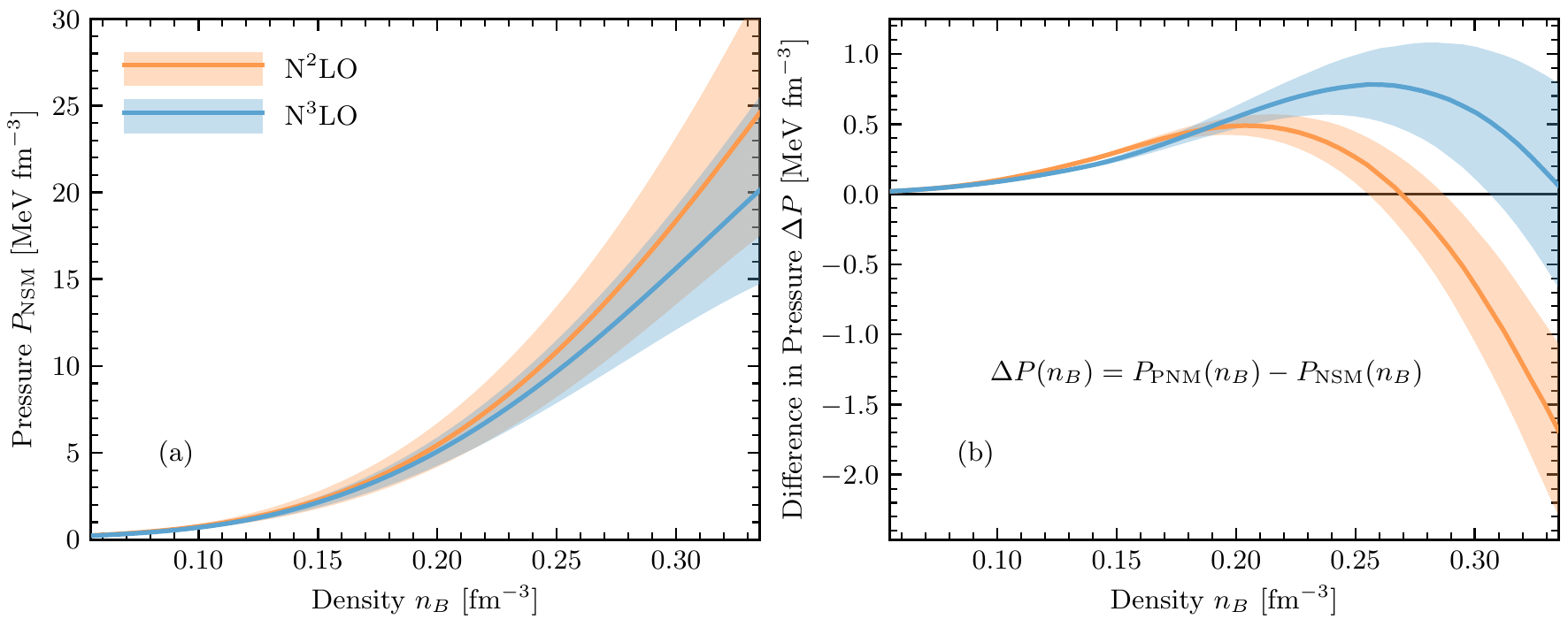}
\end{center}
\caption{Panel (a): pressure of neutron-star matter (NSM) in the outer core 
as a function of the 
baryon number 
density at \NNLO (orange-shaded band) and \NNNLO (blue-shaded band) in the chiral expansion; panel (b): differences of the PNM and NSM 
in the outer core with the same notation. 
Uncertainty bands depict $1\sigma$ confidence regions.}
\label{fig:P_nB-NSM}
\end{figure*}

To construct the EOS of charge-neutral, beta-equilibrated NSM in the outer core between $n_{cc}$
and $\sim 2.0\,\nsat$, we use the standard approximation of keeping only the
quadratic term in the nuclear energy expanded in the isospin asymmetry parameter
$\beta = 1-2x$, where $x=n_p/\nb$ is the proton fraction, and $n_p$ the proton
density. The total energy per baryon of NSM is then
\beq 
\label{eq:quad_exp}
    E_{\rm NSM} = E_{\rm PNM} (1-2x)^2 + E_{\rm SNM} \, 4x (1-x) +E_e+E_\mu ,
\eeq
where $E_{\rm PNM}$ and $ E_{\rm SNM}$ ($E_e$ and $E_\mu$) are the energies per baryon of PNM and SNM (electrons and muons),
respectively. Microscopic calculations of asymmetric matter based on chiral
NN and 3N interactions at $\nb \lesssim \nsat$
have confirmed that the quadratic expansion~\Eqn{eq:quad_exp} is a reasonable
approximation of the full isospin dependence of the
EOS~\cite{Drischler:2013iza,Drischler:2015eba,Kaiser:2015qia,Wellenhofer:2016lnl,Somasundaram:2020chb}. Beta equilibrium follows then from the condition that the total charge-neutral energy be minimized with
respect to $x$, i.e.,
\beq 
\label{eq:NSM_min}
\frac{\partial E_{\rm NSM}}{\partial x} = 0 ,
\eeq
or in terms of the associated chemical potentials
\beq
\mu_n-\mu_p = 4 E_{\rm sym} (1-2x) =\mu_e=\mu_\mu .
\eeq

Using the Jupyter notebooks~\cite{BUQEYEgithub} provided by the BUQEYE collaboration we extract the mean values, standard deviations (encoding the EFT truncation errors), and correlation information of the energy per particle, pressure and speed of sound in PNM and SNM, and also the symmetry energy. These data sets form the microscopic input of our interpolation to NSM.

Specifically, BUQEYE's EFT truncation error analysis~\cite{Drischler:2020hwi, Drischler:2020yad} is based on recent order-by-order MBPT calculations in PNM and SNM with chiral NN and 3N interactions up to \NNNLO~\cite{Drischler:2017wtt, Leonhardt:2019fua,Drischler:2020hwi}. More details on the underlying nuclear interactions can be found in Appendix~\ref{app:chiral_interactions}. The range in density covers $\nb = 0.05-0.34 \fmiq$. These calculations significantly improved previous MBPT studies in PNM at \NNNLO~\cite{Tews:2012fj,Kruger:2013kua,Drischler:2016djf}, and assessed, for the first time, the SNM EOS with NN and 3N interactions order-by-order up to \NNNLO. The high-order MBPT calculations were performed by the novel Monte Carlo framework introduced in Ref.~\cite{Drischler:2017wtt}, which enables MBPT calculations of the EOS with controlled many-body uncertainties for these \chiEFT interactions.
The statistical analysis indicates that the EFT truncation error is strongly correlated. In other words, perturbing the EOS at one point in the density (or the proton fraction) perturbs neighboring points as well.  In general, the range of these correlations, called the correlation length, depends on the density
and the underlying nuclear interactions.
Chiral 3N forces make for important contributions to the EOS in PNM and SNM at $\nb \gtrsim \nsat$, and  typically have a markedly different density-dependence than NN contributions.  The correlation lengths inferred are comparable to the $\kf$ associated with $\nsat$ in PNM and SNM, respectively.  Without including these correlations, uncertainties in derived quantities of the EOS, such as the nuclear symmetry energy, can be overestimated. 

Our approach considers correlations between the EOSs in PNM and SNM explicitly, neglecting correlations in density.\footnote{Such correlations could be implemented in future work by directly sampling from the GPs, which the BUQEYE collaboration uses to model the correlated EFT truncation errors.}  The extracted observables are then given by independent normal distributions sampled on a fine grid in density using the GPs; e.g.,
\begin{align}
E_{\rm PNM} &\sim \normal\left(\mu_{\rm PNM}, \sigma_{\rm PNM}^2\right),\\
E_{\rm SNM} &\sim \normal\left(\mu_{\rm SNM}, \sigma_{\rm SNM}^2\right).
\end{align}
The nuclear symmetry energy is defined as
\beq
\label{eq:def_Esym}
E_{\rm sym} = E_{\rm PNM} - E_{\rm SNM} \sim \normal\left(\mu_{\rm NSM},
\sigma_{\rm NSM}^2\right)  ,
\eeq
and, hence, has mean and variance (see, e.g.,~Ref.~\cite{bevington2003}):
\begin{align}
\mu_{\rm sym} &= \mu_{\rm PNM}- \mu_{\rm SNM}, \\
\begin{split}
\sigma_{\rm sym}^2 &= \sigma_{\rm PNM}^2+\sigma_{\rm SNM}^2
-2\rho \sigma_{\rm PNM} \sigma_{\rm SNM} , \label{eq:sigma_sym}
\end{split}
\end{align}
where $\rho$ is the correlation coefficient between the energies per particle in PNM and SNM. For subsequent discussion, we introduce 
here also 
the usual 
parameters $S_v$ and $L$ in the density expansion of the nuclear symmetry energy~\Eqn{eq:def_Esym}, 
\beq 
\label{eq:Esym_expanded}
E_{\rm sym}  = S_v + \frac{L}{3} \left( \frac{\nb-\nsat}{\nsat} \right) + \ldots . 
\eeq

The correlation between the coefficients in the \chiEFT expansions for the PNM and SNM energy per particle was quantified to be $\rho^* = 0.934$, corresponding to \emph{very strong} correlations~\cite{Evans1996,Asuero:2006abc}. A detailed discussion can be found in Sec.~IV~A of Ref.~\cite{Drischler:2020yad}. We have checked that $\rho \simeq \rho^*$ by comparing $E_{\rm sym}$ against the values obtained in Ref.~\cite{Drischler:2020yad}: the maximum deviation between the mean values of two approaches is $37\keV$ ($340 \keV$ for its $\pm1\sigma$ bounds) at the highest density, $\nb=0.34\fmiq$, which is negligible compared to the overall EFT truncation error at that density.

We also found that numerical integration of the pressure of PNM and SNM agreed well with the energy found in the GP approach, the maximum deviation of the mean values being $3 \keV$ and $1 \keV$ for PNM and SNM, respectively ($290 \keV$ and $500 \keV$ for their respective $\pm1\sigma$ bounds) at the highest density. 
There are mainly two related reasons why finite differencing for the pressure, discrete integration for the energy, and subtraction for the symmetry energy, works so well. 
First, the correlation length of the EOS is much longer than the length scale used for finite differencing. That means numerical differentiation follows closely the curves $\mu \pm \sigma$, which are two realizations of the underlying GP\@.
Secondly, the raw EOS data has already been preprocessed by BUQEYE's truncation error model. Numerical noise
from the many-body method has been smoothed out, and the EOS 
has been sampled on a fine grid in density using the GP interpolant. This underlines that GP interpolants are efficient tools for analyzing \chiEFT calculations of the EOS. 


 Propagating the EFT uncertainties
to $E_{\rm NSM}$ associated with \Eqn{eq:quad_exp} is straightforward
because of the condition~\eqref{eq:NSM_min}. We obtain
\bea
\sigma_{E_{\rm NSM}}^2 &=&
\left(\frac {\partial E_{\rm NSM}} {\partial E_{\rm PNM}}   \right)^2
\sigma_{E_{\rm PNM}}^2 +
\left(\frac {\partial E_{\rm NSM}} {\partial E_{\rm SNM}}   \right)^2
\sigma_{E_{\rm SNM}}^2 \nonumber \\
&+& 2 \rho \frac {\partial E_{\rm NSM}} {\partial E_{\rm PNM}} \frac {\partial E_{\rm NSM}} {\partial E_{\rm SNM}} 
\sigma_{E_{\rm PNM}}  \sigma_{E_{\rm SNM}} ,
\label{eq:s2NSM}
\eea
with the derivatives $\partial E_{\rm NSM}/\partial E_{\rm PNM}=(1-2x)^2$ and $\partial E_{\rm NSM}/\partial E_{\rm SNM}=4x(1-x)$. 

Figure~\ref{fig:P_nB-NSM} (a) shows the pressure of NSM (including contributions from the leptons) $P_{\rm NSM} = \nb^2 (dE_{\rm NSM}/d\nb)$ 
in the outer core. 
The blue (orange) uncertainty band corresponds to the \NNNLO (\NNLO) results at the $1\sigma$ level. Panel~(b) displays the difference in pressures between PNM and NSM. The zero crossings indicate where the pressure of NSM equals that of PNM. 
Depending on the chiral order, these crossings occur at $n \approx 1.6-2.1 \, \nsat$. 
They are due to a softening of $E_{\rm sym}$ at the higher densities; nevertheless, $E_{\rm NSM}$ is always less than that of $E_{\rm PNM}$.
In no case does $x$ exceed about $0.055$ for $\nb \leq 0.34 \fmiq$.

\section{Results}
\label{sec:Results}

\subsection{Minimum and maximum radius bounds with \chiEFT and causality}
\label{sec:rmaxandlambda}

\begin{figure*}[htb]
\includegraphics[width=\textwidth]{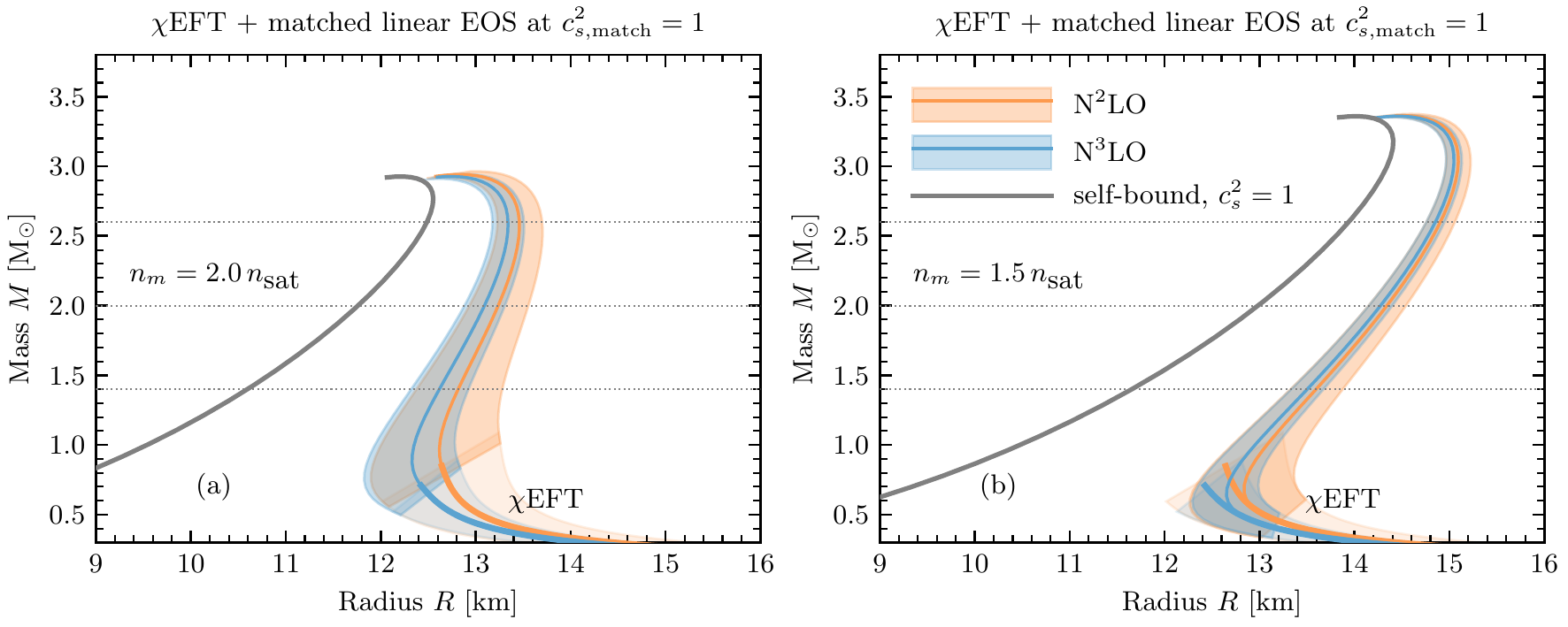}
\caption{Panel (a): $M$--$R$ diagram for NSM based on MBPT calculations shown in 
Fig.~\ref{fig:P_nB-NSM}~(a), including \NNLO-NSM (orange-shaded band) and 
\NNNLO-NSM (blue-shaded band) 
in the outer core 
matched to a linear causal ($\csqmat=1.0$) EOS 
at $\nmat= 2.0 \,\nsat$, and the maximally compact 
EOS for self-bound stars with the same value of $\Mmax$ (black solid). Horizontal lines 
indicate $M=1.4$, 2.0, $2.6\,\Msolar$. The colored bands above $\nmat=2.0\,\nsat$ represent
upper bounds on the NS radius for a given mass, as the high-density matter is assumed 
maximally stiff without discontinuities in the overall EOS (see detailed discussions in 
Sec.~\ref{sec:rmaxandlambda}). 
Panel (b): similar to (a) but with a lower matching density, $\nmat=1.5\,\nsat$.
}
\label{fig:MR_match-EFT}
\end{figure*}

Earlier work has shown that canonical-mass ($1.1-1.7\,\Msolar$) neutron star radii are most sensitive to the EOS in the density interval $1.5-3.0\,\nsat$~\cite{Lattimer:2000nx}, and this is further quantified in Appendix~\ref{sec:Sensitivity}.\footnote{
Appendix C also
quantifies the sensitivity of the key observables $\Rtyp$, $\Rtwo$, and $\Mmax$ to the pressure as a function of density $P(\nb)$.  The highest correlations, i.e., the most sensitive regions, involve the density ranges  $1.0-3.0\,\nsat$, $1.5-4.0\,\nsat$, and $2.0-6.0\,\nsat$, respectively.} 
As a result, calculations up to $\lesssim2.0\,\nsat$ are adequate to place stringent bounds on the NS radius~\cite{Hebeler:2010jx,Gandolfi:2011xu,Tews:2018iwm}. We assume a typical crust EOS~\cite{Negele:1971vb,Baym:1971pw} below $0.5\,\nsat$, the EOS for NSM based on MBPT-\chiEFT calculations~\cite{Drischler:2020yad} 
in the outer core, and
a matching 
linear EOS 
$P(\ep) = \pmat+ \csq\,(\ep-\emat)$ 
characterized by $\csq$ 
in the inner core.  

Figure~\ref{fig:MR_match-EFT}~(a) shows the $M$--$R$ relation for \NNNLO-NSM, \NNLO-NSM outer core EOSs in Fig.~\ref{fig:P_nB-NSM}~(a) matched at $\nmat=2.0 \,\nsat$ to the stiffest linear EOS ($\csq=1$); the solid colored curves refer to the central values and the color-shaded bands refer to $\pm 1\sigma$ uncertainties 
in the MBPT-\chiEFT calculations.   
Results for matching at a lower density $\nmat=1.5\,\nsat$ are shown in Fig.~\ref{fig:MR_match-EFT}~(b). 
As expected, for a given value of $\nmat$, 
the largest radii result from the largest matching pressure $P_{\rm m}$, and thus \NNLO+$1\sigma$; note that  
\NNLO-$1\sigma$ shows little difference compared to \NNNLO-$1\sigma$. 
In general, the lower is $\nmat$, the larger are the maximum radii (Figure~\ref{fig:MR_match-EFT}~(b)). 
Any discontinuities in the energy density 
for $\nb\ge\nmat$, such as from a phase transition, would serve to decrease $R(M)$, emphasizing the results in these figures as being upper bounds.
The extreme case, described in Appendix~\ref{sec:Bounds}, self-bound (crustless) stars with $P_{0}=0$ and $\csq=1$ for a given $\Mmax$ represent the ``maximally compact'' configurations that exhibit the smallest possible radii at all masses, and for comparison their mass-radius relations are also displayed (black solid lines). The causal limit $\csq=1$ 
in all cases shown 
leads to  maximum masses as high as $\approx2.93\,\Msolar$, 
as predicted by  
\beq \Mmax\simeq4.09~\sqrt{\esat\over\ep_0}\Msolar, \label{eq:mmax_approx} \eeq
using $\ep_0=\emat\simeq2.0\,\esat$ (see derivation in \Eqn{eq:mmax0}). 
Differences at low densities, e.g., between \NNNLO and \NNLO, have negligible effects on $\Mmax$, as already noted in the crustless case of Appendix~\ref{sec:Bounds}. 
For a given value of $\csqmat$, $\Mmax$ is essentially determined by $\nmat$ and is relatively insensitive to $P_{\rm m}$. 
With smaller values of $\csqmat$ for a given $\nmat$, the maximum mass decreases.
It can be seen that the upper bounds on $\Rtyp$ (where the bands intersect with the $M=1.4\,\Msolar$ horizontal line) are about $12.9 \km$ ($13.6 \km$) if $\nmat= 2.0\,\nsat$ ($\nmat= 1.5\,\nsat$). 
Although $M_{\rm max}$ is not sensitive to the low-density EOS (see also Appendix \ref{sec:Sensitivity}), $R_{\rm  max}(M)$ for canonical-mass neutron stars ($1.1-1.7\,\Msolar$) is. The relatively soft \NNNLO EOS up to $2.0 \,\nsat$ guarantees that the typical NS radius $\lesssim 13 \km$, even with very stiff matter at higher densities 
that can lead to $\Mmax> 2.6\,\Msolar$.

In contrast to $R_{\rm max}$, it is possible to deduce a minimum radius $R_{\rm min}$ for a given low-density ($\nb<\nmat$) EOS by introducing a finite discontinuity in the energy density $\De\emat$ at $\nmat$.  Above the density $\emat+\De\emat$, the EOS is assumed to be the causal EOS with $\csq=1$. 
The larger is $\De\emat$, the smaller is the resulting value of $\Mmax$, which has a one-to-one relation with it. 
If the pressure at $\nmat$ is vanishingly small, this effectively gives the $R_{\rm min,c} (M)$ relation for the maximally compact EOS 
of self-bound stars 
as described in Appendix~\ref{sec:Bounds} but with $\ep_0=\emat+\De\emat$. 
With finite pressure at $\nmat$ based on \chiEFT calculations, $R_{\rm min} (M)$ is larger and is the minimum radius for normal NSs.  In the self-bound case, the magnitude of $\De\emat$ is related to the maximum mass according to \Eqn{eq:mmax_approx} by imposing $\ep_0=\emat+\De\emat$. Even in the case with a crust, since the  maximum mass is reached at very high densities, this relation remains relatively accurate.  For $\Mmax=2.0\,\Msolar$, we find that $\De\emat \approx \ep_{\rm nuc}(\nmat)\simeq2.0\,\esat$. 
To accommodate a maximum mass of $2.6\,\Msolar$, for example, requires a much smaller discontinuity, $\De\emat \approx 0.25 \,\ep_{\rm nuc}(\nmat)$. Furthermore, all the trajectories within any $\pm2\sigma$ band for each value of $\Mmax$ have nearly identical values of $\De\emat$ 
resulting from the fact that $P_{\rm m}\ll\emat$. 
The relation between $\De\emat$ at $2.0\,\nsat$ and $\Mmax$ is indeed relatively insensitive to the low-density EOS.

\begin{figure}[h]
\includegraphics[width=.95\columnwidth,clip=true]{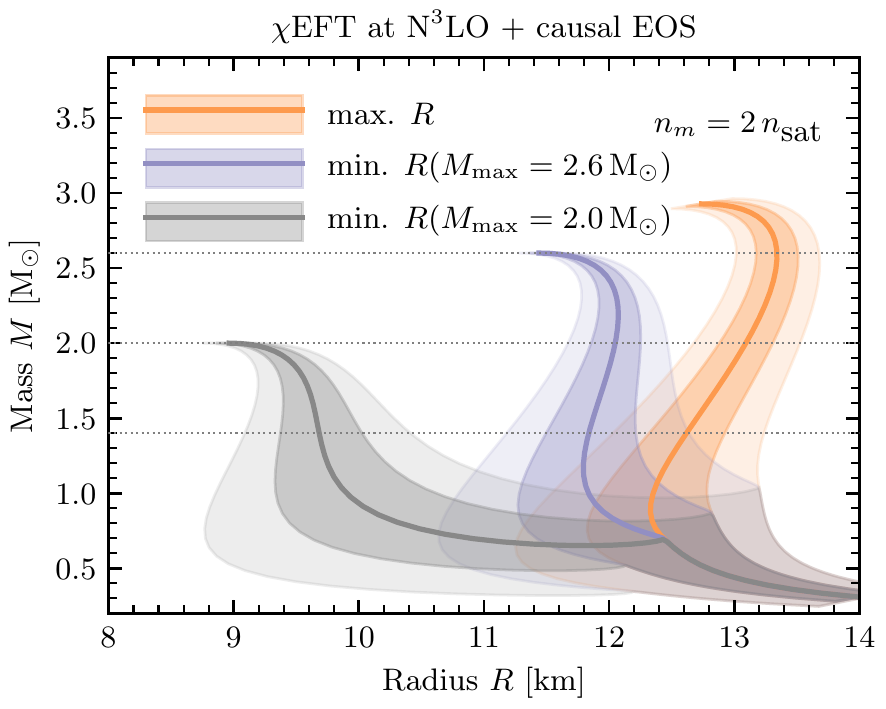}
\caption{Radius bounds obtained by combining \NNNLO-\chiEFT predictions up to $\nmat=2.0\,\nsat$ and maximum-mass information is shown. The orange bands show the upper bound on the NS radius, 
while 
the black and purple bands 
depict 
the lowers bounds corresponding to $\Mmax=2.0\,\Msolar$ and $\Mmax=2.6\,\Msolar$, respectively.
}
\label{fig:rbound}
\end{figure}

Figure~\ref{fig:rbound} shows the combined minimum and maximum radius bounds. 
The central values of the minimum radii $R_{\rm min}(M)$ for $\Mmax=2.0\,\Msolar$ and $\Mmax=2.6\,\Msolar$ are shown as black and purple solid curves, respectively, while the darker and lighter bands reflect $1\sigma$ and $2\sigma$ uncertainties, 
respectively. 
To $2\sigma$ confidence, the minimum radius of a $1.6\,\Msolar$ star ranges from 
$9.2-12.2 \km$  
as $\Mmax$ is varied from $2.00\,\Msolar$ to $2.93\,\Msolar$; roughly, the minimum value of $R_{1.6}\propto\Mmax^{3/4}$.  Similarly, the minimum values of $\Rmax$ vary from 
$9.0 -12.6 \km$. 
It is interesting to compare these results with claims that $R_{1.6}>10.68 \km$ and $\Rmax>8.6 \km$ from observations of GW170817~\cite{Bauswein:2017vtn} 
using empirical relations established in hydrodynamical simulations that relate $R$, $\Mmax$, and the threshold binary mass $M_{\rm thres}$ for prompt collapse of a merger remnant.  We can therefore provide a more restrictive bound for $\Rmax$ since $\Mmax$ is believed to be $\geq2.0\,\Msolar$.

Figure~\ref{fig:rbound} demonstrates how
future 
discoveries 
of NSs with large masses could constrain the radii of all NSs. Several interesting insights can be gleaned from this figure. A striking, albeit expected, feature is the convergence of the upper and lower radius bounds with increasing $\Mmax$.  This is in accordance with the facts that the discontinuity $\De\emat$ leading to the minimum radii has to decrease to achieve a higher $\Mmax$~\cite{Alford:2015gna} and that the limit $\De\emat\to 0$ defines the maximum radii.
For example, the uncertainty in theoretical predictions for the radius of a $1.4\,\Msolar$ NS would be reduced from about $3 \km$ when $\Mmax=2.0\,\Msolar$ to about $0.5 \km$ when $\Mmax=2.6\,\Msolar$. Another feature worth noting is the evolution of the $2\sigma$ lower bound on the NS radius. It increases by about $2\km$, from $9.2 \km$ for $\Mmax=2.0~\Msolar$ to $11.2\km$ when $\Mmax=2.6~\Msolar$. Comparing the black and purple bands shows that the radii of heavier neutron stars are even more tightly constrained with increasing $\Mmax$.  Future observational constraints on NS radii in the mass range $1.4-2.0~\Msolar$ could be valuable in this regard since X-ray and GW observations are best suited to provide radius information at the level of 5\% uncertainty in this mass range~\cite{Watts:2016uzu}. 
Results in Fig.~\ref{fig:rbound} also demonstrate that an upper bound of about $13 \km$ for $\Rtyp$ obtained from GW170817 is consistent with NSs with $\Mmax\simeq 2.6~\Msolar$.

The trends seen in Fig.~\ref{fig:rbound} also have important implications for the EOS of matter at the highest densities encountered in the NS inner core. Our results imply that $\Mmax>2.5\,\Msolar$ and/or radii $>12.5\km$ for neutron-star 
masses 
$\simeq 1.4\,\Msolar$ can only be achieved if $\csq \simeq 1$ over a wide density range encountered in the NS core. We emphasize here that this insight relies on the relatively soft EOS 
predicted 
by \NNNLO-\chiEFT calculations. Improving the EOS, especially the EFT truncation errors in the vicinity of $\nb \simeq 2.0\,\nsat$, will be critical in extracting better constraints on the EOS at higher densities in the core if future observations favor these large radii or masses. Supporting  $\csq \simeq 1$ from $2-5\,\nsat$ requires a form of strongly interacting relativistic matter that poses significant challenges for dense-matter theory and QCD~\cite{McLerran:2018hbz}.

\begin{figure}[hb]
\includegraphics[width=.95\columnwidth,clip=true]{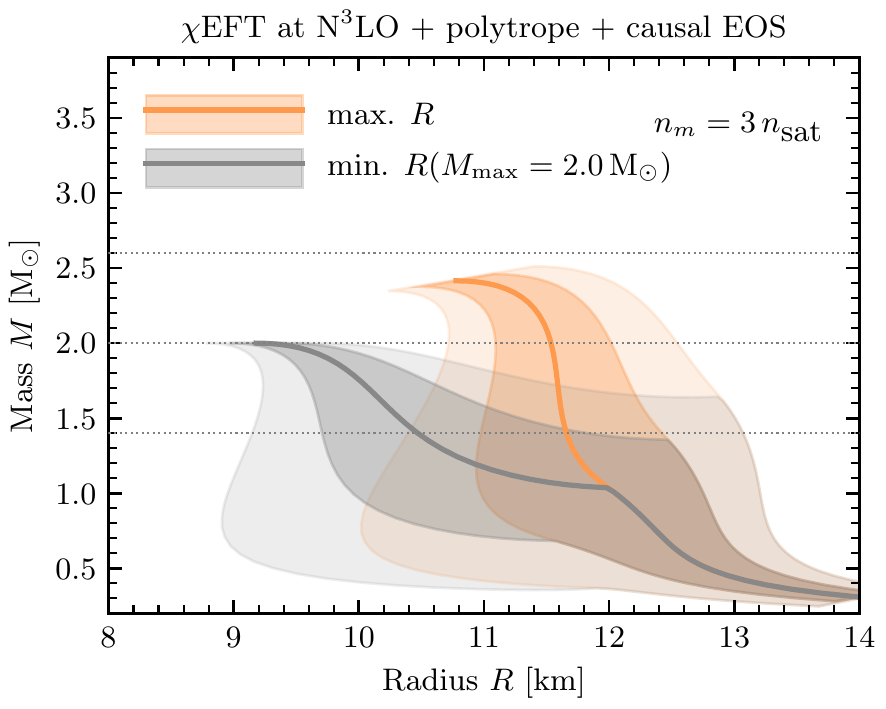}
\caption{Similar to Fig.~\ref{fig:rbound} but obtained using the polytropic extrapolation of the \chiEFT EOS up to $\nmat=3.0\,\nsat$.
}
\label{fig:rbound3}
\end{figure}

\subsection{Consequences of 
increasing $\nmat$ or decreasing $\csq$
}

\begin{figure}[htb]
\parbox{\hsize}{
\includegraphics[width=\hsize]{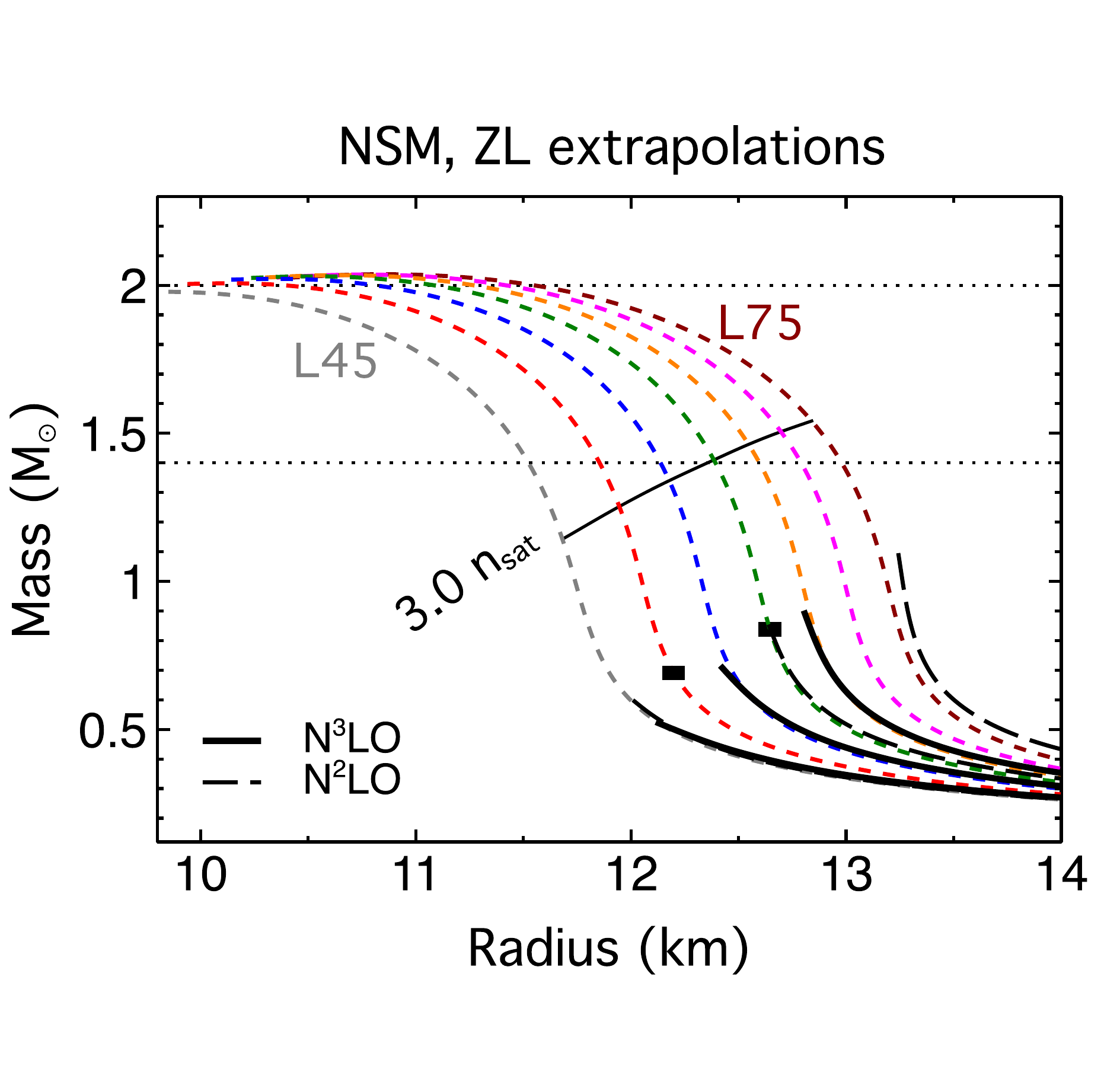}\\[-3ex]
}
\caption{$M$--$R$ relations for NSM EOSs extrapolated to $\nb\geq 2.0\,\nsat$ beyond the \chiEFT calculations  using the ZL parametrization~\cite{Zhao:2020dvu}; the thin black line indicates where the NS central densities are $3.0\,\nsat$. From left to right, the colored dotted curves represent $L=45\MeV$ to $L=75\MeV$ in increments of $5\MeV$, and the black-solid (black-dashed) curves refer to \chiEFT-\NNNLO (\NNLO) with $\pm1\sigma$ uncertainties. 
The $L=50\MeV$ (red) and $L=60\MeV$ (green) ZL EOSs are used in Fig.~\ref{fig:Mmax_csq_nB} because they best represent $\pm1\sigma$ bounds.  
}
\label{fig:MR_ZL}
\end{figure}

Encouraged by the apparent convergence of \chiEFT calculations over the density interval $1-2\,\nsat$, it is natural to ask if a nuclear physics based description of dense matter can be extended to higher density. Extrapolating the EOS from $2.0\,\nsat$ to $3.0\,\nsat$ will be model-dependent, even in the absence of phase transitions to non-nucleonic matter, since we presently do not have reliable calculations at higher densities. We climb this rung of the density ladder with some reservation to motivate and explore the impact of future calculations of the EOS in this density interval.

\begin{figure*}[htb]
\begin{center}
\includegraphics[width=0.999\textwidth]{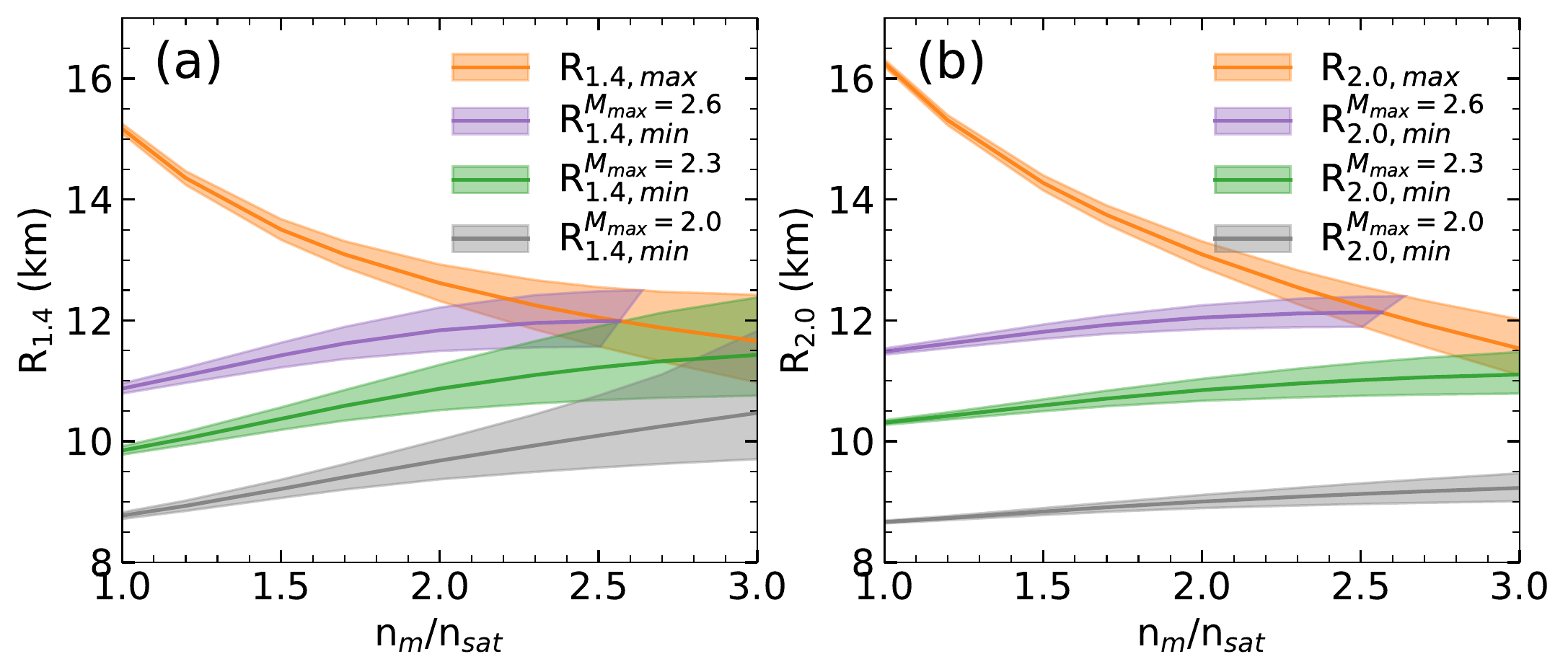}\\[-2ex]
\end{center}
\caption{ 
Similar to Figs.~\ref{fig:rbound} and \ref{fig:rbound3}, but displaying the minimum and maximum radii of $1.4\,\Msolar$ (panel (a)) and $2.0\,\Msolar$ (panel (b)) stars as a function of the matching density $\nmat=1.0-3.0\,\nsat$. 
Additionally, $R_{\rm min}$ contours and uncertainty bands for the case $\Mmax=2.3\,\Msolar$ are shown. 
}
\label{fig:R_nm}
\end{figure*}

We first consider a polytropic model where $P= \kappa  \ep^\gamma$ in which the parameters $\kappa$ and $\gamma$ are determined by fitting to the behavior predicted by \chiEFT calculations in the density interval $1.9-2.1\, \nsat$ to extrapolate the EOS from $2.0\,\nsat$ to $3.0\,\nsat$. This choice is somewhat arbitrary and is chosen to approximately capture the key features of the density dependence of the EOS predicted by \chiEFT. The resulting radius bounds are shown in Fig.~\ref{fig:rbound3}. 
We have also found 
that an alternative parametrization~\cite{Zhao:2020dvu} of NSM matter, which has a single parameter corresponding to the symmetry energy coefficient $L$, to be a convenient extrapolation tool, referred to hereafter as the ZL parameterization.  Figure~\ref{fig:MR_ZL} shows $M$--$R$ curves for the ZL EOSs together with a standard crust. For example, $L=45 \MeV$ ($65 \MeV$) successfully tracks \NNNLO, while $L=45 \MeV$ ($75 \MeV$) tracks \NNLO, for $-\sigma$ ($+\sigma$).  
We have checked that alternate extrapolations using the polytropic model, with parameters chosen to suitably match the \chiEFT results at $2.0\,\nsat$, do not significantly alter our conclusions.

A comparison between the results shown in Fig.~\ref{fig:rbound} with those in Fig.~\ref{fig:rbound3} reveals the following insights. First, the increase in $\nmat$ does not alter the bounds on $R_{\rm min}(M)$ (including $\Rmax$), as a function of $\Mmax$, except that in the extrapolated case $M$ and $\Mmax$ cannot exceed about $2.5\,\Msolar$.  These bounds are therefore particularly robust for $M<2.5\,\Msolar$.

The increase in $\nmat$ results in more stringent upper bounds on the NS radius for masses in the range $1.4-2.5~\Msolar$. For example, the polytropic extrapolation to $3.0\,\nsat$ predicts $R_{\rm max}(1.4 \,\Msolar) = 11.6^{+ 0.8}_{ - 0.6} \km$, which is to be contrasted with $R_{\rm max}(1.4 \,\Msolar) = 12.5^{+ 0.3}_{ - 0.2} \km$ obtained using $\nmat=2.0\,\nsat$. This reduction has implications for the interpretation of future radius measurements which aim for an accuracy of better than 5\%~\cite{Watts:2016uzu}. If these observations favor NSs in this mass range to have radii $>12 \km$, it would require new mechanisms to rapidly stiffen the EOS below $3.0\,\nsat$. 

It is also apparent, if the secondary component in GW190814 were to be confirmed to be a massive NS, new mechanisms would also be implicated 
at a 
low density, since the extrapolated EOS
up to $3.0\,\nsat$ 
predicts $\Mmax$ in the range $2.32-2.53\,\Msolar$ at $\pm2\sigma$.

The results shown in Figs.~\ref{fig:rbound} and~\ref{fig:rbound3} are summarized in Fig.~\ref{fig:R_nm} for the specific cases of $\Rtyp$ and $\Rtwo$, 
with a broader range of $\nmat$ explored between $1.0-3.0\,\nsat$. 

\begin{figure}[htb]
\includegraphics[width=\hsize]{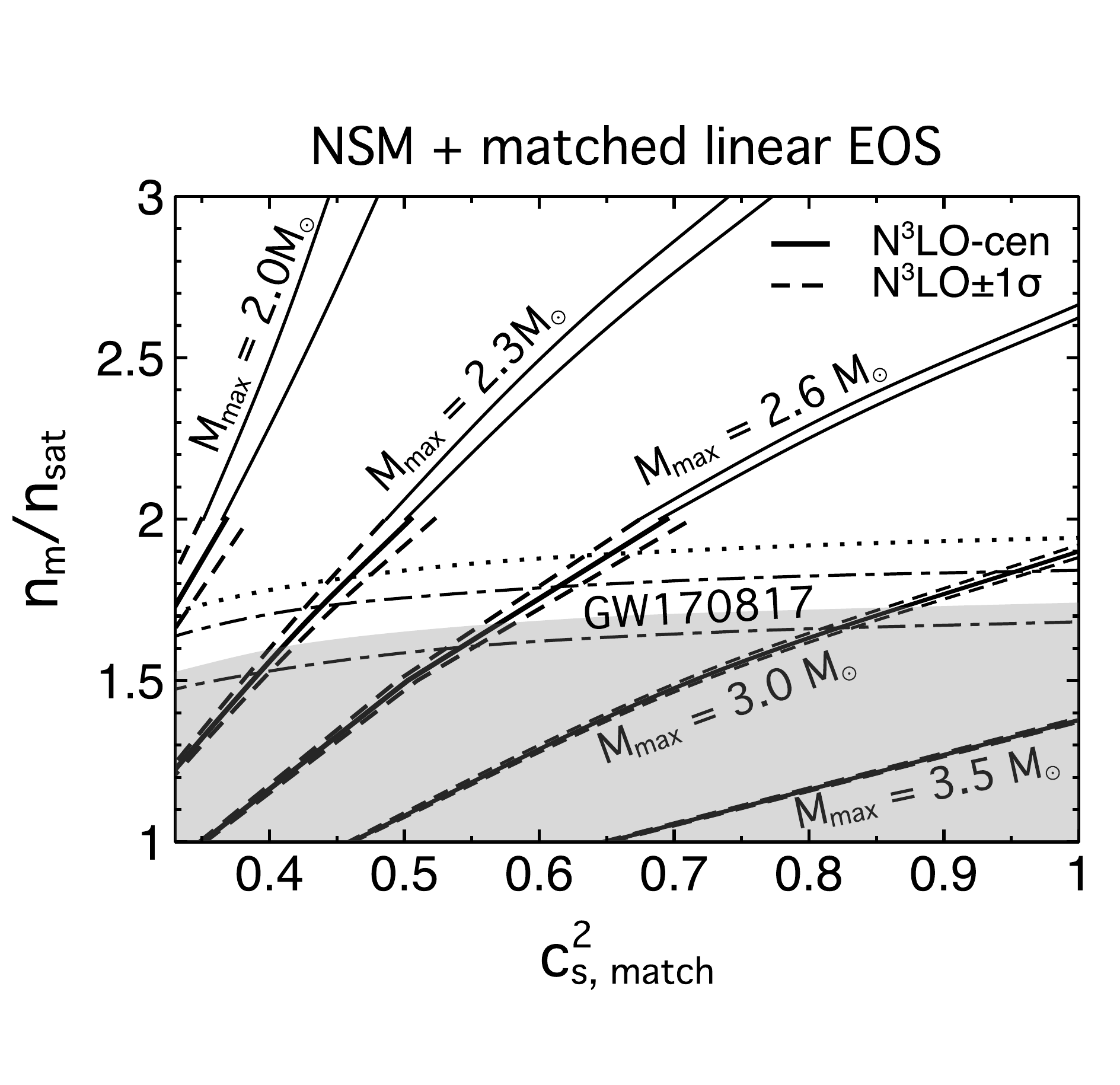}\\[-4ex]
\caption{$\Mmax$ contours on the $(\csqmat,\nmat)$ plane, obtained when $\De\emat=0$.  For each value of $\Mmax$, the central solid curve shows results with the central value of \chiEFT-\NNNLO; dashed lines indicate $\pm1\sigma$ bounds.  Extensions to $\nmat>2.0\,\nsat$ for the central and $+\sigma$ bound are also shown as solid curves.  The grey-shaded region 
is excluded by the binary tidal deformability constraint $\tilde\La_{1.186}\leq 720$ from GW170817 
at the 
90\% credibility level~\cite{Abbott:2018wiz} if \NNNLO-cen is assumed;
the dot-dashed lines refer to constraints with the \NNNLO$\pm 1\sigma$ boundaries. 
The GW170817 bounds will be shifted downwards if there is a first-order transition at such low densities. 
}
\label{fig:Mmax_contour}
\end{figure}

\begin{figure*}[htp]
\begin{center}
\includegraphics[width=0.999\textwidth]{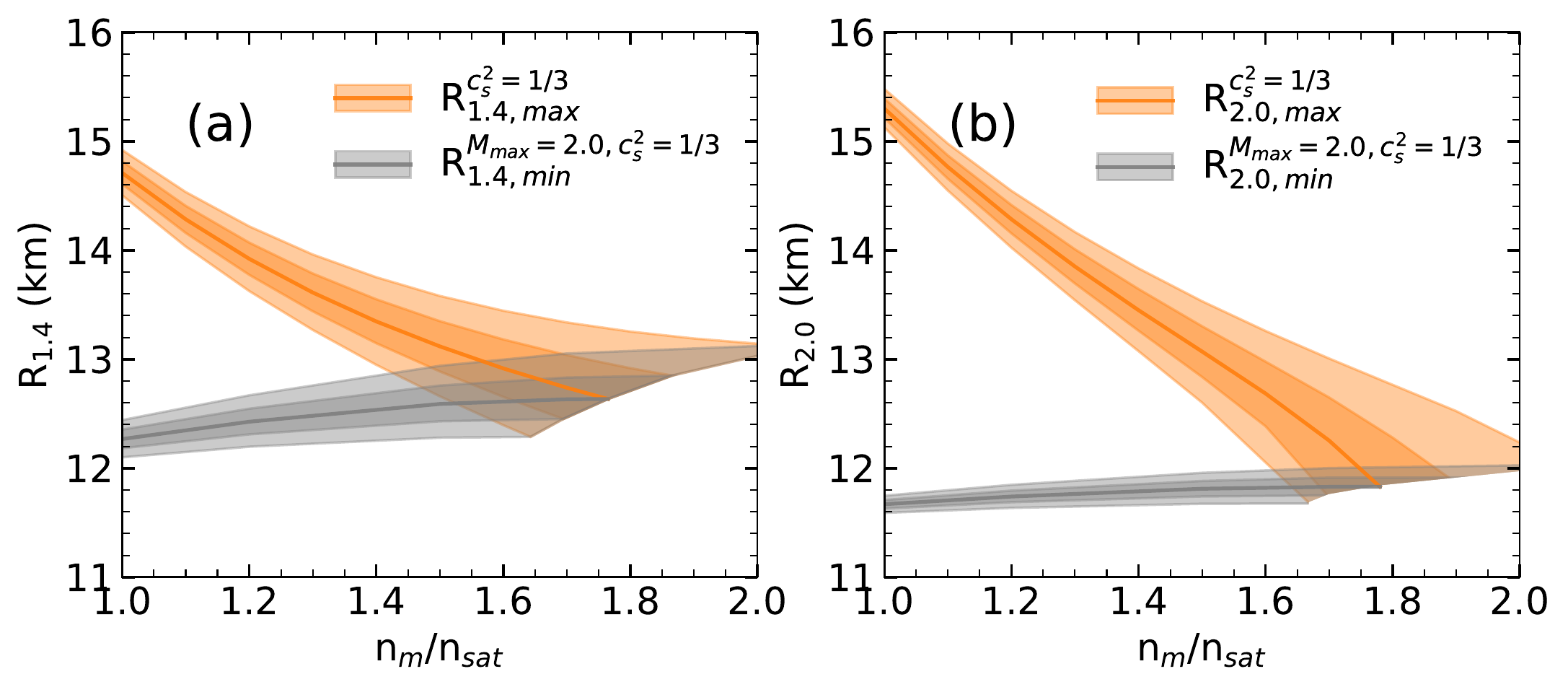}\\[-4ex]
\end{center}
\caption{The maximum (orange) and minimum (black) bounds on $\Rtyp$ and $\Rtwo$ assuming $\csq\leq1/3$ above $\nb=\nmat$; \chiEFT-\NNNLO uncertainties are indicated (darker bands: $\pm1\sigma$; lighter bands: $\pm2\sigma$). The bands merge and terminate at critical matching densities above which $\Mmax<2.0\,\Msolar$. 
}
\label{fig:R_nm_csq3}
\end{figure*}

Figure~\ref{fig:R_nm} 
also 
conveniently illustrates the dramatic effect of increasing the lower bound on $\Mmax$ for the allowed ranges between $R_{\rm min}(M)$ and $R_{\rm max}(M)$, 
which improves (shrinks) the $\Rtyp\, (\Rtwo)$ bounds by an average $3 \km/\Msolar$ ($5\km/\Msolar$); 
these limits could be
further 
restricted by 
forthcoming 
observations. 
We note that 
$\Rtyp$ or $\Rtwo$ 
$<10.7\km$ would be incompatible with $\Mmax>2.3\,\Msolar$ (assuming $\nmat=2.0\,\nsat$). 
In addition, if future measurements from different sources and messengers, e.g., X-ray data from QLMXBs or PREs (or GW detections of mergers by LIGO) vs NICER targets, were to exhibit discrepancies in the radius inference close to or larger than the gaps between the minimum and maximum bands shown on this figure, then 
these are hints of a 
large energy-density discontinuity $\De\emat$ in the EOS (accompanied with high-density stiff matter) 
occurring at $\nb\lesssim\nmat$~\cite{Han:2020adu}.

It is important to recall that $\Mmax$ depends monotonically on $\nmat$ (or $\emat+\De\emat$) for fixed $\csqmat$.  So far, we have only considered causal EOSs ($\csqmat=1$).  However, it is almost certain that the EOS in this high-density region will be subluminal.  To keep the discussion straightforward, we now consider the consequences of fixing the sound speed in this region to a constant value $\csqmat\le1$.    Therefore, assuming a crust EOS, the validity of \chiEFT up to $\nmat$, and a constant sound speed for the highest density region, implies that $M$-$R$ trajectories, and $\Mmax$, will depend on three quantities: $\nmat$, $\De\emat$ and $\csqmat$.  Instead of using the polytropic parameterization, we extend the nucleonic EOS to $3.0\,\nsat$ with the ZL parametrization. 
We find 
that the ZL EOSs 
corresponding to $L=60\MeV$ and $50\MeV$, respectively, smoothly join the $\Mmax(\nmat)$ relations for the \NNNLO$+1\sigma$ and 
\NNNLO-cen EOSs, even though those corresponding to $L=65$ MeV and $L=45$ MeV seem to match the $\pm1\sigma$ $M$-$R$ results below $2.0\,\nsat$\footnote{Note that $\Mmax(\nmat)$ for the extrapolated EOSs will eventually bend upwards at sufficiently large $\nmat$, 
which is a generic feature whenever a ``standard'' 
nucleonic-like EOS 
(i.e. gradually increasing $\csq$ without kinks or discontinuities that naturally extends from low-density e.g. \chiEFT calculations) 
is switched to a linear EOS at some critical density, with or without discontinuities in $\ep$ (see, e.g., Fig.~5 in Ref.~\cite{Alford:2015gna}).  
However, we limit our studies to $\nmat\lesssim 3.0 \,\nsat$, as there is little guidance for the validity of nucleonic degrees of freedom at higher densities from theory.}; 
the reason is that the masses of stars with central density $\ncent=2.0\,\nsat$ are similar in both cases. 
If \chiEFT-\NNNLO is assumed valid up to $2.0\,\nsat$, the upper and lower bounds on NS radii are substantially tightened in comparison with using \chiEFT-\NNNLO only up to $\nsat$: for example if $\Mmax\geq 2.0\,\Msolar$ then $\Rtyp$ must lie between $12.5^{+ 0.3}_{ - 0.2} \km$ and $9.7^{+ 0.4}_{ - 0.3} \km$ at the $1\sigma$ level, which is consistent with earlier studies in Ref.~\cite{Tews:2018iwm}. Radius constraints are further tightened if \chiEFT-\NNNLO is assumed valid to higher densities, but there is a diminishing return. 

Figure~\ref{fig:Mmax_contour} shows how $\Mmax$ depends on $\nmat$ and $\csqmat$. We find, for example, that $\Mmax\ge2.6\,\Msolar$ requires 
$\csqmat>0.35$ (i.e., 
the conformal limit $\csq\leq1/3$ is violated) 
if $\nmat=\nsat$, and $\csqmat>0.7$ if $\nmat=2.0\,\nsat$.
The conformal limit 
is also violated for $\nmat>1.7\,\nsat$, even if $\Mmax$ is as low as $2.0\,\Msolar$. 
If $\Mmax>2.45\,\Msolar$, $\nmat$ must not exceed $3.0\,\nsat$ no matter what the value of $\csqmat$ is. 
The calibrated uncertainties in \chiEFT-\NNNLO lead to relatively small uncertainties, less than $0.1\,\Msolar$, in $\Mmax(\nmat,\csqmat)$.

There has been speculation that the speed of sound in QCD at finite baryon density may be bounded by the conformal limit which requires $\csq< 1/3$~\cite{Cherman:2009tw}. This speculation is in part  based on strong-coupling calculations of $\text{SU}(N_c)$ gauge theories for which a holographic or gravity dual exist. In these theories the speed of sound can be calculated at finite baryon density in the large-$N_c$ limit using classical supergravity methods in a curved spacetime~\cite{Maldacena:1997re}, and for a large class of such theories (for exceptions, see Refs.~\cite{Ecker:2017fyh,Ishii:2019gta}) $\csq <1/3$ \cite{Cherman:2009tw,Hohler:2009tv}. In addition, at finite temperature and zero baryon density, where lattice QCD calculations provide reliable predictions, $\csq<1/3$ at all temperatures. The sound speed increases rapidly in the hadronic phase (dominated by pions) reaching a maximum value $\csq \simeq 0.2$, then decreases across hadron-quark cross-over region, corresponding to temperatures in the range $100-200\MeV$, and eventually increases again to reach its asymptotic value of $\csq \simeq 1/3$ at $T\simeq 500\MeV$~\cite{Romatschke:2017ejr}.

Motivated by the discussion above, we briefly comment on the astrophysical implications of the conjecture that $\csq <1/3$ in QCD \cite{Cherman:2009tw} in light of our results. It was already noted in Refs.~\cite{Bedaque:2014sqa,Tews:2018kmu} that it is difficult to accommodate $\csq <1/3$ at high density and $\Mmax > 2.0~\Msolar$ while still allowing for a soft EOS at intermediate density needed to ensure that $\Rtyp < 13 \km$. This is also evident from Fig.~\ref{fig:Mmax_contour} which shows that when $\csq <1/3$, it is impossible, at the $1\sigma$ level, to simultaneously satisfy the tidal deformability constraint from GW170817 and $\Mmax>2.0 ~\Msolar$ if \chiEFT-\NNNLO is valid beyond $1.8~\nsat$.

\begin{figure*}[htp]
\begin{center}
\includegraphics[width=0.999\textwidth]{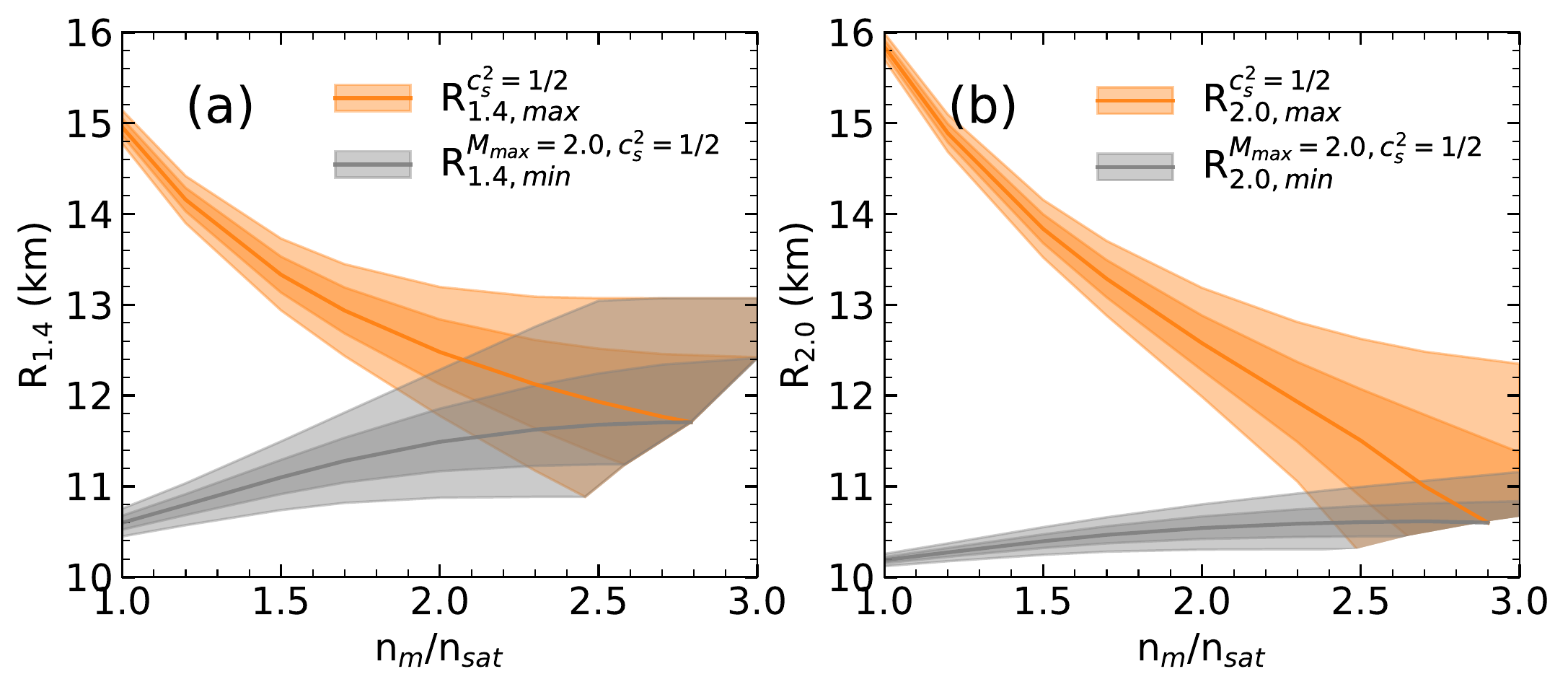}\\[-4ex]
\end{center}
\caption{Similar to Fig.~\ref{fig:R_nm_csq3}, but assuming $\csq\leq1/2$ above $\nb=\nmat$. 
}
\label{fig:R_nm_csq2}
\end{figure*}

Figure~\ref{fig:R_nm_csq3} 
shows how the bounds on the radius are influenced when $\csq <1/3$ at high density. The rapid decrease in the maximum value of $\Rtyp$ with $\nmat$ is striking and implies that if $\csq <1/3$ and \chiEFT-\NNNLO is valid up to $1.5~\nsat$, then $\Rtyp$ must lie between $12.4^{+ 0.2}_{ - 0.2} \km$ and $13.1^{+ 0.3}_{ - 0.3} \km$ at the $1\sigma$ level. Further, requiring that $\Mmax > 2.0~\Msolar$ excludes a significant fraction of the \chiEFT-\NNNLO predicted range for the pressure for densities between $1.5-2.0$ $\nsat$. A tiny sliver of high pressure close to the edge of the $2\sigma$ boundary remains, and implies that $\Rtyp =13.1 \pm 0.1 \km$! 
Predictions for $\Rtwo$ are shown in the right panel. 
In Fig.~\ref{fig:R_nm_csq2} 
we show the maximum and minimum bounds on $\Rtyp$ and $\Rtwo$ obtained by imposing an intermediate limit of $\csq\leq1/2$. 
In this case for $\nmat=2.0\,\nsat$, we find that $11.5^{+ 0.3}_{ - 0.3} \km<\Rtyp<12.5^{+ 0.3}_{ - 0.2} \km$ and $\Mmax< 2.29\pm0.04~\Msolar$ (Fig.~\ref{fig:Mmax_csq_nB}~(b)), to $1 \sigma$ confidence. 
The corollary to this implies that measurements of $R_{\sim 1.4}$ that are smaller than $11.2\km$ would favor a stiff EOS with $\csq\geq 1/2$ above $2.0\,\nsat$, or that $\nmat<2.0\,\nsat$. 
This is particularly interesting because a recent analysis of the tidal deformability constraints from GW170817 in Ref.~\cite{Capano:2019eae} suggests $11.0^{+0.9}_{-0.6} \km$ (90\% credible interval).

\subsection{Tidal deformability constraints}
\label{sec:tidal}
\begin{figure*}[htb]
\parbox{0.5\hsize}{
\includegraphics[width=\hsize]{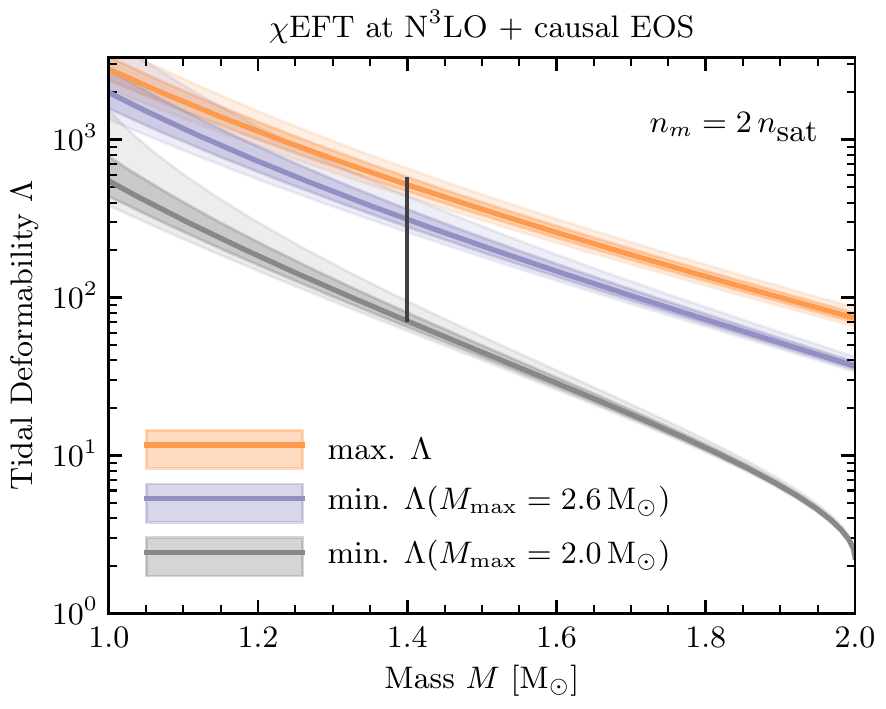}\\[-3ex]
}\parbox{0.5\hsize}{
\includegraphics[width=\hsize]{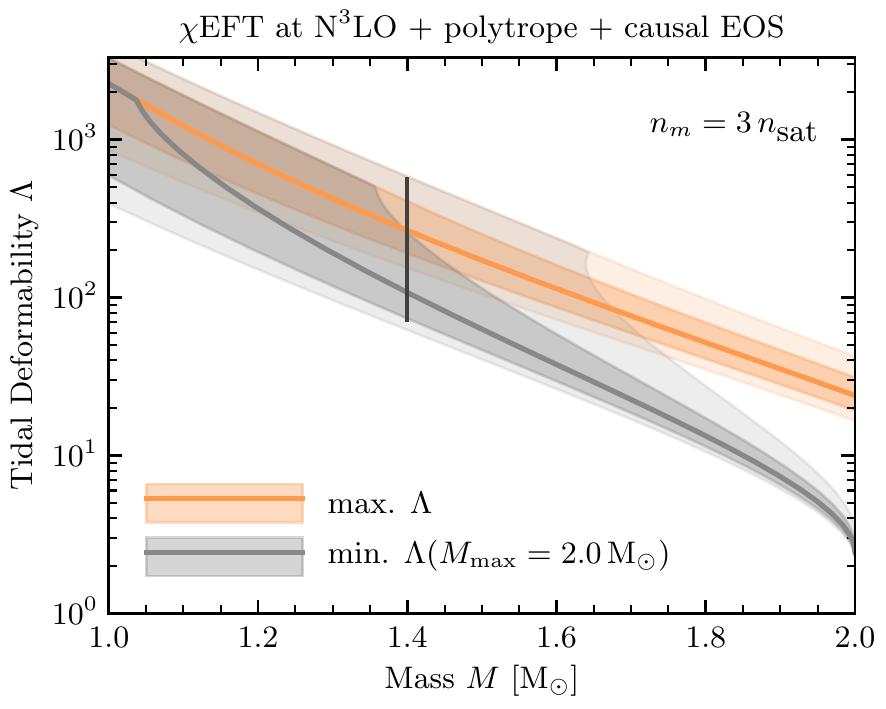}\\[-3ex]
}
\caption{The left panel shows bounds on the tidal deformability $\Lambda$ obtained using \chiEFT \NNNLO EOS up to $\nmat=2.0\,\nsat$, and the right panel extends the low-density EOS to $\nmat=3.0\,\nsat$ using the polytropic extrapolation. As in Fig.~\ref{fig:rbound}, the orange bands show the upper bound, 
while 
the lower bounds corresponding to $\Mmax=2.0\,\Msolar$ and $\Mmax=2.6 \,\Msolar$ are shown by the black and purple bands, respectively. The vertical solid line depicts the constraint inferred from GW170817, $70\leq\La_{1.4}\leq580$.  When $\nmat=3.0\,\nsat$, $\Mmax<2.6\,\Msolar$.}
\label{fig:Lam_bound}
\end{figure*}

\begin{figure*}[htb]
\parbox{0.5\hsize}{
\includegraphics[width=\hsize]{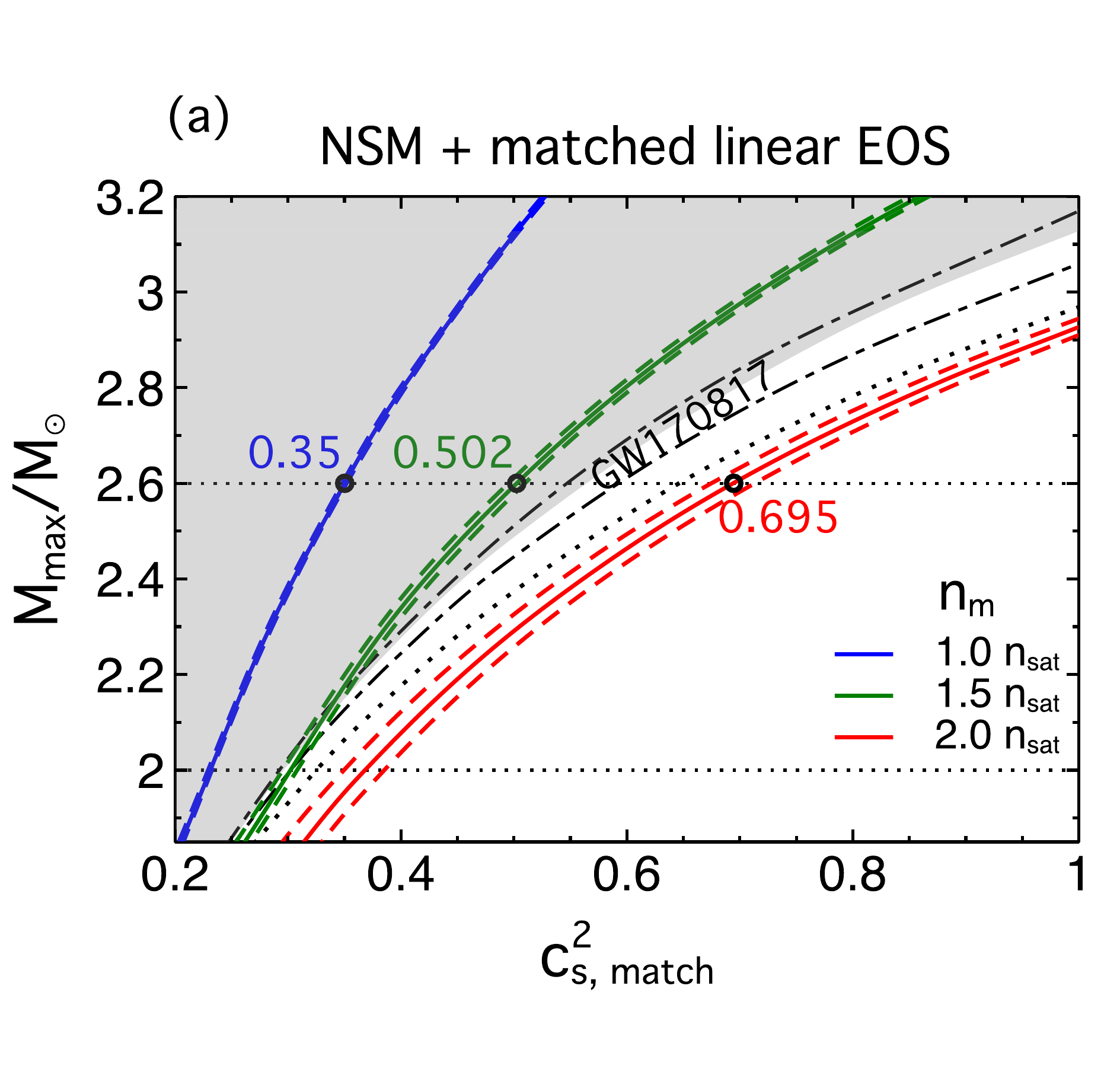}\\[-4ex]
}\parbox{0.5\hsize}{
\includegraphics[width=\hsize]{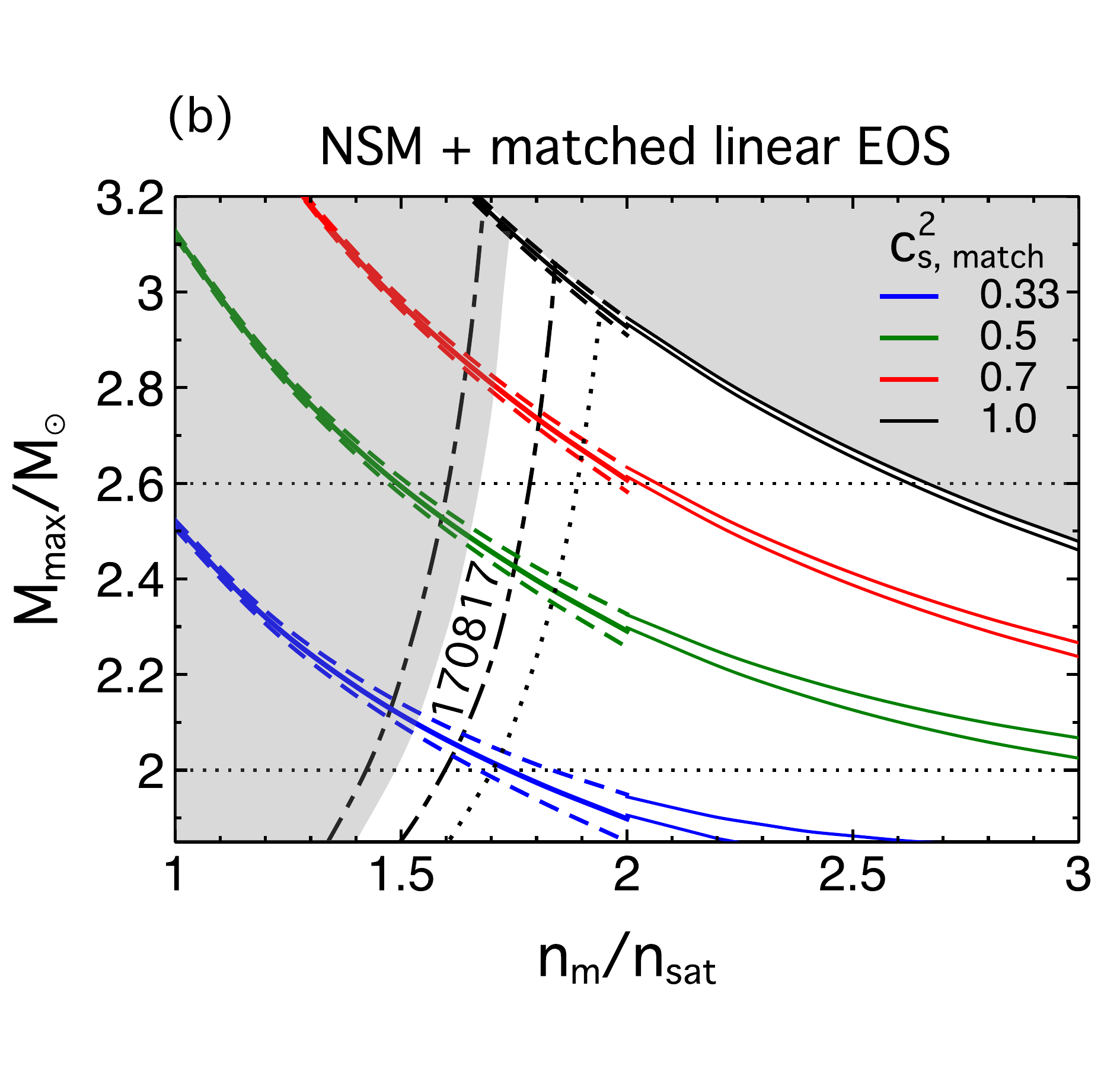}\\[-4ex]
}
\caption{
Panel~(a): the solid lines show contours of $\nmat$ in the $\Mmax$--$\csqmat$ plane, and the dashed lines bracket \NNNLO$\pm 1\sigma$ uncertainties. 
The upper horizontal line indicates $\Mmax=2.6\,\Msolar$; see also examples later 
in  Fig.~\ref{fig:MR_csq_nB-M26}.  
The grey-shaded region 
is excluded by the binary tidal deformability constraint $\tilde\La_{1.186}\leq 720$ from GW170817 
at the 
90\% credibility level~\cite{Abbott:2018wiz} if \NNNLO-cen is assumed;
the dot-dashed lines refer to constraints with the \NNNLO$\pm 1\sigma$ boundaries. 
The thin dotted line indicates a lower upper bound with \NNNLO-cen and $\tilde\La_{1.186}\leq 600$. 
Panel~(b): same as panel (a), except that contours of $\csqmat$ are displayed in the $\Mmax$--$\nmat$ 
plane; 
the upper-right 
grey-shaded region is excluded by causality. 
For $\nmat\in[2.0, 3.0]\,\nsat$, extrapolations from \chiEFT using ZL models with $L=50\MeV$ and $L=60\MeV$ are applied (see Fig.~\ref{fig:MR_ZL}).
}
\label{fig:Mmax_csq_nB}
\end{figure*}

Gravitational waveform fitting using the standard PhenomPNRT model~\cite{Dietrich:2018uni,Abbott:2018wiz} directly sets constraints on the binary chirp mass $\Mchirp=1.186\pm0.001\,\Msolar$ and the binary tidal deformability $\tilde\La\leq 720$ (90\% credibility). 
In what follows, we will denote this constraint as ${\tilde\La}_{1.186}\leq720$, where the chirp mass is
\begin{equation}
{\cal M}={m_1^{3/5}m_2^{3/5}\over(m_1+m_2)^{1/5}},
\label{eq:chirp}
\end{equation}
and the binary tidal deformability is defined as
\begin{equation}
\tilde\Lambda={16\over13}{\Lambda_1m_1^4(m_1+12m_2)+\Lambda_2m_2^4(12m_1+m_2)\over(m_1+m_2)^5}.
\label{eq:deform}
\end{equation}
Here $\Lambda_1$ and $\Lambda_2$ refer to the individual deformabilities of the binary components with masses $m_1$ and $m_2$, respectively.   It can be shown~\cite{Zhao:2018nyf} that $\Lambda$ is approximately proportional to $(R/M)^6$ and $\tilde\Lambda$ is approximately proportional to $(\bar R/{\cal M})^6$, where $\bar R$ is the average radius of stars with masses constrained by $\tilde\Lambda\simeq1.2\,\Msolar$ and $q>0.7$, where $q=m_2/m_1$, i.e., the component masses are confined to the interval between $1.1\,\Msolar$ and $1.6\,\Msolar$.  Therefore, the maximum radius $R_{\rm max}(M)$ bound is tantamount to a maximum $\tilde\Lambda$ bound, and vice-versa.
The $\tilde\La$--$\Mchirp$ constraint can be translated to a constraint on $\La$ at the mass $M$, $\La_M$, but it is subject to small additional uncertainties from the poorly determined mass ratio $q$ of GW170817 and EOS systematics. 
Using the resulting quasi-universal EOS relation $\La_1=q^6\La_2$, which is valid to $10\%-20\%$ for $\Mchirp=1.186\,\Msolar$ and $q>0.7$~\cite{Zhao:2018nyf}, one finds 
\beq
\La_M\simeq2^{6/5}(\Mchirp/M)^6\tilde\La_\Mchirp,
\label{eq:lambda-m}
\eeq
valid to a few percent. 
Absolute bounds from causality on the tidal deformability $\Lambda$ can be derived in the same way as radius bounds: upper bounds are determined by smoothly matching a low-density EOS to a causal EOS at $\nmat$~\cite{VanOeveren:2017xkv}, whereas lower bounds are determined by introducing a discontinuity $\De\emat$ (which lowers $\Mmax$)~\cite{Zhao:2018nyf,Han:2018mtj}. The bounds for the \NNNLO-\chiEFT EOS with $\nmat=2.0\,\nsat$ are shown in Fig.~\ref{fig:Lam_bound}. The role of $\Mmax$ is clear from comparison of the $\Mmax=2.0\,\Msolar$ and the $\Mmax=2.6\,\Msolar$ cases. 
This figure also shows the effects of increasing $\nmat$ using the polytropic extrapolation 
from 
\chiEFT from $2.0\,\nsat$ to $3.0\,\nsat$.  In this case, $\Mmax<2.6\,\Msolar$. 
The fact that uncertainties in the GW170817 constraint of $\La$ 
extend almost precisely between the lower ($\Mmax=2.0\,\Msolar$ with a large discontinuity $\De\emat$ at $\nmat$) and upper bounds ($\csqmat=1$ without discontinuity) to within $2\sigma$ for both $\nmat=2.0\,\nsat$ and $\nmat=3.0\,\nsat$ cases is not a coincidence. 
It is a consequence of the fact that for those values of $\nmat$, $\tilde\La_{1.186}<720$ is always satisfied for all values of $\csqmat\le1$ (see Fig.~\ref{fig:Mmax_csq_nB}).

A comparison between the results shown in Fig.~\ref{fig:Lam_bound} provides quantitative insights into how access to the EOS at higher density will impact predictions for  the tidal deformability $\Lambda$, especially for more massive NSs. It illustrates how constraints on $\Lambda$ from future GW detections from binaries with massive NSs can provide insights on the evolution of $\csq$ in the density interval $2-3~\nsat$. For example, if $\Lambda_{2.0} \gtrsim 100$, it would pose a serious challenge for \chiEFT predictions even in the density interval $1-2~\nsat$, and $\Lambda_{2.0} \gtrsim 50$ would be difficult to accommodate without new mechanisms to significantly stiffen the EOS in the density interval $2-3~\nsat$. On the other hand,  if $\Lambda_{1.4} \lesssim 100$, it would imply a soft EOS between $1-3~\nsat$, a near-causal EOS at higher densities, and $\Mmax$ not significantly larger than $2\,\Msolar$. 

Results for $\Mmax$ using subluminal sound speeds for the high-density EOS are shown  in Fig.~\ref{fig:Mmax_csq_nB}~(a) for the cases $\nmat=1,1.5,2.0\,\nsat$. This figure, in $\Mmax-\csqmat$ space, is a permutation of Fig.~\ref{fig:Mmax_contour} that instead shows $\Mmax$ contours in $\nmat-\csqmat$ space.  The dotted curve at $2.6\,\Msolar$ intersects the contours for those cases for $\csqmat=0.35, 0.502$ and 0.695, respectively.

The derived bounds on $\nmat$ and $\csqmat$ illuminate the importance of
including nuclear-matter calculations in the density range $1-3\,\nsat$. 
Standard extrapolations based on nucleonic models, similar to the ZL parametrization,  are usually associated with a more gradual 
profile of $\csq(\nb)$ at low-to-intermediate densities, 
which cannot reconcile the small radii and/or small tidal 
deformabilities inferred for canonical-mass NSs with large maximum masses. The necessary rapid change in the sound speed guided by the simple matching scheme 
serves to indicate 
the breakdown of such extrapolations at 
high densities.
A very high NS mass, e.g., $\gtrsim 2.45\,\Msolar$ ($2.6\,\Msolar$), would be in conflict with causality and standard extrapolation up to $3.0\,\nsat$ ($2.66\,\nsat$); 
therefore indicating something unusual in the EOS 
should be taking place near this density.  This is consistent with the findings of Refs.~\cite{Tan:2020ics,Lim:2020zvx}.

A more conservative estimate for the maximum mass, such as $2.2-2.3\,\Msolar$, increases the allowed range for $\nmat$ and $\csqmat$ to be consistent with data; the generic trend is shown in Fig.~\ref{fig:Mmax_csq_nB}.
Specifically, 
Fig.~\ref{fig:Mmax_csq_nB}~(a) demonstrates how $\Mmax$ scales with $\csqmat$ using the \NNNLO-NSM EOS for $\nmat=1.0, 1.5, 2.0 \,\nsat$. The solid curves correspond to results for 
\NNNLO-cen  
and the dashed ones with $\pm 1\sigma$ uncertainties. 
The dots indicate the intersections of the central curves with $\Mmax=2.6\,\Msolar$ for the same EOSs as shown 
later in Fig.~\ref{fig:MR_csq_nB-M26}~(b). The \chiEFT uncertainties at the respective densities only slightly broaden these correlations. Together with GW170817, the constraint $\Mmax\geq 2.1\, \Msolar$ rules out very weakly-interacting matter ($\csq\approx 0.33$) at high densities, whereas $\Mmax\geq 2.5\, \Msolar$ rules out matter with $\csq\lesssim 0.5$. \\

The third permutation of Fig.~\ref{fig:Mmax_contour} is displayed in Fig.~\ref{fig:Mmax_csq_nB}~(b). 
It is noteworthy that the GW170817 boundary (edge of the grey-shaded region) for \NNNLO-cen is nearly parallel to the $\nmat$ contours. 
For matching densities $\lesssim1.5-1.8\,\nsat$, all constructed EOSs result in $\tilde\La_{1.186}> 720$ and can be therefore considered ruled out by GW170817 (see also examples 
later in Fig.~\ref{fig:LamM_Lamwt_Mc}~(a)). If an even lower upper bound on $\tilde\La_{1.186}$ were to be established, the excluded region would become larger, increasing the threshold of minimally allowed $\nmat$.

\begin{figure*}[htb]
\parbox{0.5\hsize}{
\includegraphics[width=\hsize]{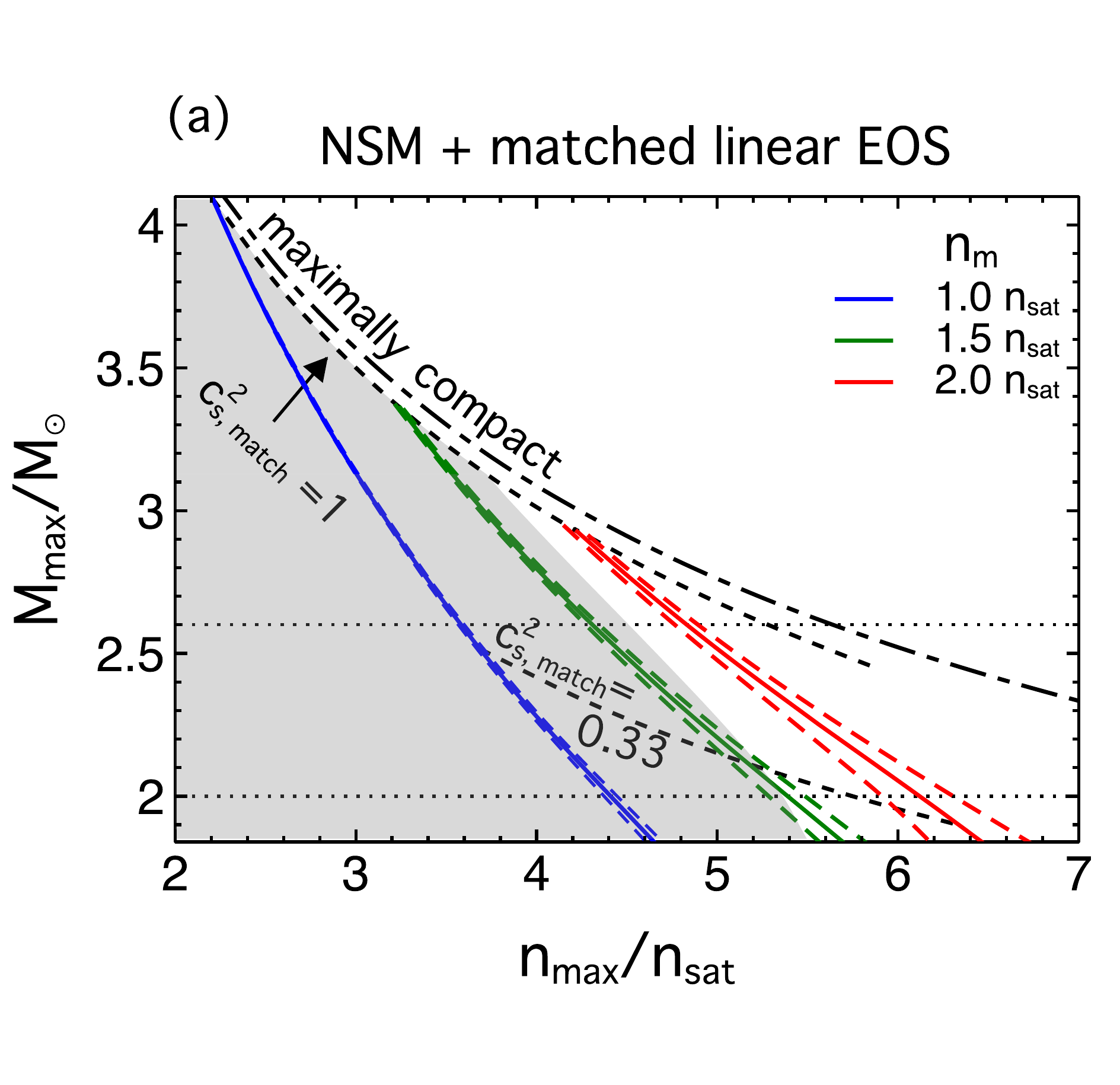}\\[-4ex]
}\parbox{0.5\hsize}{
\includegraphics[width=\hsize]{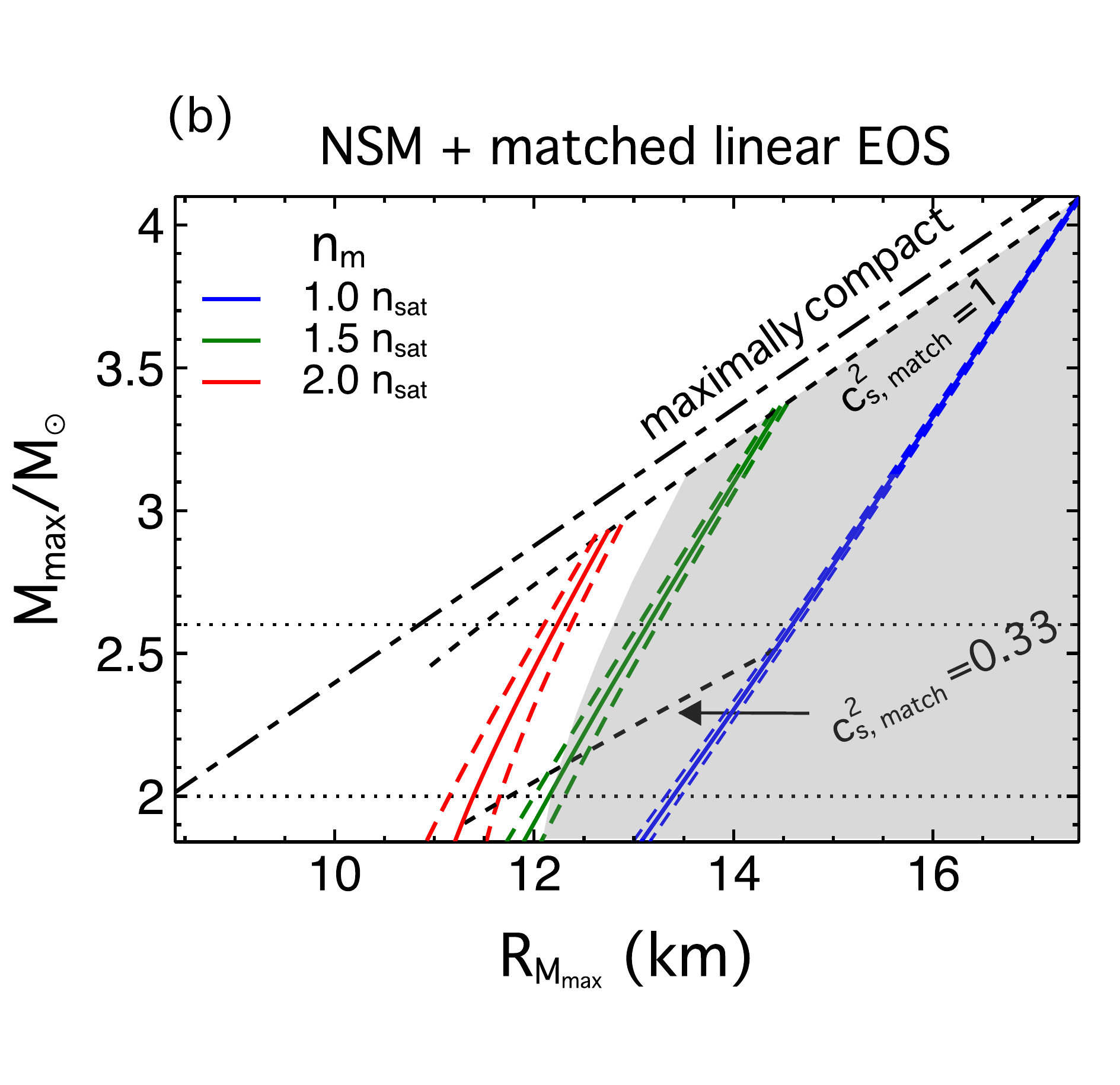}\\[-4ex]
}
\caption{Panel~(a): scaling relations between $\Mmax$ and $\nmax$; panel~(b): scaling relations between $\Mmax$ and $\Rmax$. Both relations, shown as dot-dashed lines, follow from the maximally compact EOS (see Appendix~\ref{sec:Bounds}). 
The black dashed curves correspond to the presence of a low-density nuclear mantle 
(crust + \NNNLO EOS) for $\nb\leq\nmat$, with fixed sound speeds $\csqmat=0.33$ and $\csqmat=1.0$ for $\nb>\nmat$. 
The grey-shaded region is excluded by GW170817 ($\tilde\La_{1.186}\leq720$ and \NNNLO-cen). The solid colored curves show contours of $\nmat=1.0, 1.5, 2.0 \,\nsat$ for \NNNLO-cen; dashed colored curves show $\pm1\sigma$ 
uncertainties. 
For EOSs that accommodate $\Mmax\geq2.6\,\Msolar$, 
the permitted ranges of $\nmax$ and $\Rmax$ are severely restricted.}
\label{fig:Mmax_nmax_Rmax}
\end{figure*}

Compatibility with GW170817 is readily satisfied if 
the 
\chiEFT calculations (with uncertainties) are assumed valid up to $2.0\,\nsat$
consistent with previous studies~\cite{Tews:2018iwm}.
The evolution of $\Mmax$ with $\csqmat$ has been known~\cite{Lattimer:1990zz,Lattimer:2015nhk,Moustakidis:2016sab,Margaritis:2019hfq},
but it was unclear how the uncertainty in the low-density EOS translates to an uncertainty in the derived upper bound. 
As shown in Fig.~\ref{fig:Mmax_csq_nB}~(b), we find that for $\nmat=2.0\,\nsat$, the uncertainty in $\Mmax$ ranges from $\approx 0.1 \,\Msolar$ for $\csqmat = 0.33$ (blue-dashed line) to $\approx 0.05 \,\Msolar$ for $\csqmat=1$ (black-dashed line) with $\NNNLO\pm 1\sigma$ inputs at low densities.

In summary, satisfying the GW170817 tidal deformability constraint $\tilde\La_{1.186}<720$ and imposing $\Mmax>2.1\,\Msolar$ requires $\nmat>1.5\,\nsat$ and $\csq>0.35$.  This limit is not very sensitive to $\Mmax$. 
Even if $\Mmax>2.6\,\Msolar$, it is required that $1.7<\nmat/\nsat<2.6$ and $\csqmat>0.55$. 
The existence of a $2.6\,\Msolar$ star evidently requires a significant change from normal hadronic EOSs 
to a much stiffer EOS between $1.7\,\nsat$ and $2.6\,\nsat$. 
In the presence of a discontinuity in $\ep$, the lower bound $\nmat\gtrsim1.7\,\nsat$ can decrease, 
whereas the upper bound $\nmat\lesssim2.6\,\nsat$ remains unaffected as it is imposed by causality.

For stars with a normal crust, refined upper limits to $\Rmax$ can 
be found using the GW10817 constraint and an assumed value for $\Mmax$, while lower limits follow from the causal EOS: 9 km $<\Rmax<12.2$ km for $\Mmax\le2.1\,\Msolar$ and 11.3 km $<\Rmax<12.8$ km for $\Mmax\le2.6\,\Msolar$.

\subsection{$\Mmax$ scalings compared to the maximally compact case}
\label{sec:compact}

In Fig.~\ref{fig:Mmax_nmax_Rmax}~(a) 
we show the absolute upper limit on $\Mmax$ (see e.g.  \Eqn{eq:nmax0}) as a function of $\nmax$, the highest possible baryon density from the maximally compact EOSs, as represented by the dot-dashed boundary.
The slightly lower black dashed boundary matches the maximally compact EOS to a low-density nuclear EOS at some density $\nmat$ varying from $\nsat$ to about $3.0\,\nsat$ (from left to right). The relatively small difference between these two boundaries suggests that effects on the absolute upper bound on $\nmax$ and $\Mmax$ from the low density EOS is small, and for $\Mmax\geq2.6\,\Msolar$,  $\nmax$ should be smaller than $5.3-5.6\,\nsat$. This is in good agreement with $\approx 5\,\nsat$ obtained in Ref.~\cite{Tan:2020ics}. For $\nmat\leq2.0\,\nsat$, we employ \chiEFT calculations with uncertainties, and the ZL parametrizations (see Fig.~\ref{fig:MR_ZL}) are applied for $\nmat$ between $2.0-3.0\,\nsat$. If the high-density matter is assumed to be much softer with $\csqmat=0.33$, matching it to the nuclear EOS at different matching densities $\nmat$ gives rise to the predicted $\Mmax$--$\nmax$ relation shown by the lower
dashed curve. 
The grey-shaded region is ruled out by tidal deformability constraints inferred from GW170817, prohibiting 
small values of $\nmat$ below $1.5-1.8\,\nsat$. 
As a result, $\csqmat\lesssim0.33$ is incompatible with $\Mmax\gtrsim2.1\,\Msolar$; see also Fig.~\ref{fig:Mmax_csq_nB}. 
Furthermore, imposing $\Mmax\geq2.0\,\Msolar$ leads to $5.23<\nmax/\nsat<5.79$.

The colored curves in Fig.~\ref{fig:Mmax_nmax_Rmax} indicate where the matching densities are fixed at $\nmat/\nsat= 1.0$ (blue), $1.5$ (green), and $2.0$ (red), 
and they track decreasing values of $\csqmat$ from $1$ to below $0.33$. In each case, the highest $\Mmax$ as well as the smallest $\nmax$ correspond to where they end at the $\csqmat=1$ upper boundary (black dashed line). The \NNNLO$\pm 1\sigma$ uncertainty at $2.0\,\nsat$ translates to $\approx0.4\,\nsat$ uncertainty in $\nmax$ 
($5.9-6.3\,\nsat$) if $\Mmax=2.0\,\Msolar$, and $\approx0.1\,\nsat$ uncertainty for $\Mmax=2.6\,\Msolar$. 
Beyond $\nmat\gtrsim2.0\,\nsat$, extrapolation 
of the \chiEFT calculations 
is needed for which the curves would move to the lower-right while remaining under the $\csqmat=1$ bound. Using the ZL parametrization to extrapolate up to $3.0\,\nsat$ (not shown), we obtain $\nmax\leq5.71-5.92\,\nsat$ and $\Mmax\leq2.45-2.48\,\Msolar$.

As discussed in Appendix~\ref{sec:Bounds},  
the maximally compact EOS with $\csq=1$ determines the smallest possible radius at a given mass.
Figure~\ref{fig:Mmax_nmax_Rmax}~(b) displays 
the absolute bound  on the radius of the maximum mass star, $\Rmax$, as well as a more realistic bound taking into account the low-density EOS below $\nmat$.
Assuming \chiEFT up to $\nmat=2.0\,\nsat$ and $\Mmax=2.0\,\Msolar$, the $\NNNLO\pm 1\sigma$ uncertainties induce an uncertainty $\approx0.5 \km$ in $\Rmax=11.14-11.66 \km$. 
For $\Mmax=2.6\,\Msolar$, an uncertainty $\approx0.3\km$
is found with $\Rmax=12.09-$12.38 km.
Extrapolating to higher densities $\nmat\gtrsim2.0\,\nsat$, $\Mmax\geq 2.6\,\Msolar$ leads to $\Rmax\geq 11.49\km$.
The tidal deformability constraint inferred from GW170817 
instead corresponds to limits on the radii of canonical-mass stars. With the simple matching condition used here, that constraint simultaneously rules out too large $\Rmax$, e.g., $\Rmax\leq12.18\km$ if $\Mmax=2.0\,\Msolar$ and $\Rmax\leq12.79 \km$ if $\Mmax=2.6\,\Msolar$.

On the other hand, introducing a finite discontinuity in 
$\ep$
would decrease $\Rmax$ and increase $\nmax$, but to reach the same $\Mmax$ necessitates the transition density to be smaller than the matching density $\nmat$ when there is no discontinuity~\cite{Han:2020adu}. The overall effect is that larger $\nmax$ and smaller $\Rmax$ are possible but must still lie within the bounds set by 
the maximally compact EOSs.

\section{Discussion}
\label{sec:Discs}
It is worth mentioning that 
so far we have 
largely avoided finite discontinuities in the energy density $\ep$, except when located at $\nmat$, which would otherwise introduce 
an additional parameter that characterizes the strength of a sharp first-order phase transition. In that scenario, the $\Mmax$ bounds will be shifted downwards due to the softening induced by the phase transition, while GW170817 boundaries 
may become more complicated depending on the possible formation of disconnected branches at intermediate densities on the $M$--$R$ diagram~\cite{Chatziioannou:2019yko,Han:2018mtj}.
However, given the systematic uncertainties involved in obtaining $\tilde\La$ from gravitational waveform data, the previously inferred bounds should still apply~\cite{Zhao:2018nyf}. In any case, as discussed in Appendix~\ref{sec:Bounds}, useful information on the minimal radii $R_{\rm min} (M)$ can be obtained from matching to the causal EOS with a discontinuity $\De\emat$ specified by $\Mmax$, and we 
have elaborated on these lower bounds on $R$ with \chiEFT inputs up to $\nmat$ 
in Sec.~\ref{sec:rmaxandlambda}.

\subsection{Current and future constraints}

To shed light on the properties of dense matter, the observational constraints used in this work are taken from (i) a handful of well measured NS masses from radio observations~\cite{Demorest:2010bx,Antoniadis:2013pzd,Fonseca:2016tux,Arzoumanian:2017puf,Cromartie:2019kug}, (ii) the chirp and combined masses as well as bounds on tidal deformabilities of NSs deduced from GW detections in the binary NS-NS merger event GW170817~\cite{LIGO:2017qsa,Abbott:2018wiz,Abbott:2018exr}, and (iii) radius estimates from NICER for a NS of mass $\simeq 1.4\,\Msolar$~\cite{Riley:2019yda,Miller:2019cac}. 
An upper bound of $\Mmax\lesssim2.3\,\Msolar$ 
on the maximum gravitational mass of a cold, spherical NS was inferred from several studies using EM and GW data from GW170817~\cite{Ruiz:2017due,Margalit:2017dij,Rezzolla:2017aly,Shibata:2019ctb,Abbott:2019eut}, 
but an upper bound on $\Mmax$ itself does not 
provide further limits 
on 
the sound speed or bounds to NS radii since the EOS could suddenly soften above $\nmat$.  

The NICER 
$M$--$R$ constraints on J0030+0451, namely, 
$R=13.02^{+1.24}_{-1.19}\km$, $M=1.44^{+0.15}_{-0.14}\,\Msolar$~\cite{Miller:2019cac} and $R=12.71^{+1.14}_{-1.19}\km$, $M=1.34^{+0.15}_{-0.16}\,\Msolar$~\cite{Riley:2019yda}, and some EM observations of GW170817~\cite{Radice:2017lry,Coughlin:2018miv,Kiuchi:2019lls} favor larger radii 
than indicated by GW observations from GW170817, $10-13\km$~\cite{Abbott:2018wiz,Abbott:2018exr}, but the degree of tension is slight.  
Joint analyses of these data yield tighter but still consistent constraints on the typical NS radius 
$\sim12.3\km$~\cite{Landry:2020vaw,Essick:2020flb,Jiang:2019rcw,Al-Mamun:2020vzu}; 
Ref.~\cite{Zhao:2020c} 
found $11.8^{+1.0}_{-0.7} \km$ to 68.3\% confidence. 

It is fortunate that NICER targets also include several pulsars for which the masses are independently measured to high precision, e.g., PSR J1614-2230 $\simeq 1.91\,\Msolar$ and PSR J0740+6620 $\simeq2.14 \,\Msolar$, 
and PSR J0437-4715~\cite{Reardon:2015kba} with mass $\approx 1.44 \,\Msolar$. The possibility to measure radii of both intermediate as well as very massive NSs 
opens up the possibility to contrast the radii of $\sim2.0\,\Msolar$ stars, $\Rtwo$, and more typical $\sim1.4\,\Msolar$ stars, $\Rtyp$, to further constrain the EOSs~\cite{Han:2020adu,Xie:2020tdo}.

\begin{figure}[h]
\includegraphics[width=.95\columnwidth,clip=true]{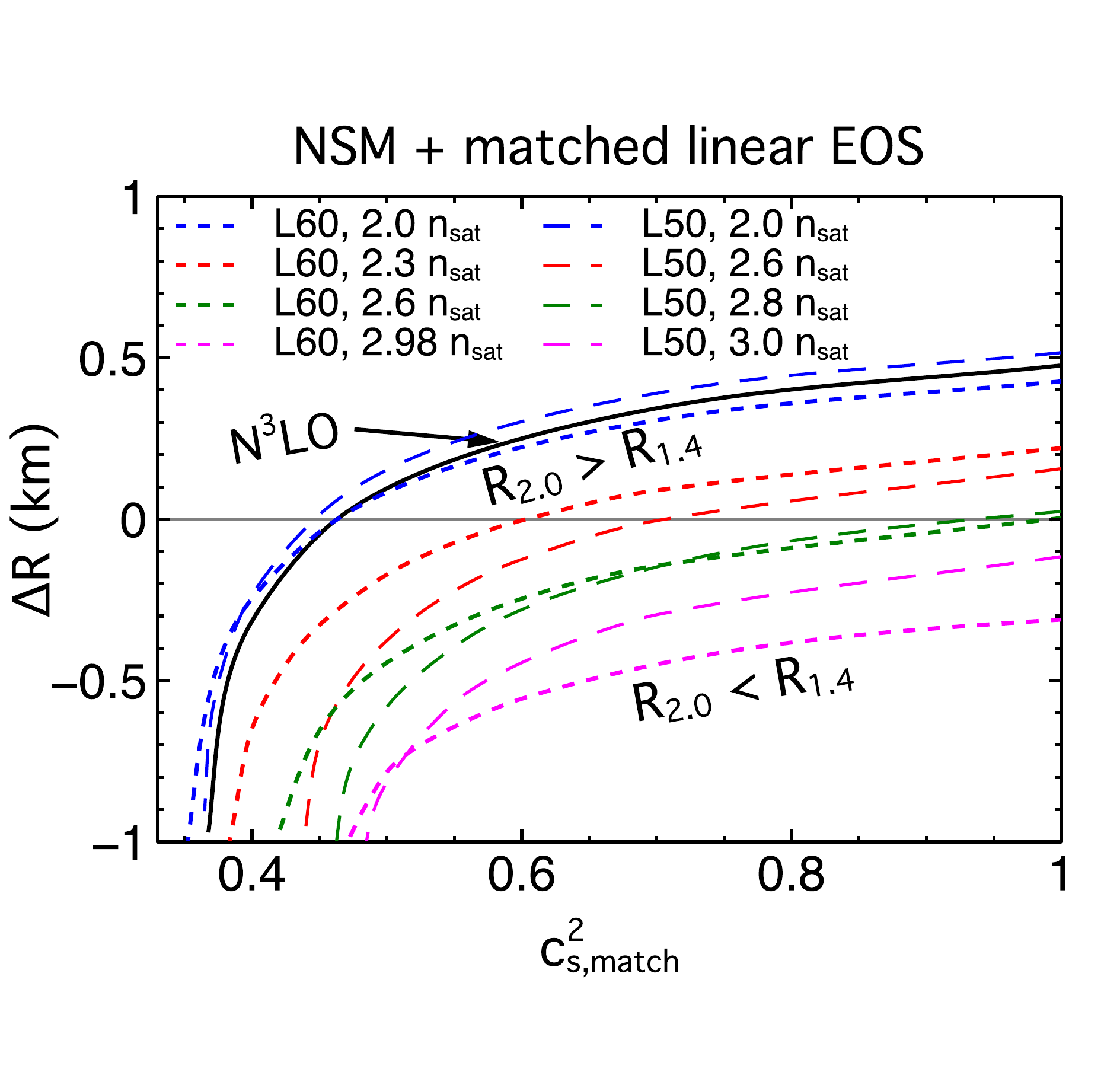}\\[-4ex]
\caption{Radius differences $\De R = \Rtwo-\Rtyp$ using the ZL extrapolations with $L=50\MeV$ and $L=60\MeV$ 
joined continuously to linear EOSs 
at 
$\nmat$ between $2.0\,\nsat$ and {$3.0\,\nsat$}. 
}
\label{fig:R20_R14}
\end{figure}

We show in Fig.~\ref{fig:R20_R14} the difference $\De R = \Rtwo-\Rtyp$ for stars with the \NNNLO EOS up to $2.0\,\nsat$, ZL EOS extrapolations up to a range of matching densities $\nmat =2.0- 3.0\,\nsat$, and various linearly matched EOSs with different $\csqmat$ at higher densities. The ZL 
extrapolation 
with $L=50\MeV$ 
indicates that roughly above $\nmat\gtrsim 2.8 \,\nsat$, all values of $\csqmat$ 
lead to $\Rtwo\leq\Rtyp$. The boundary between positive and negative $\De R$ shifts a bit when using the slightly stiffer ZL extrapolation  with $L=60\MeV$: in this case
$\nmat\gtrsim2.6 \,\nsat$ will guarantee $\Rtwo\leq\Rtyp$; note that $2.98 \,\nsat$ is already the central density of a $1.4\, \Msolar$ star. 
We also checked radii differences between $2.1\,\Msolar$ and $1.4\,\Msolar$ stars, $\De R^{'}=R_{2.1}-\Rtyp$, and found that $\De R^{'}$ is generally less than $\De R$, with 
the largest 
decreases 
of  
a few tenths of a km occurring 
for the smaller values of 
$\csqmat$. 
For $\csqmat\gtrsim0.7$, there are negligible  
differences. 
$\De R$ or $\De R^{'}$ being negative is typical when extrapolations to even higher densities are applied, or if there is additional softening in the EOS before reaching the central density of the maximum-mass star. 
Should observations suggest $\Rtwo>\Rtyp$ or $R_{2.1}>\Rtyp$,
standard extrapolations such as ZL-models predict some unusual stiffening should occur below $\lesssim2.6-2.8 \,\nsat$. 
Furthermore, if $\De R$ turns out to be greater than $0.5\km$, then we should expect that this stiffening occurs for $\nmat\lesssim2.0 \,\nsat$, 
which suggests a 
very high $\Mmax$ and less compatibility with radius constraints from GW170817; see Fig.~\ref{fig:Mmax_csq_nB}~(b). 
However, NICER observations may not achieve 
the 
needed 
$\mathcal{O}(0.5 \km)$ resolutions in the 
near future. 
Since central densities of $\sim2.0\,\nsat$ correspond to $0.5-1.0\,\Msolar$ within $1\sigma$ uncertainties of \chiEFT calculations (Fig.~\ref{fig:rbound}), it will be greatly helpful if 
radii of very low-mass NSs $\sim1.1\,\Msolar$ can be obtained through X-ray observations, or tidal deformability measurements of binary systems with very low chirp masses. 

From a different perspective, 
more accurate experimental determinations of $S_v$ and $L$ at $\nsat$ from e.g., 
PREX, CREX, and 
FRIB/MSU, will be important to test \chiEFT predictions of properties of neutron-rich matter. 
At the present time, $S_v$ and $L$ are believed to be understood to the 10\% and 40\% levels, respectively~\cite{Lattimer:2012xj}.  
For $\nb>\nsat$, constraints from the analyses of the collective flow of matter in HICs 
could be informative. 

The best available information for the present comes from the analysis 
of HICs 
of \isotope{Au} nuclei using Boltzmann-type kinetic equations. The elliptic and 
sideways
flow observables from these collisions are sensitive to the mean-field potential and to in-medium NN collisions 
at central densities of $2-5\,\nsat$, and suggest SNM pressures of $7.5\MeV \fmiq$ to 
$14\MeV \fmiq$ 
at $2.0\,\nsat$~\cite{Danielewicz:2002pu}. 
In comparison, \NNNLO calculations for SNM predict somewhat larger pressures of $10.5\MeV \fmiq$ to $18.5\MeV \fmiq$ at $2.0\,\nsat$~\cite{Drischler:2020hwi}, which are, nevertheless, 
consistent within their stated $1\sigma$ uncertainties. However, 
the predictions from HICs involve model-dependent assumptions concerning the density- and momentum-dependencies of the assumed nuclear interactions, which have not been systematically explored; see Ref.~\cite{Constantinou:2015mna} and references therein for the relevance of single-particle potentials in HICs. 
In addition to these 
uncertainties, 
HICs probe nearly symmetric matter, and to apply their observables to NSM requires an additional 
extrapolation involving the symmetry energy at supra-nuclear densities. 

To improve the current status, heavy-ion facilities across the world, such as RHIC, FAIR, NICA, J-PARC, and HIAF, have launched programs to map out the QCD phase diagram 
of strongly interacting matter. 
The study of more neutron-rich matter in HICs, together with improved, systematic, modeling 
would be very valuable for dense-matter physics, not only for cold neutron stars, but also for understanding mergers involving NSs. 
As the analyses of HIC data have largely been done with nucleonic degrees of freedom, it would be also interesting and desirable to extend such analyses to include quark degrees of freedom and their subsequent hadronization as in RHIC and CERN experiments at higher energies. 

\subsection{$2.6\,\Msolar$ neutron stars and the nature of the components of GW190425 and GW190814} 

\begin{figure*}[htb]
\parbox{0.5\hsize}{
\includegraphics[width=\hsize]{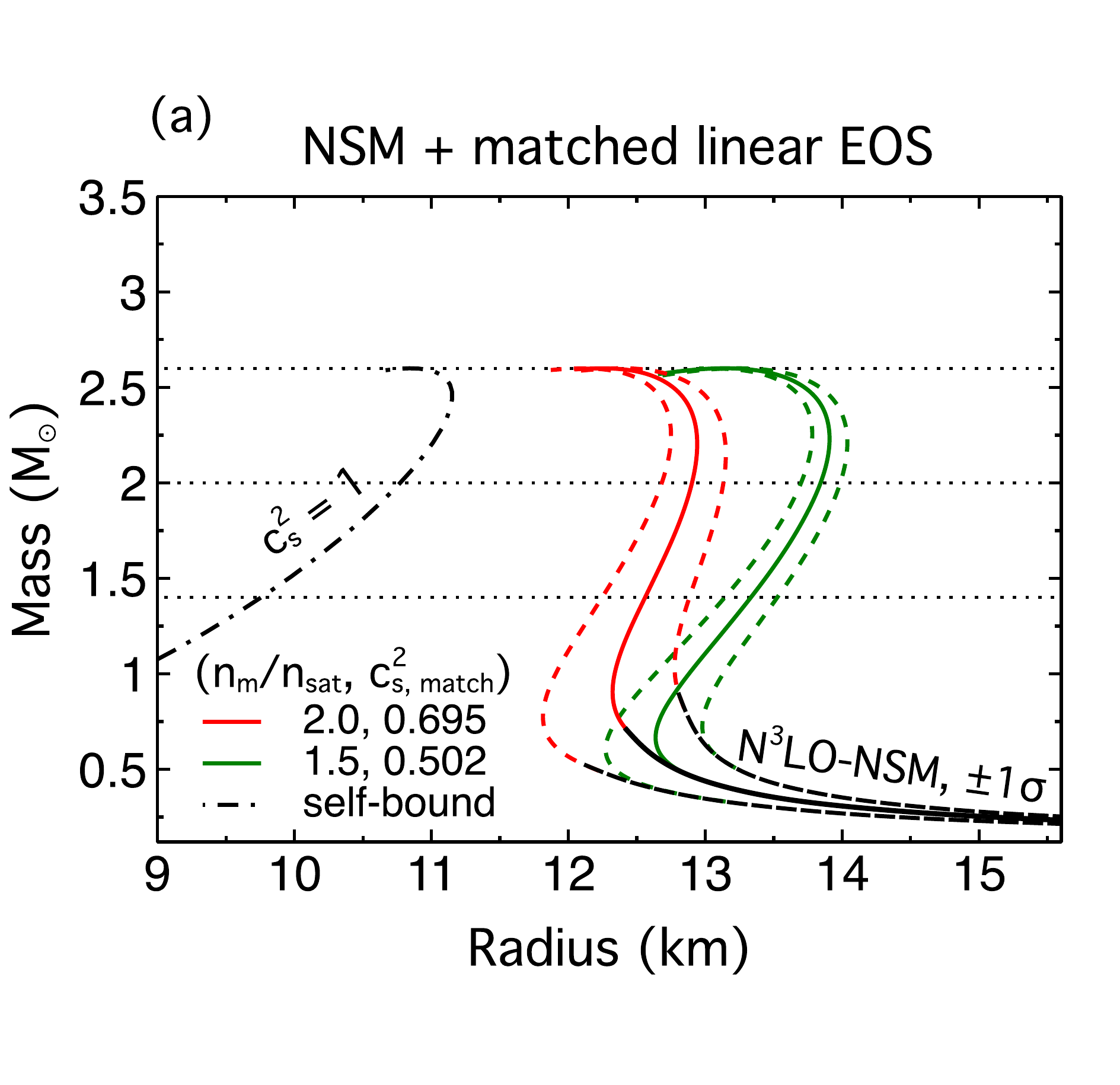}\\[-4ex]
}\parbox{0.5\hsize}{
\includegraphics[width=\hsize]{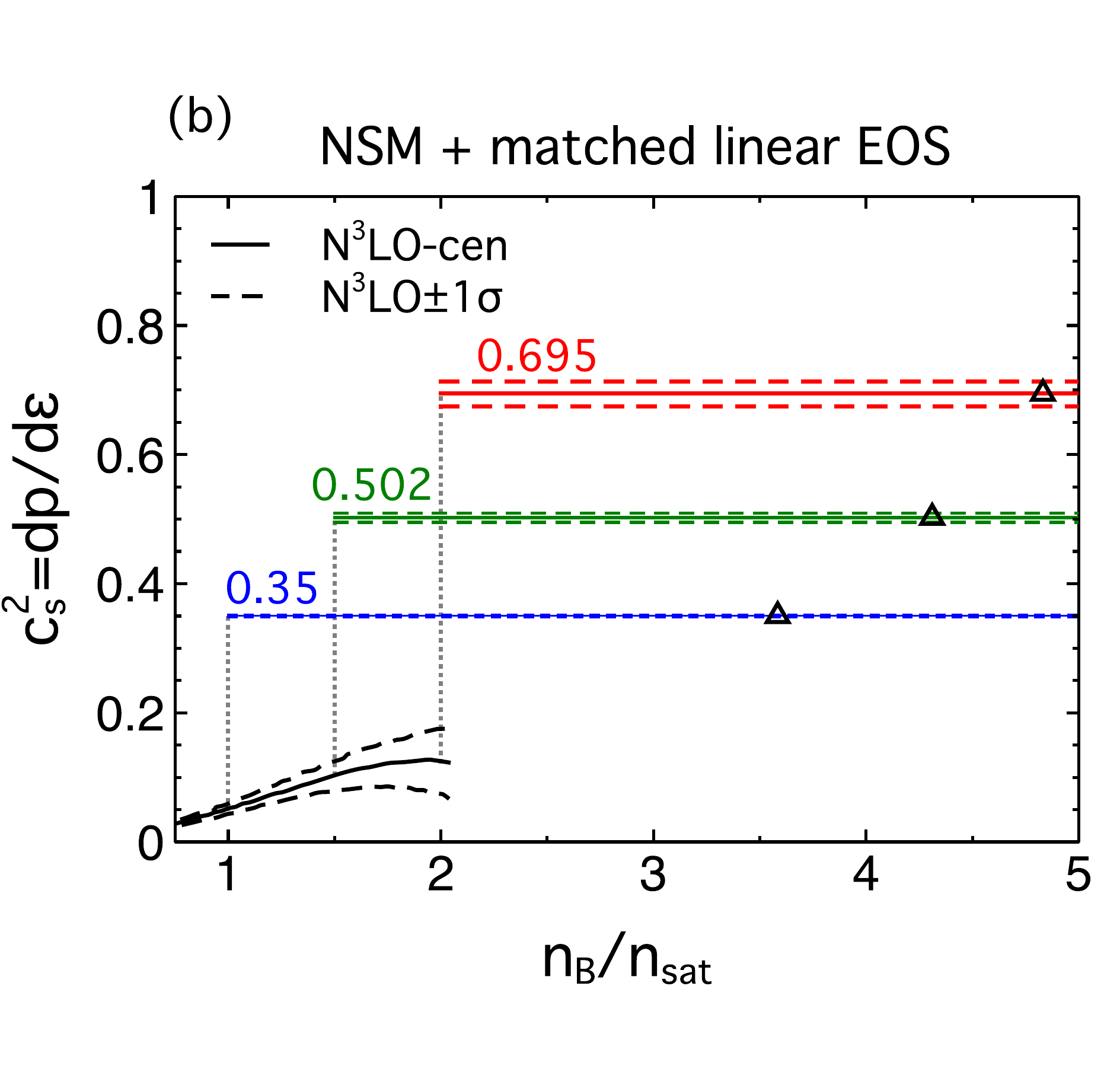}\\[-4ex]
}
\caption{Panel (a): $M$--$R$ diagram for matched linear EOSs that give rise to $\Mmax=2.6\,\Msolar$ with \NNNLO-NSM ($\pm 1\sigma$) 
applied for low densities $\leq 2.0 \,\nsat$. Corresponding values of $\nmat$ and $\csqmat$ are indicated (see also Fig.~\ref{fig:Mmax_csq_nB} (a)). Panel (b): sound speed profiles $\csq(\nb)$ for \NNNLO-NSM only (black-solid for the central value and black-dashed for $\pm 1\sigma$ uncertainties), and matched linear EOSs with different values of $\csqmat$ associated with $\Mmax=2.6\,\Msolar$ in panel (a) (colored horizontal lines). The open triangles mark the central densities of the maximum-mass stars $\Mmax=2.6\,\Msolar$.
}
\label{fig:MR_csq_nB-M26}
\end{figure*}

It is also of interest to examine what 
matching conditions relating $\nmat$ and $\csqmat$ ensue from a restriction such as $\Mmax=2.6\,\Msolar$. Fig.~\ref{fig:MR_csq_nB-M26}~(a) depicts the $M$--$R$ relations for 
$\nmat$ and $\csqmat$ that lead to $\Mmax=2.6\,\Msolar$, and the corresponding $\csq$ profiles are explicitly shown in panel~(b). The required values of $\csqmat$ are indicated in the plot (solid horizontal lines for the \NNNLO-central (denoted as \NNNLO-cen) and dashed for $\pm 1\sigma$ uncertainties), which increase with the matching density $\nmat$. At fixed matching density indicated by the vertical dotted lines, the variation in $\csqmat$ above $\nmat$ is consistent with the uncertainties in $\csq$ from \chiEFT calculations 
at $\nmat$, and a softer EOS (smaller $\csq$) at low densities is compensated by a stiffer EOS (larger $\csqmat$) at higher densities.

The simple linear parametrization of high-density EOS used here can be viewed as a guide to assess the stiffness required at higher densities to achieve $\Mmax\geq2.6\,\Msolar$. Assuming \chiEFT-\NNNLO is valid up to $\nmat = 2.0 \,\nsat$ ($1.5 \,\nsat$), to reach  $2.6 \,\Msolar$ the ``averaged'' $\csq$ above $2.0\,\nsat$ ($1.5 \,\nsat$) has to be greater than $\sim 0.7$ ($\sim 0.5$).
This is probably not achievable by using standard extrapolations of nonrelativistic nucleonic models 
(for which 
$\csq$ is gradually increasing) 
without violating causality 
below the central density of the maximum mass star. 

\begin{figure*}[htb]
\parbox{0.5\hsize}{
\includegraphics[width=\hsize]{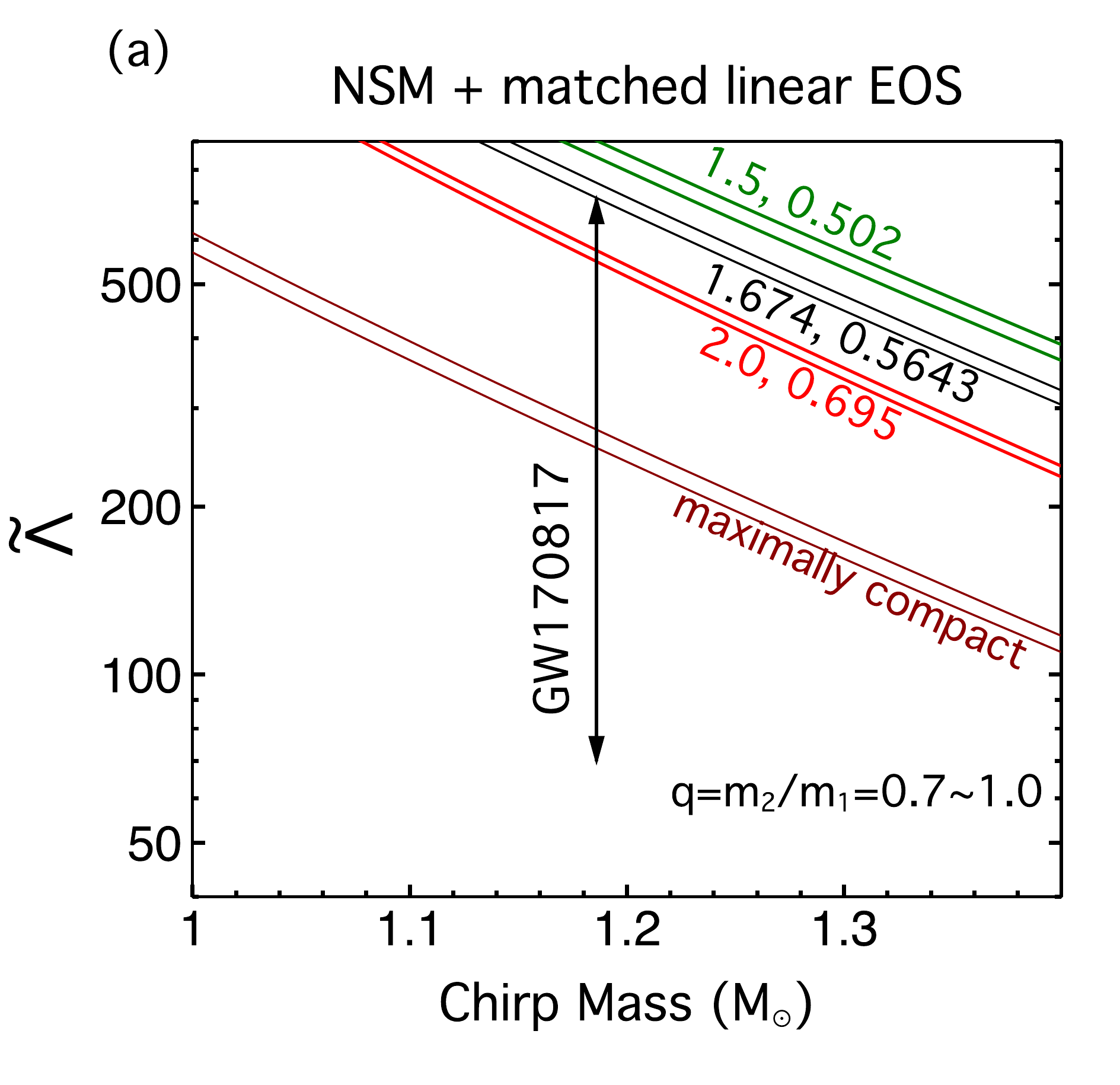}\\[-4ex]
}\parbox{0.5\hsize}{
\includegraphics[width=\hsize]{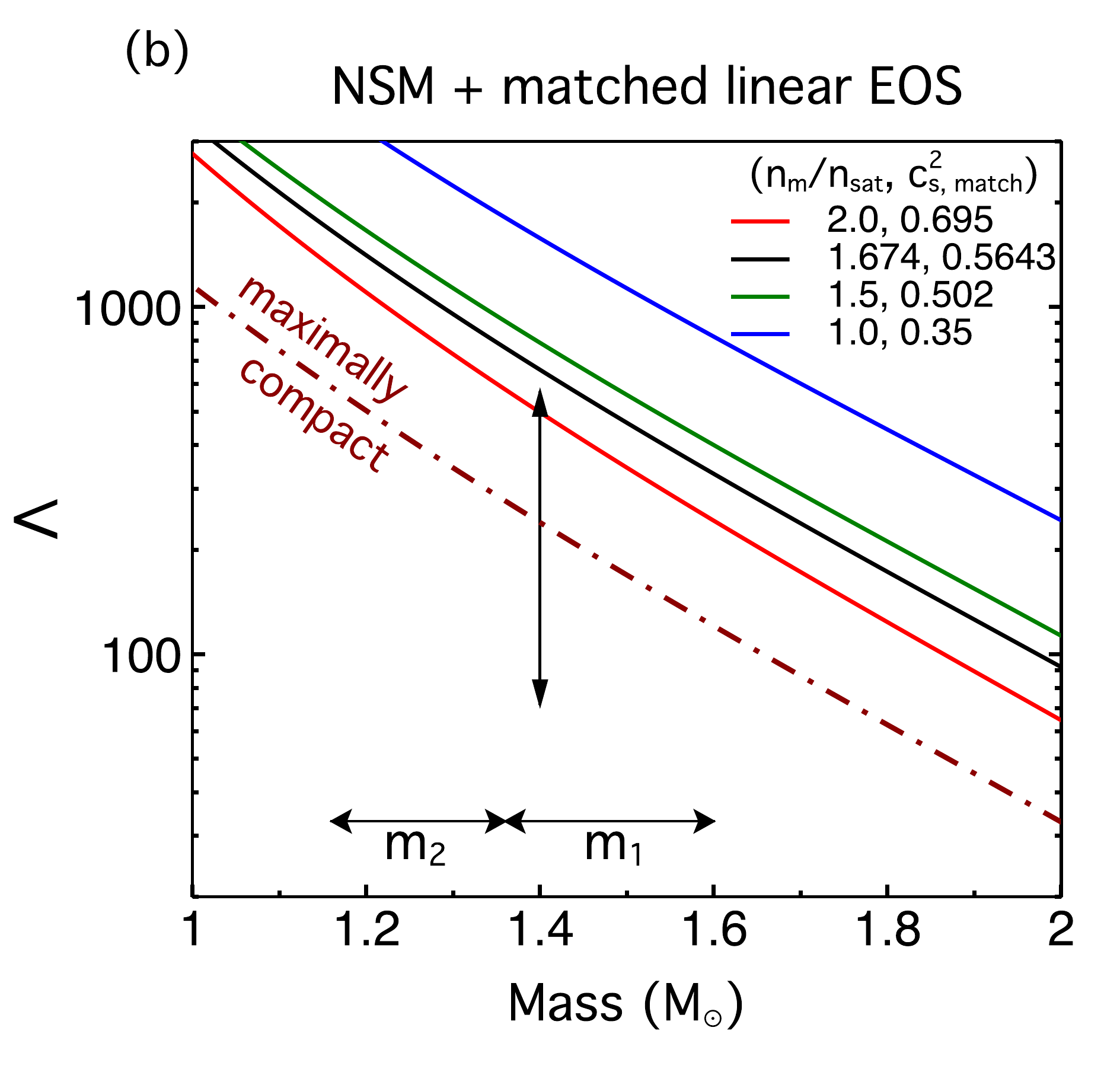}\\[-4ex]
}
\caption{$\tilde\La$--$\Mchirp$ and $\La$--$M$ relations confronted with constraints from GW170817~\cite{Abbott:2018wiz,Abbott:2018exr} (vertical lines with arrows), with fixed $\Mmax=2.6\,\Msolar$ as an example.
Parameters for matched EOSs are the same as in Fig.~\ref{fig:MR_csq_nB-M26}, except for the special case with $\nmat=1.674\,\nsat$ (with $\csqmat=0.5643$), which refers to the minimum matching density that survives $\tilde\La (\Mchirp=1.186\,\Msolar)\leq720$.
}
\label{fig:LamM_Lamwt_Mc}
\end{figure*}

Fig.~\ref{fig:LamM_Lamwt_Mc}~(a) shows an application of the deformability constraints from GW170817 
(see discussions in Sec.~\ref{sec:tidal}) in the case that $M_{\rm max}$ is fixed to $2.6\,\Msolar$. 
As mentioned before, a small matching density $\nmat$ results in a large radius for a given $\csqmat$. 
An EOS stiffening drastically from \NNNLO below 
$1.5\,\nsat$  ends up violating ${\tilde\La}_{1.186}\leq720$ 
if $\Mmax\gtrsim2.6\,\Msolar$ (green band in Fig.~\ref{fig:LamM_Lamwt_Mc}~(a)).

Even considering the $q$ and EOS uncertainties, one sees that $\nmat\lesssim1.5\,\nsat$ violates the GW170817 constraint (Fig.~\ref{fig:LamM_Lamwt_Mc}~(b)). 
There exists a minimum 
$\nmat\approx 1.7\,\nsat$ for \NNNLO-NSM to survive the $\tilde\La\leq720$, or $\Lambda\leq580$ constraint (when $\Mmax\geq2.6\,\Msolar$ is assumed), and an even smaller upper bound e.g., $\tilde\La\simeq 600$~\cite{De:2018uhw,Landry:2018prl,Capano:2019eae}  which would increase the minimum required $\nmat$. 
It is noteworthy that the posteriors of $\tilde\La$ for GW170817 suggest a peak value around $\approx 225$, noticeably smaller than the upper bound of 720 (90\% credible level).

The ranges of $1.6-2.5\,\Msolar$ in  GW190425~\cite{Abbott:2020uma} and $2.59_{-0.09}^{+0.08} \,\Msolar$ in GW190814~\cite{Abbott:2020khf} for one of the components in these merger events have raised the possibility that those compact objects could be NSs as opposed to being low-mass BHs. 
The data from GW190425 was inconclusive concerning the nature of the inspiralling binary~\cite{Abbott:2020uma}, but some works  favored the scenario in which the more massive component is a BH instead of a very heavy NS~\cite{Foley:2020kus}.  If it is \emph{a priori} assumed that $\Mmax\lesssim 2.3\,\Msolar$, a possibility motivated by 
EM and GW 
data from GW170817, the interpretation that it was a BNS merger instead statistically favors masses of approximately $1.5\pm0.2\,\Msolar$ and $1.9\pm0.2\,\Msolar$, while a 
neutron-star-black-hole (NSBH)
merger interpretation favors a $1.3\pm0.1\,\Msolar$ NS and a $2.2\pm0.2\,\Msolar$ BH~\cite{Foley:2020kus}. 
While both scenarios are statistically equally likely, the fact that the BNS masses are incompatible with those of observed galactic BNS systems, while the NS mass in the NSBH scenario is compatible, seems to favor the NSBH interpretation. 
However, in either scenario according to this analysis, GW190425 would likely not contain a NS 
$>2.1\,\Msolar$.  In the case of GW190814, there is no additional information, aside from one's assumption about $\Mmax$, to decide if the primary is a high-mass NS or a low-mass BH.
However, 
statistical analyses suggest that the probability of its secondary being a NS is very low~\cite{Abbott:2020khf,Tews:2020ylw,Essick:2020ghc}.  
If either GW190425 or GW190814 contains a $\sim2.5-2.6\,\Msolar$ NS, 
questions to address are:  
\emph{What is the physical state of dense matter that could support such a heavy NS, and what radius constraints would follow?}

The scenario that GW190814's secondary component 
was an approximately $2.6\,\Msolar$ NS does not itself violate theoretical limits from causality and the GW170817 constraint that $\tilde\La<720$ for $\Mchirp=1.186\,\Msolar$, but challenges 
remain finding physical 
mechanisms
that can connect very stiff high-density matter with the  
relatively soft nuclear matter at $\lesssim 2.0\,\nsat$ predicted from modern \chiEFT calculations. 
As Fig.~\ref{fig:Mmax_csq_nB} shows, 
the conformal limit $\csq\leq1/3$ must be violated~\cite{Landry:2020vaw} below the central density of the maximum-mass star even by the requirements from pulsar timing that $\Mmax\gtrsim2.1\,\Msolar$ and from GW170817's tidal deformability constraint. 
Standard extrapolations that assume 
gradually increasing 
$\csq$ profiles are unlikely to be compatible with $\Mmax\geq2.6\, \Msolar$~\cite{Abbott:2020khf}. 

In particular, the requirement that 
$\csq$ remains 
above $\sim 0.6$ 
for a wide range of densities $\gtrsim2.0\,\nsat$ is hard to explain. 
Extrapolations of non-relativistic potential models generally result in steadily increasing sound speeds with density, and it becomes problematic to prevent them from becoming acausal within NSs. 
At densities relevant to the center of very massive NSs, it is reasonable to expect the emergence of exotic degrees of freedom. A sharp first-order transition to stiff quark matter at some intermediate density is capable of reconciling small radii and high masses $\gtrsim2.4\,\Msolar$ (see examples of $R_{\rm min} (M)$ in Sec.~\ref{sec:rmaxandlambda}). With an increasing lower bound on $\Mmax$ and/or smaller assumed values of $\csq$ at high densities, the transition threshold has to be pushed downward approaching $1.5-2.0\,\nsat$ (similar to the results shown in Fig.~\ref{fig:Mmax_csq_nB} but involving a discontinuity $\De\emat$ that further decreases $\Mmax$ and favors lower values of $\nmat$~\cite{Han:2020adu}). 

For most 
microscopic quark-matter models, for example the original MIT bag model~\cite{Baym:1976yu}, 
the original Nambu-Jona--Lasinio (NJL) model~\cite{Nambu:1961tp}, 
and their variations, 
perturbative QCD matter~\cite{Kurkela:2009gj}, and quartic polynomial parametrizations~\cite{Alford:2004pf}, the speed of sound turns out to be weakly density-dependent. 
To be consistent with massive pulsars $\sim2\,\Msolar$, strong repulsive interactions that stiffen the quark EOS, possibly reaching $\csq\geq 0.4$, have been implemented~\cite{Kojo:2014rca,Klahn:2015mfa,Gomes:2018eiv}. The maximally achievable $\csq$ is model-dependent, and requiring $\csq\gtrsim 0.6$ on average in quark matter is expected to push model parameters to extreme values. 

In contrast to sharp phase transitions, hadron-to-quark crossovers as in quarkyonic models~\cite{McLerran:2018hbz,Han:2019bub,Zhao:2020dvu} or with interpolation schemes~\cite{Baym:2017whm} provide a natural stiffening to support high masses, but 
can 
also induce large radii. 
Quarkyonic models generate large values of $\csq$ by restricting the nucleonic momentum phase space when quarks appear, and in some cases are capable of simultaneously reaching $>2.5\,\Msolar$ and satisfying the GW170817 constraint $\tilde\La_{1.186}<720$.  Some versions~\cite{Zhao:2020dvu}, in which quarks come to rapidly dominate the composition, leading to a high, but narrow, $\csq$ peak behavior, cannot jointly satisfy these conditions, reaching at most $\Mmax\simeq2.4\,\Msolar$. 
However, we find that other versions~\cite{McLerran:2018hbz,Han:2019bub}, in which the quark abundances grow more slowly and that can retain large abundances of nucleons at high density, can simultaneously achieve these conditions. 

Using extrapolation functions in terms of $\csq$ and $\mu$, 
Annala~\etal~\cite{Annala:2019puf} 
found that the risk of hadronic EOSs violating causality at high-enough densities ($\gtrsim4.0\,\nsat$) to achieve high masses is remedied if a transition to perturbative QCD-like (soft $\csq\approx1/3$) quark matter occurs at high densities. 
However, considering that current calculations in perturbative QCD itself are only valid at densities $\nb\gtrsim40\,\nsat$, interpolations down to NS densities are problematic. The main feature of such a transition can be reproduced by simply requiring $\csq\to1/3$ for $\nb\gtrsim6\,\nsat$, but 
at intermediate densities 
the conformal limit $\csq\leq1/3$ 
being violated is strongly favored~\cite{Landry:2020vaw}. 
Moreover, despite the fact that hadronic matter breaking the causal limit is never a necessity, it is
nearly impossible to distinguish such high-density transitions using observations of the $M$--$R$ relation or tidal deformabilities due to the masquerade problem~\cite{Alford:2004pf}.

\subsection{Comparison with other works} 
%
As noted earlier, the uncertain nature of the less compact object in GW190814 with mass $\simeq 2.6\,\Msolar$ has piqued the interest of the dense-matter and nuclear-physics communities. 
Below we briefly discuss how our study differs from or complements the findings of several other recent articles~\cite{Tan:2020ics,Lim:2020zvx,Tews:2020ylw,Essick:2020ghc,Tsokaros:2020hli,Fattoyev:2020cws,Godzieba:2020tjn,Kanakis-Pegios:2020jnf} that have addressed the implications of the possible existence of NSs with such high masses.

Several of these articles, including Refs.~\cite{Tan:2020ics,Tews:2020ylw, Tsokaros:2020hli}, have relied on nuclear physics based EOSs to describe matter in the crust and outer core to show that the existence of a $2.6~\Msolar$ NS would require
$\csq \geq 0.6$ in the inner core. The authors of Ref.~\cite{Kanakis-Pegios:2020jnf} use the upper bound on the tidal deformability of NSs set by GW170817 to further strengthen the need for a large $\csq$ in the inner core. 
Most notably, Ref.~\cite{Godzieba:2020tjn} derives strict upper bounds on the maximum mass of NSs that depend only on bulk properties of NSs, such as the radii and the tidal deformabilities to find that  a NS in GW190814 would not be inconsistent with present astronomical constraints if $\csq$ is large in the inner core. Our finding 
suggests that
a $2.6~\Msolar$ NS would require 
$\csq \geq 0.55-0.6$ (see Fig.~\ref{fig:Mmax_csq_nB}~(a)) 
in the inner core, which is in general agreement with these earlier studies. A unique feature of our study is the use of the \NNNLO-\chiEFT EOS that allows us to properly incorporate EFT truncation errors at $\nb \leq 2.0\, \nsat$.

Lim~\etal~\cite{Lim:2020zvx} combine nuclear models valid in the vicinity of normal nuclear densities and a maximally stiff EOS at higher density to show that $2.5-2.6\,\Msolar$ NS can exist without strongly affecting the properties such as radius, tidal deformability, and moment of inertia of canonical NSs with mass $\sim1.4\,\Msolar$. They argue that properties of NSs with masses $\sim 2\,\Msolar$ 
such as $R_{\sim 2.14}$ 
would be significantly different depending on whether the secondary 
component of GW190814 
was a black hole or a NS. 
Our results support these findings, but go beyond by delineating how the lower and upper bounds on the radii of NSs in the mass range $1.4-2\,\Msolar$ would be constrained if future observations were to confirm the existence of NSs with masses $\simeq 2.5-2.6\,\Msolar$.

Using FSU-type relativistic mean field-theoretical (RMFT) models, Fattoyev~\etal~\cite{Fattoyev:2020cws} found that the rapid increase in pressure with density required to support a $2.6~\Msolar$ NS, while 
barely 
accommodating the 
deformability 
constraint from 
the first analysis of GW170817 data that indicates $\La_{1.4}\leq 800$~\cite{LIGO:2017qsa} but not the updated bounds $70\leq\La_{1.4}\leq580$~\cite{Abbott:2018exr} 
(see Ref.~\cite{Huang:2020cab} for a similar study), is 
inconsistent with 
energy density functionals tuned to reproduce properties of nuclei 
and flow data from HICs. 
Note that Fattoyev~\etal~\cite{Fattoyev:2020cws} only applied $\La_{1.4}$ constraint without a comparison of the binary tidal deformability $\tilde\La$. 
We have confirmed that FSU-like RMFT interactions cannot 
accommodate both 
${\tilde\La}_{1.186}\leq720$ 
and $\Mmax\geq2.54\,\Msolar$~\cite{Zhao:2020c}.

Other recent works studied hyperonic matter in the EOS and/or rapid rotations that stabilize more massive stars than non-rotating configurations, which may or may not be consistent with GW190814~\cite{Most:2020bba,Zhang:2020zsc,Dexheimer:2020rlp,Sedrakian:2020kbi}; we do not consider these effects in the present paper.

\section{Conclusion and Outlook}
\label{sec:Concs}

We determined the NSM EOS in beta-equilibrium from MBPT calculations of PNM and SNM up to \NNNLO in \chiEFT. For a given $\nb$, the NSM EOS always has a lower $\ep$ than the PNM EOS. The pressure of NSM is less than PNM at the same $\nb$, 
typically by $<1\MeV \fmiq$, except for $\nb\gtrsim2.0\,\nsat$ when it becomes greater (Fig.~\ref{fig:P_nB-NSM} (b)). 
The proton fraction below $2.0\,\nsat$ never exceeds the critical minimum value required for the direct URCA process of enhanced neutrino emission~\cite{Boguta:1981mw,Lattimer:1991ib}. 

The existence of the NS crust together with a nucleonic EOS below a matching density $\nmat$ establishes $R_{\rm max}(M)$.
Extremes are again found by assuming $\csqmat=1$ for densities above $\nmat$, 
for which the EOS is now $\ep=\emat+P-P_{\rm m}$.  
Assuming 
$\emat=\esat$, 
and that $P_{\rm m}$ 
is given by \chiEFT-\NNNLO,  
the upper bounds are 
$R_{1.4, \rm max}\approx15.1\km$ and $R_{2.0, \rm max}\approx16.2 \km$ 
(see Fig.~\ref{fig:R_nm} where $\nmat=\nsat$), which are nearly identical to the case shown in Fig.~\ref{fig:caus} with a slightly different value of $P_{\rm m}$ at $\emat=\esat$. 
These values are not in tension with observations, 
and with increasing $\nmat$, the corresponding upper bounds on $\Rtyp$ and $\Rtwo$ decrease. 
For the same $\emat$ or $\nmat$, 
$\Mmax$ is not sensitive to the value of $P_{\rm m}$ or the nucleonic EOS between the crust and $\nmat$, and is close to that of the case $P_0=0$ (self-bound stars) for the causal EOS; see also Fig.~\ref{fig:MR_match-EFT}.

The merger events GW190425 and GW190814 are each consistent with at least one component $\gtrsim2.5\,\Msolar$ which could be either a massive NS or a low-mass BH, although GW190425 could instead involve two $\sim1.7\,\Msolar$ NSs. 
Should either system contain a NS with $M\gtrsim2.5\,\Msolar$, the 
implications would be that the conformal limit 
$\csq\leq 1/3$ 
is almost certainly violated (since $\nmat$ is likely larger than $\nsat$);
if $\nmat>1.5\,\nsat~(2.0\,\nsat)$, the average $\csq$ above $\nmat$ should be $>0.5~(0.67)$. 
More importantly, 
in order to also satisfy the small binary tidal deformability inferred from GW170817, $\nmat\gtrsim1.65\,\nsat$ 
(could be lowered if there is sudden softening in the EOS induced by a strong first-order transition) 
and $\csqmat\gtrsim0.6$ are necessary. 
These conditions are 
typically 
not satisfied by most microscopic
quark models 
unless parametrizations with explicit large sound speeds, or 
some 
crossover-like transitions 
that can be realized in, e.g., 
quarkyonic matter, are assumed. 
Even in the crossover scenario, 
severe constraints 
would follow 
and require fine-tuning of model parameters. 

Assuming $\Mmax\ge2.6\,\Msolar$, we find $R_{\rm min}(1.4\,\Msolar)>9.75 \km$ and $R_{\rm min}(2.0\,\Msolar)>10.8 \km$ 
(Table~\ref{tab:scale}). If instead an upper limit $\csq<1$ is assumed so that $\ep=\ep_0+P/\csq$, then $R_{\rm min}(M)$ and $\Mmax$ depend sensitively on
$\csq$ and decrease with it. 
For the case $\csq=1/3$ and $\ep_0=\esat$, for example, $\Mmax=2.48 \,\Msolar$, $R_{\rm min}(1.4\,\Msolar)=12.8 \km$, and $\Rmax=13.3 \km$ (Fig.~\ref{fig:caus}). 

We showed that positive values of $\De R=\Rtwo-\Rtyp$, potentially possible with NICER, would  indicate low matching densities $\lesssim
2.0-2.5\,\nsat$ and relatively large values of $\csqmat\gtrsim 0.45-0.6$, which would also imply large values of $\Mmax$. 
In the absence of a dramatic stiffening of the EOS near $2.0\,\nsat$, the expectation is that $\De R<0$. 
This is usually the case if extrapolations based on nucleonic-like models are used up to even higher densities and/or there is extra softening below $\Mmax$. 

Our studies have highlighted the interplay of $\Mmax$, the radii of NSs, and the role of the nucleonic EOS for densities beyond $\nsat$. We also have illustrated that systematic order-by-order calculations up to \NNNLO in the \chiEFT expansion provide an EOS for NSM up to $\sim 2.0\,\nsat$ whose EFT truncation errors~\cite{Drischler:2020hwi,Drischler:2020yad} are small enough to  have relatively minor influence on our major conclusions. 
Nevertheless, our results also reveal that theoretical studies at $\nb \gtrsim 2\,\nsat$ can have a significant impact on NS properties, especially on the correlation between $\Mmax$ and the NS radii. Detailed studies of EFT truncation errors at these higher densities and for a wide range of chiral interactions   would be valuable. This requires the development of improved order-by-order \chiEFT NN and 3N potentials within different regularization schemes~\cite{Hoppe:2019uyw,Huther:2019ont,Epelbaum:2019kcf}. Further, models that include additional degrees of freedom such as pions, hyperons, and quarks (while still being able to accommodate massive NSs) can provide new insights but need to be improved. Work along these lines is in progress. The advances in nuclear-matter calculations from \chiEFT at low densities (see, e.g., Refs.~\cite{Drischler:2017wtt,Lonardoni:2019ypg}) combined with Bayesian uncertainty quantification (see, e.g., Refs.~\cite{Wesolowski:2015fqa,Wesolowski:2018lzj,Drischler:2020hwi}) will enable astrophysical applications over a wide range in density and proton fraction, which would soon be confronted with X-ray, radio, and GW observations.

\begin{acknowledgments}

We thank R.~J. Furnstahl, B.-A. Li, J.~A. \mbox{Melendez}, and D.~R. Phillips for useful discussions, and the Network for Neutrinos, Nuclear Astrophysics, and Symmetries (\href{https://n3as.berkeley.edu}{N3AS}) for encouragement and support.
C.D. acknowledges support by the Alexander von Humboldt Foundation through a Feodor-Lynen Fellowship and the U.S. Department of Energy, the Office of Science, the Office of Nuclear Physics, and SciDAC under awards DE-SC00046548 and DE-AC02-05CH11231. 
S.H. is supported by the National Science Foundation, Grant PHY-1630782, and the Heising-Simons Foundation, Grant 2017-228. 
J.M.L. and T.Z. acknowledge support by the U.S. DOE under Grant No. DE-FG02-87ER40317 and by NASA's NICER mission with Grant 80NSSC17K0554. 
M.P.'s research was supported by the Department of Energy, Grant No. DE-FG02-93ER40756. 
The work of S.R. was supported by the U.S. DOE under Grant No. DE-FG02-00ER41132. 

\end{acknowledgments}

\appendix

\section{Bounds Imposed by Causality}
\label{sec:Bounds}

The assumption of causality, i.e., that the maximum sound speed $c_s=\sqrt{dP/d\ep}$ is unity in units of $c$, 
can establish relations limiting both minimum and maximum radii, as functions of mass, for NS.  These limits will explicitly depend on assumptions concerning the NS maximum mass $\Mmax$. 
These causal bounds can be improved with the consideration of nuclear physics inputs as will be discussed in Sec.~\ref{sec:Results}.
The causality limit is imposed by using the EOS
\beq 
\label{eq:ceos}
P(\ep) = P_0+ (\ep-\ep_0)
\eeq
for 
the pressure 
$P>P_0$ and 
the energy density 
$\ep>\ep_0$.

The minimum radius as a function of mass $R_{\rm min}(M)$ for any EOS is conjectured~\cite{Koranda:1996jm} to result from using \Eqn{eq:ceos} with 
$P_0=0$, $P=0$ for $\ep\leq\ep_0$ (i.e., a self-bound star). In this case, the EOS has a single parameter ($\ep_0$) and solutions of the Tolman-Oppenheimer--Volkoff (TOV) equation~\cite{Tolman:1939jz,Oppenheimer:1939ne} scale with it.  Letting $m$ be the mass enclosed within the radius $r$, one can define
\beq
r=x{c^2\over\sqrt{G\ep_0}},\quad m=y{c^4\over\sqrt{G^3\ep_0}},\quad \text{and} \quad
P=z\ep_0,
\label{eq:dim}
\eeq
where $y(x)$ and $z(x)$ are dimensionless functions, with 
the boundary conditions 
$y_c=y(x=0)=0$ and $z_c=z(x=0)>0$ at the stellar center, and $y_s=y(x=x_s)$ and $z(x=x_s)=0$ at the stellar surface $x_s$. The quantities $y_s$ and $x_s$ depend on $z_c$.
For small $x_s$, $y_s\propto x_s^3$, as expected. 
It should also be noted that the EOS~\Eqn{eq:ceos} implies 
that 
the baryon number density is
\beq
\nb=n_0\sqrt{\ep+P\over\ep_0+P_0},
\eeq
with 
$n_0=(\ep_0+P_0)/\mu_0$ 
and 
$\mu_0$ being the baryon chemical potential at $\ep_0$. 

In the case that $P_0=0$, the central baryon density is $\ncent=n_0\sqrt{1+2z_c}$. 
Also, the maximum mass configuration occurs for $dy_s/dx_s=0$, or when $x_{\rm max,s}$ = 0.2405, $y_{\rm max,s}=0.08513$, and $z_{\rm max,c}=2.023$ (and therefore $n_{\rm max,c}/n_0=2.246$). The maximum mass can 
then 
be expressed as
\beq
\Mmax={y_{\rm max,s}c^4\over\sqrt{G^3\ep_0}}\simeq4.09~\sqrt{\esat\over\ep_0}\Msolar,
\label{eq:mmax0}
\eeq
and the radius of the maximum mass configuration is 
\beq
R_{\Mmax}={x_{\rm max,s}c^2\over\sqrt{G\ep_0}}\simeq17.1\sqrt{\esat\over\ep_0}{\rm~km}.
\label{eq:rmax0}
\eeq
The central energy density for the maximum mass configuration is $\ep_{\rm max,c}=(z_{\rm max,c}+1)~\ep_0$, or using \Eqn{eq:mmax0} to eliminate $\ep_0$,
\beq
\ep_{\rm max,c}\simeq50.8\left({\Msolar \over \Mmax}\right)^2\esat,
\eeq
where $\esat\simeq 150 \MeV \fmiq$ is the energy density at~$\nsat$.  This must be the largest energy density found in any NS and it scales with $\Mmax^{-2}$. The maximum baryon density is 
\beq
n_{\rm max,c}\simeq37.6{m_{\rm B}\over\mu_0}\left({\Mmax\over \Msolar}\right)^2\,\nsat,
\label{eq:nmax0}
\eeq
where $\mu_0\sim m_{\rm B}$, the baryon mass.  As an example, if one assumes that $\Mmax=2.6\,\Msolar$ and $\mu_0=m_{\rm B}$, it is found that $\ep_0=2.475\,\esat$,  $\ep_{\rm max,c}=7.48\,\esat$ and $n_{\rm max,c}=5.56\,\nsat$.

\begin{figure}[htbp]
\begin{centering}
\includegraphics[width=1.04\hsize]{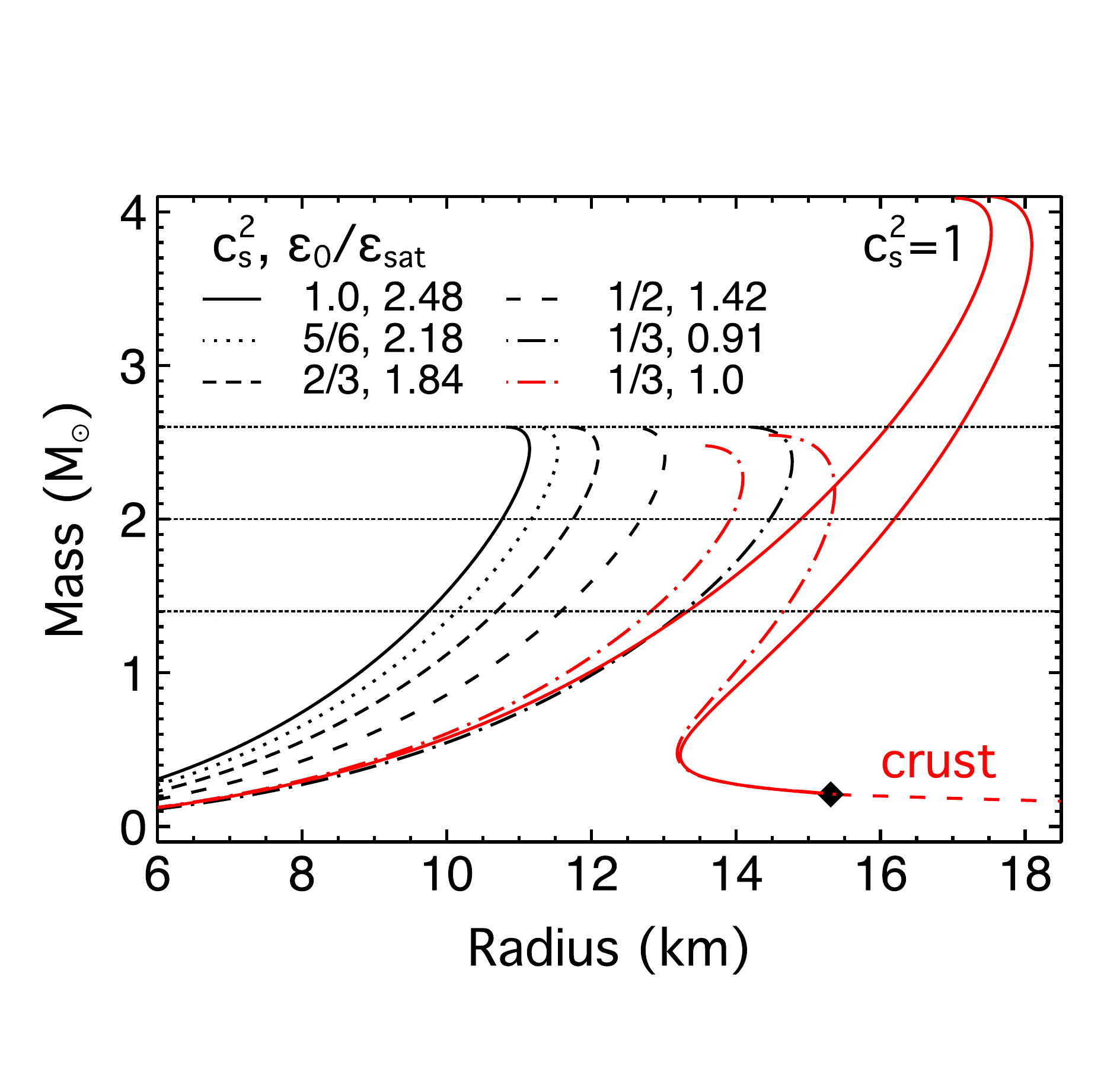}\\[-1ex]
\end{centering}
\caption{
The mass as a function of the radius for the EOS~\Eqn{eq:cseos} with
$P_0=0$, $P=0$ for $\ep<\ep_0$, and various values for $\csq$ with fixed $\Mmax=2.6\,\Msolar$, are shown as five black curves (see legend). 
These curves correspond to the minimum possible radius $R_{\rm min}(M)$, for different maximum values of the sound speed. 
The four red curves correspond to 
$\ep_0=\esat$, either $\csq=1$ (and $\Mmax\simeq4.09\,\Msolar$) or $\csq=1/3$ (and $\Mmax\simeq2.48\,\Msolar$) for $P>P_0$, and either $P_0=0$ (self-bound) or  $P_0=0.02\,\ep_0\simeq3\MeV \fmiq$ and a normal crust EOS for $P<P_0$ 
(maximum possible radii $R_{\rm max}(M)$); 
the configuration where $\ep_c=\esat$ is indicated by a diamond. 
}
\label{fig:caus}
\end{figure}

The dimensionless $M$--$R$ curve for the causal self-bound configuration is thus defined by $y_s(x_s)$.  Its \emph{dimensionful} radius, as a function of mass, is conjectured to be  the minimum radius for any configuration, $R_{\rm min}(M)$.  It scales with $\ep_0$ and therefore with the assumed value of the maximum mass:
\beq
\hspace*{-0.4cm}R_{\rm min}={G\Mmax x_s\over y_{\rm max,s}c^2}=
\frac {G\Mmax}{y_{\rm max,s}c^2}\,
y_s^{-1}\!\left(y_{\rm max,s} \frac {M}{\Mmax}\right),
\eeq
where $y_s^{-1}=x_s$ is the inverse function.  $R_{\rm min}$ increases as $\Mmax$ increases. 
The cases with $\Mmax=2.6\,\Msolar$ and $\Mmax=4.09\,\Msolar$ for which $\ep_0$ is $2.48\,\esat$ and $\esat$, respectively, 
are shown in Fig.~\ref{fig:caus}. 
For the case that $\Mmax=2.0\,\Msolar$ for which $\ep_0=4.2\,\esat$, we obtain $R_{\rm min}(1.4\,\Msolar)=8.2\km$ and $\Rmax=8.4\km$.

If the assumed maximum sound speed is less than $c$, 
$R_{\rm min}(M)$ will increase. Assuming the sound speed never exceeds a 
given 
value of 
$c_s$, $R_{\rm min}(M)$ can be found using
\beq
P=P_0+\csq(\ep-\ep_0),
\label{eq:cseos}
\eeq
with $P_0=0$ and $P=0$ for $\ep<\ep_0$. Once again, the TOV equation can be rendered into dimensionless form using \Eqn{eq:dim}. Now, however, the baryon number density becomes
\beq
\nb= n_0\left({P+\ep\over P_0+\ep_0}\right)^{1/(1+\csq)}
\eeq
and
\beq
\ncent=n_0\left[1+z_c\left(1+c_s^{-2}\right)\right]^{1/(1+\csq)}.
\eeq
The dimensionless $M$--$R$ curve $y_s(x_s)$ changes, as do the properties of the maximum mass configuration $x_{\rm max,s}$, $y_{\rm max,s}$ and $z_{\rm max,c}$.  Figure~\ref{fig:caus} shows $M$--$R$ solutions for $\csq=1,5/6,2/3,1/2$, and $1/3$, all scaled so that $\Mmax=2.6\,\Msolar$. 
In the case $\csq=1$, one finds
\bea
R_{\rm min,1.4}=9.75 \km
\,\, {\rm and} \,\, 
R_{\rm min,2.0}=10.8 \km \,.
\eea
Approximately, the minimum radii for smaller values of $c_s$ scale as $c_s^{-1/2}$~\cite{Lattimer:1990zz}, and for $\csq=1/3$, one finds that
\bea
R_{\rm min,1.4}\simeq 13.3~ {\rm km}
\,\, {\rm and} \,\, 
R_{\rm min,2.0}\simeq 14.5~{\rm km}\,.
\eea
$\ep_{\rm max,c}/\esat$ is proportional to $z_{\rm max,c}+1$,
which for $\csq<1$, is seen to scale roughly as $c_s^{3/2}$.
Relevant properties of these solutions are given in Table~\ref{tab:scale}.

Stars with $P_0=0$ are often referred to as \emph{self-bound} stars. 
In contrast, normal 
NSs have a low-density crust 
with $P_0 > 0$.
For normal stars, 
$R_{\rm min}(M)$ will be larger than those 
shown in Fig.~\ref{fig:caus}. Generally, the radius will increase with the assumed values of 
$\ep_0$ and $P_0$ for a given value of 
$c_s$, and, to a lesser degree, will also depend on the crust EOS for $P<P_0$. Most importantly, since $\Mmax$ and $\ep_0$ remain closely related, $R_{\rm min}(M)$ will be very sensitive to the lower limit to $\Mmax$.   Details and implications are discussed in Sec.~\ref{sec:rmaxandlambda}.

\begin{table}[tb]
\caption{Maximum mass solutions for the EOS~\Eqn{eq:cseos} with $P_0=0$. The last two columns give the minimum radii in km for $1.4\,\Msolar$ and $2.0\,\Msolar$ stars, respectively, assuming  $\Mmax=2.6\,\Msolar$.
} 
\begin{center}
\begin{ruledtabular}
\begin{tabular}{cccccc}
$\csq$ & $x_{{\rm max},s}$ & $y_{{\rm max},s}$ & $z_{{\rm max},c}$ & $R_{{\rm min},1.4}$ &$R_{{\rm min},2.0}$\\[2pt] \hline
1 & 0.2405 & 0.08513 & 2.023 &9.75 &10.8\\
5/6 &0.2329 & 0.07992 & 1.884&10.1 &11.2 \\
2/3 & 0.2234 & 0.07328 &  1.705 &10.7 &11.7\\
1/2 & 0.2105 & 0.06439 &1.499 &11.6 &12.7\\
1/3 & 0.1908 & 0.05169 & 1.277 &13.3 &14.5
\label{tab:scale}
\end{tabular}
\end{ruledtabular}
\end{center}
\end{table}

Ironically, the maximum radius as a function of mass $R_{\rm max}(M)$ can also be found by appending the same EOS~\Eqn{eq:cseos} at a matching density $\nmat$ or $\emat$ onto an assumed lower-density (crust) EOS. 
This is because \Eqn{eq:cseos} is the stiffest possible EOS for an assumed maximum value of the sound speed $c_s$. Although the same EOS is used, the $R_{\rm min}(M)$ bound involves a finite surface energy density $\ep_0=\emat$, while the $R_{\rm max}(M)$ bound is assumed to lack a discontinuity in $\ep$ when appending the crust\footnote{Note that if a discontinuity in $\ep$ is assumed at $\emat$, a smaller $R_{\rm max}$ trajectory is obtained, but one with a correspondingly smaller $\Mmax$ as well.  This situation is briefly discussed in Sec.~\ref{sec:compact}.}. 
The resulting $R_{\rm max}(M)$ trajectory, and $\Mmax$, will depend on the matching density $\emat$ and pressure $P_{\rm m}$, the crust EOS, and assumed maximum sound speed $c_s$, 
and both roughly scale as $\emat^{-1/2}$. Since there is no evidence that a transition to a non-hadronic EOS occurs for densities smaller than $\esat$, a limiting set of $R_{\rm max}(M)$ curves is found assuming $\emat=\esat$. 
As the matching pressure $P_{\rm m}$ is not negligibly small, 
$P_{\rm m}\simeq0.02\,\esat$ for $\emat\simeq\esat$, 
the $M-R$ curve is considerably altered, and forms a maximum radius trajectory 
$R_{\rm max}(M)$ 
which 
lies at a larger radius for each mass than $R_{\rm min}(M)$, as 
can be 
seen by comparing the two solid red curves for $\csq=1$ in Fig.~\ref{fig:caus}. 
$R_{\rm max}$ for $\csq=1$ can be safely assumed to give, approximately, the largest possible radii for normal NS (it varies with the assumed EOS below $\emat$). 
It is interesting that 
the maximum masses 
with 
$\emat=\esat$ for a self-bound star (left red solid curve) 
and 
for a normal star with a crust 
(right red solid curve) 
are nearly identical and are substantially larger than $2.6\,\Msolar$, for example. Lower maximum masses are obtained if the matching density is increased, which decreases $R_{\rm max}(M)$ as well. An observed upper limit on $\Mmax$ below $4.09\,\Msolar$ will automatically alter the $R_{\rm max}$ boundary, however, because in this case either $\emat$ would have to increase or $c_s$ would have to decrease to correspondingly reduce $\Mmax$.

The situation is similar if a lower 
fixed sound speed is assumed. Figure~\ref{fig:caus} also displays $R_{\rm min}(M)$ and $R_{\rm max}(M)$ trajectories for $\csq=1/3$ for the self-bound and realistic crust cases (the left and right red 
dot-dashed 
curves, respectively), which have smaller radii and $\Mmax$ values than for $\csq=1$. 
Note that $R_{\rm max}(M)$ for $\csq=1/3$ 
(right red dot-dashed curve) 
can become smaller than $R_{\rm min}(M)$ for $\csq=1$ and $P_0=0$ (left red solid curve) for $M\gtrsim2.3\,\Msolar$, 
suggesting that $\csq=1/3$ is incompatible with the assumption that $\Mmax=2.6\,\Msolar$; the maximum value of $\csq$ must be larger than $1/3$ in the interior of a $2.6\,\Msolar$ star, or $P_0>0$ (i.e., there is a crust), or $\Mmax<2.6\,\Msolar$.


A more 
realistic maximum radius boundary will depend on both the matching density and the EOS below 
that 
density.  In the next section we discuss realistic constraints on this portion of the EOS stemming from theoretical studies of NSM. 

\section{Chiral interactions used and their nuclear saturation properties}
\label{app:chiral_interactions}

The chiral nuclear interactions this work is based on were constrained in Ref.~\cite{Drischler:2017wtt} as follows: NN potentials by Entem, Machleidt, and
Nosyk~\cite{Entem:2017gor} up to \NNNLO were combined with 3N forces at the same
order and momentum cutoff so as to construct a set of order-by-order NN and
3N interactions. The two 3N low-energy couplings $c_D$ and $c_E$, which govern
the intermediate- and short-range 3N contributions, respectively, at \NNLO were
constrained by the triton binding energy and the empirical saturation point of
SNM\@. Several combinations of $c_D$ and
$c_E$ with reasonable saturation properties could be obtained at \NNLO and
\NNNLO for the momentum cutoffs $\Lambda = 450$ and $500\MeV$. 
A momentum cutoff is a typical scale in the regulator function that is applied to \chiEFT interactions to suppress contributions from high-momentum modes. 
Note that the EFT breakdown scale $\Lambda_b$ is a physical scale inherent to the EFT, whereas the results should not be sensitive to the artificial scale $\Lambda$; in practice, however, this has not yet been achieved in \chiEFT for infinite matter.
The BUQEYE collaboration found that their results do not significantly dependent on which $c_D$ and $c_E$ combination is chosen for a given momentum cutoff. Furthermore, the 3N contributions proportional to $c_D$ and $c_E$ vanish in PNM for nonlocal regulator functions~\cite{Hebeler:2009iv}. Consequently, they considered only one combination for each cutoff, and focused their analysis on the Hamiltonian with $\Lambda = 500 \MeV$, while the results for the $\Lambda = 450 \MeV$ interaction were provided in the Supplemental Material there.

We follow this strategy 
here, 
and note that the residual cutoff dependence is well within the EFT truncation-error estimates at the $1\sigma$ level; i.e., 
for $\Lambda = 450 \MeV$, $P_{\rm PNM}(2.0\,\nsat) = 17.29 \pm 4.56 \MeV \fmiq$ and $E_{\rm PNM}(2.0\,\nsat) = 42.86 \pm 5.01 \MeV$, whereas for $\Lambda = 500 \MeV$, $P_{\rm PNM}(2.0\,\nsat) = 18.53 \pm 5.14 \MeV$ and $E_{\rm PNM}(2.0\,\nsat) = 41.55 \pm 5.77  \MeV \fmiq$. 

Experimental validation of \chiEFT predictions for the EOS of bulk matter relies on comparisons to the empirical saturation point, and constraints on the nuclear symmetry energy and its derivative with respect to density at $\nsat$. While the region in the $S_v$--$L$ plane predicted by the nuclear interactions used in this work are well within the joint experimental constraint~\cite{Drischler:2020hwi}, the $\Lambda = 500 \MeV$ Hamiltonians---as discussed in Ref.~\cite{Drischler:2017wtt}---actually do not saturate inside the empirical range for the saturation point, $\nsat = 0.164 \pm 0.007 \fmiq$ with $(E/A)_{\rm sat} = -15.86 \pm 0.57 \MeV$. Note, however, that this empirical range was obtained in Refs.~\cite{Drischler:2015eba, Drischler:2017wtt} from a set of energy density functionals, and thus only has limited statistical meaning. The predicted $2\sigma$ confidence ellipses for the nuclear saturation point at \NNLO and \NNNLO are shown in Fig.~9 of Ref.~\cite{Drischler:2020yad}. 

In contrast to the properties of neutron-rich NSM EOS, nuclear saturation in SNM is sensitive to the short- and intermediate-range 3N interactions at \NNLO that do not contribute to the PNM EOS; e.g., the 3N contact interaction ($\propto c_E$) is Pauli-blocked in PNM~\cite{Hebeler:2009iv}. Together with the fact that the proton fraction is small, this means that the nuclear saturation properties are of relatively minor importance for constructing the NSM EOS\@. 
Nonetheless, a better understanding of nuclear saturation properties may help identify and quantify systematic uncertainties in the nuclear interactions. This might also lead to a better understanding of the link between (saturation) properties of infinite matter and medium-mass to heavy nuclei~\cite{Hoppe:2019uyw,Huther:2019ont} to explain why \chiEFT potentials generally tend to underestimate charge radii~\cite{Binder:2013xaa, Lapoux:2016exf, Epelbaum:2019kcf}. In this context, it is worth noting that systematic EFT calculations of the EOS of NSM, which is characterized by a small proton fraction, would obviate the need to rely on the quadratic expansion~\Eqn{eq:quad_exp} (see, e.g., Ref.~\cite{Roggero:2014lga} in which the energy of adding a proton to PNM was calculated). When such calculations become available one can gauge the extent to which the EOS of NSM is correlated with the empirical properties of SNM. 

\section{Sensitivity to EOS density ranges}
\label{sec:Sensitivity}

It is apparent that the limits to NS radii and tidal deformabilities are sensitive to the EOS in the density range $1-3\,\nsat$, precisely where the restrictions from \chiEFT are important.  This is not surprising given the tight correlation between $\Rtyp$ and the NSM pressure for $1-2\,\nsat$ discovered by Ref.~\cite{Lattimer:2000nx}. 
However, up to this point, we have assumed fixed sound speeds above $\nmat$. 
In this section, we 
demonstrate that this correlation is insensitive to the details of the assumed EOS at all relevant densities; furthermore, we 
quantify this correlation and extend it to 
include the quantities 
$\Rtwo$ and $\Mmax$.

We evaluate these correlations by considering several parametrization schemes  to construct families of high-density NSM EOSs at densities larger than about $0.5\,\nsat$, the assumed core-crust boundary. All configurations are assumed to have a crust modeled with the SLy4 EOS~\cite{chabanat1998skyrme}. Each EOS is given as a function of $\nb$ 
only 
and is implicitly considered to represent beta-equilibrium matter.  The parameters for each parametrization scheme are constrained to ensure causality, $\csq\ge0$, a minimum value $\Mmax=2.0\,\Msolar$, a lower limit to the neutron-matter energy and pressure suggested by the unitary-gas conjecture~\cite{Tews:2016sjl} at all supra-nuclear densities, and upper limits to the NSM energy and pressure at $\nsat$ implied by experimental limits of $S_v=36 \MeV$ and $L=80 \MeV$~\cite{Lattimer:2012xj}.  Note that the latter two constraints are broader than the NSM-\chiEFT $\pm1\sigma$ constraints, so that the correlations we find are conservatively expressed. Also, for each parametrization, we have ensured a minimum of 15,000 realizations that satisfy our constraints. 
We quantify a correlation in terms of the covariance 
between two quantities $A$ and $B$, 
\beq
\cov (A,B)=\sum_{i,j}\frac{(A_i-\bar A)(B_j-\bar B)}{\sigma_A \sigma_B}.
\eeq
The $\sigma$'s represent standard deviations.  We take $A=P(\nb)$ and $B=\Rtyp$, $\Rtwo$, or $\Mmax$. 
Here, $j$ ranges over all realizations of a given parameterized EOS and $i$ 
over all values of $\nb$ smaller than the central density of the relevant configuration for $B$.

\begin{figure}[htb]
\includegraphics[width=.95\columnwidth,clip=true]{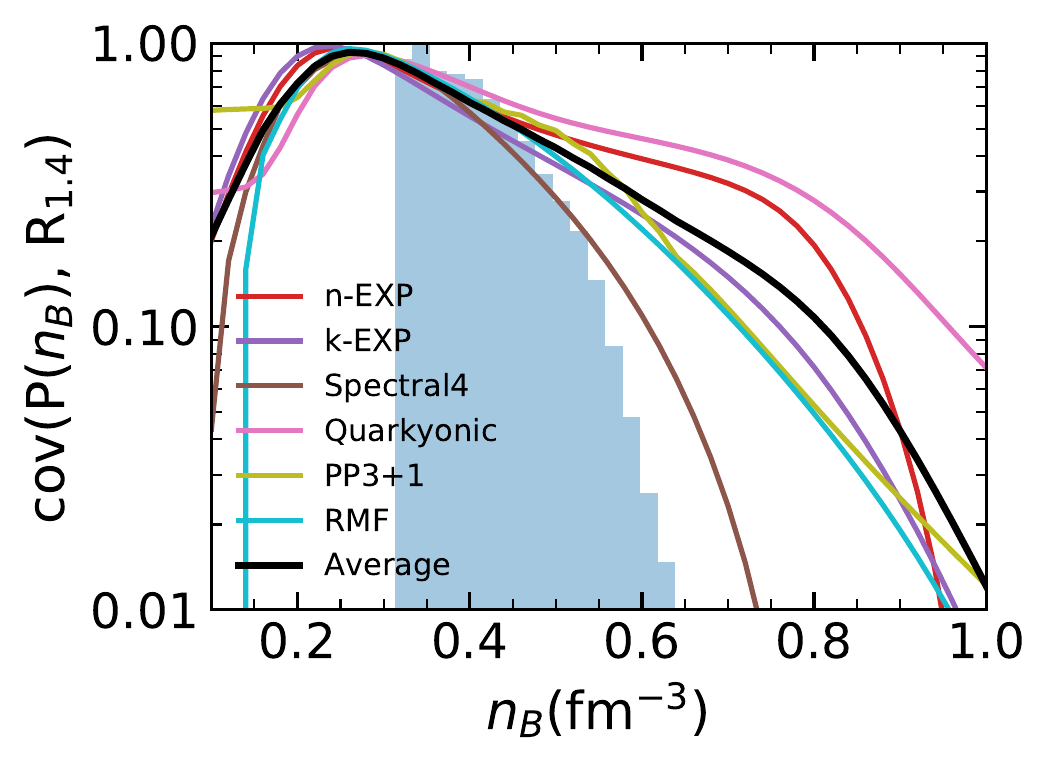}
\includegraphics[width=.95\columnwidth,clip=true]{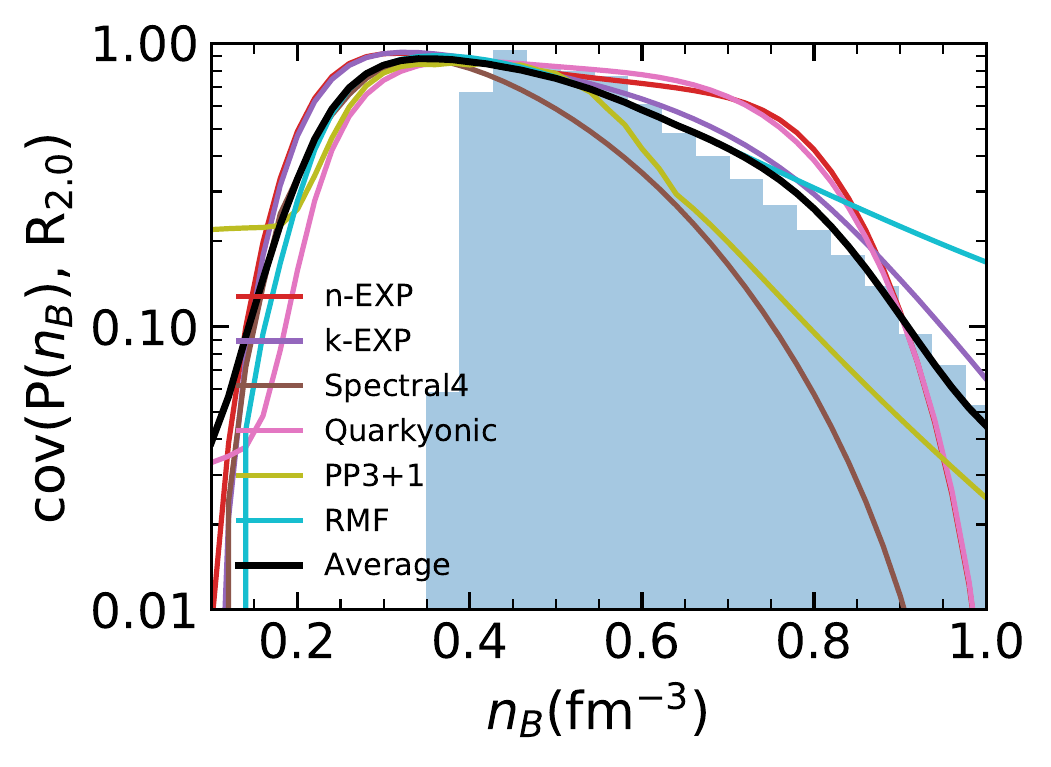}
\includegraphics[width=.95\columnwidth,clip=true]{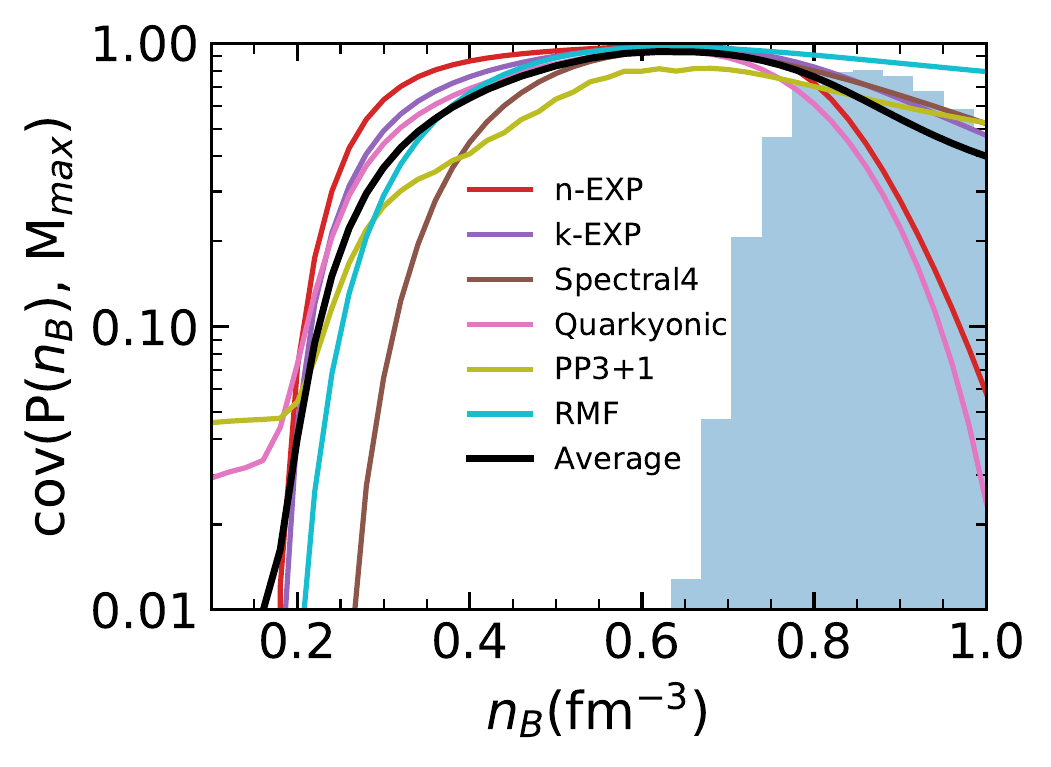}
\vspace*{-.25cm}
\caption{Correlations among $P(\nb)$, $\Rtyp$, $\Rtwo$ and $\Mmax$ for 6 EOS parametrizations (see text for details). 
``Average'' 
refers to 
the mean of all models. Blue histograms show the summed distributions of the central densities of the relevant stars.}
\label{fig:cov-P}
\end{figure}

Figure~\ref{fig:cov-P} 
shows 
the correlations between the pressure $P(\nb)$ and 
$\Rtyp$, $\Rtwo$ and $\Mmax$ 
as functions of 
the baryon number 
density 
$\nb$ 
for a variety of NSM parametrizations in common use. The parametrizations ``n-EXP'' and ``k-EXP'' are three-parameter Taylor expansions of the NSM energy in terms of $\nb$ and $\nb^{1/3}$~\cite{Tews:2016sjl}, respectively. 
``n-EXP'' is commonly used to model the nuclear energy around saturation; we take a Taylor expansion up to the 
fourth-order 
term $\left[(\nb-\nsat)/\nsat\right]^{4}$. 
Two 
of the coefficients 
are set to match the crust EOS, leaving 
three 
free parameters.
``k-EXP'' contains a kinetic term $\propto(\nb/\nsat)^{2/3}$ and a higher-order term up to $(\nb/\nsat)^{7/3}$. It also has three  
free parameters after using 
two 
coefficients to match the crust EOS. ``Spectral4'' is the 
four-parameter 
spectral decomposition method~\cite{Lindblom:2010bb,Lindblom:2012zi,Lindblom:2013kra}. ``Quarkyonic''  has two parameters, $\Lambda$ and $\kappa$, specifying the quarkyonic momentum shell thickness and the transition density, and one parameter (effectively controlling $L$) for the nucleon potential~\cite{McLerran:2018hbz}. ``PP3+1'' is a 
four-parameter 
piecewise-polytrope with 
three  
segments appended to the crust~\cite{Read:2008iy}. The density $n_1$ separating the first two segments is a parameter, while $n_2$ and $n_3$ are chosen to scale as $n_2=2\,n_1$ and $n_3=2\,n_2$.  The corresponding bounding pressures $P_1$, $P_2$, and $P_3$ are the other 
three 
free parameters\footnote{The additional parameter $n_1$ greatly increases the flexibility of PP3+1 compared to the three-parameter 
($P_1, P_2,P_3$) set PP3 often employed~\cite{Read:2008iy}.}.  ``RMF'' is a relativistic mean field model based on the FSU2 EOS~\cite{Horowitz:2001ya} and contains $\sigma$, $\omega$, and $\rho$ meson exchanges. It has 
seven 
coupling constants, of which 
three 
are fixed by saturation properties of SNM; the remaining 
four 
free parameters can be mapped to $S_v$, $L$, the effective nucleon mass at the saturation 
density, 
$M^*$, and the $\omega$ self-interaction coupling~$\zeta$. 

The covariance parameter 
$\cov(P(\nb),\Rtyp)$ peaks around $\nb=1.65^{+1.32}_{-0.68} \,\nsat$, whereas $\cov(P({\nb}),\Rtwo)$ and $\cov(P({\nb}),\Mmax)$
peak around $\nb=2.17^{+2.14}_{-0.81} \,\nsat$ and  $\nb=3.90^{+2.00}_{-1.81} \,\nsat$, respectively. The uncertainties correspond to 50\% of the peak covariance. 
Figure~\ref{fig:cov-P} also quantifies the extent to which the central baryon densities, and the width of their distributions, increase with the NS mass.  Notably, the central baryon number densities peak at about 30\% higher density than do the peak covariance in all three cases, but the widths of the central density distributions rapidly increase with NS mass.

The correlation between the pressure $P(\nb)$ and $\Rtyp$ is strongest between $\nsat$ and $3.0\,\nsat$, as expected, and that between the pressure and $\Rtwo$ is strongest at about 40\% higher densities. Significantly, these results appear to be relatively insensitive to the details of the parametrizations. The standard deviations of both $\cov(P(\nb),\Rtyp)$ and $\cov(P(\nb),\Rtwo)$ for the 
six 
parametrizations are small, being $\sigma_{\cov,R}<0.2$ for all densities and $\sigma_{\cov,R}<0.05$ near the covariance peaks. The bottom line is these results demonstrate, at present, that \chiEFT greatly constrains $\Rtyp$ and, to a slightly lesser degree, $\Rtwo$.  
The situation is somewhat different for $\Mmax$, where pressures at densities between $2.0\,\nsat$ and $6.0\,\nsat$ dominate.  In addition, the standard deviation of cov($P(\nb),\Mmax$) among the 
six 
parametrizations are somewhat larger, being $\sigma_{\cov,\Mmax}<0.25$ at all densities and $\sigma_{\cov,\Mmax}<0.1$ near the covariance peak.  Thus, the $\Mmax$ results are more model-dependent, and the significant densities 
likely 
lie above the validity range for \chiEFT. 
However, further refinement of EFT techniques at high densities combined with Bayesian uncertainty quantification might change that situation by providing improved constraints on all three quantities, although the EFT truncation error increases rapidly beyond $\nsat$. 

\bibliography{paper_v2}
\end{document}